\documentclass[a4paper,11pt]{article}
\pdfoutput=1
\usepackage{jcappub}
\usepackage{caption}
\usepackage{hyperref}
\usepackage{amsmath}
\usepackage{amstext} 
\usepackage{mathtools}
\usepackage{tablefootnote}
\usepackage{soul}
\usepackage{ulem}
\usepackage{url}
\usepackage{color}
\usepackage{verbatim}
\usepackage{float}
\usepackage{graphicx}
\usepackage{subcaption}
\usepackage[usenames,dvipsnames,svgnames,table]{xcolor}
\usepackage{makecell}

\newcommand{\dd}{{\rm d}}
\newcommand{\LCDM}{$\Lambda$CDM }

\newcommand{\deltadm}{\delta_{\rm dm}}
\newcommand{\thetadm}{\theta_{\rm dm}}
\newcommand{\deltade}{\delta_{\rm de}}
\newcommand{\thetade}{\theta_{\rm de}}
\newcommand{\cs}{c_{\rm s}}

\newcommand{\rhode}{\rho_{\rm de}}
\newcommand{\rhodm}{\rho_{\rm dm}}

\title{J-PAS: Forecasts for dark matter - dark energy elastic couplings} 

\author[a]{David Figueruelo,}
\emailAdd{davidfiguer@usal.es}

\author[b]{Miguel Aparicio Resco,}
\emailAdd{migueapa@ucm.es}

\author[a]{Florencia A. Teppa Pannia,}
\emailAdd{f.a.teppa.pannia@usal.es}

\author[a]{Jose Beltr\'an Jim\'enez,}
\emailAdd{jose.beltran@usal.es}

\author[a]{Dario Bettoni,}
\emailAdd{bettoni@usal.es}

\author[b]{Antonio L. Maroto,}
\emailAdd{maroto@ucm.es}

\author[c]{L. Raul Abramo,}

\author[d]{Jailson Alcaniz,}

\author[f,k]{Narciso Benitez,}

\author[g,h,i]{Silvia Bonoli,}

\author[j]{Saulo Carneiro,}

\author[g]{Javier Cenarro,}

\author[g]{David Cristóbal-Hornillos,}

\author[d,l,m]{Renato A. Dupke,}

\author[n]{Alessandro Ederoclite,}

\author[g]{Carlos López-Sanjuan,}

\author[g]{Antonio Marín-Franch,}

\author[o,p,q]{Valerio Marra,}

\author[n]{Claudia Mendes de Oliveira,}

\author[f,g]{Mariano Moles,}

\author[n]{Laerte Sodr\'e Jr.,}

\author[r]{Keith Taylor,}

\author[g]{Jes\'us Varela,}

\author[g]{and H\'ector V\'azquez Rami\'o}

\affiliation[a]{Departamento de F{\'i}sica Fundamental and IUFFyM, Universidad de Salamanca, E-37008 Salamanca, Spain.}

\affiliation[b]{Departamento de F\'isica Te\'orica and Instituto de F\'isica de Part\'iculas y del Cosmos IPARCOS, Universidad Complutense de Madrid, 28040 Madrid, Spain.}

\affiliation[c]{Instituto de Física, Universidade de São Paulo, Rua do Matão 1371, CEP 05508-090, São Paulo, Brazil}

\affiliation[d]{Observatório Nacional,  Rua  General  José  Cristino,  77,  São  Cristóvão, 20921-400, Rio de Janeiro, Brazil}
\affiliation[e]{Departamento de Astronomia, Instituto de Física, Universidade Fed-eral do Rio Grande do Sul (UFRGS), Av. Bento Gonçalves 9500, Porto Alegre, R.S, Brazil}

\affiliation[f]{Instituto de Astrofísica de Andalucía - CSIC, Apdo 3004, E-18080, Granada, Spain}

\affiliation[g]{Centro de Estudios de Física del Cosmos de Aragón (CEFCA), Unidad Asociada al CSIC, Plaza San Juan, 1, E-44001, Teruel, Spain}
\affiliation[h]{Donostia International Physics Center (DIPC), Manuel Lardizabal Ibilbidea, 4, San Sebastián, Spain}
\affiliation[i]{Ikerbasque, Basque Foundation for Science, E-48013 Bilbao, Spain}

\affiliation[j]{Instituto de Física, Universidade Federal da Bahia, 40210-340, Salvador, BA, Brazil}

\affiliation[l]{Department of Astronomy, University of Michigan, 311West Hall, 1085 South University Ave., Ann Arbor, USA}
\affiliation[m]{Department of Physics and Astronomy, University of Alabama, Box870324, Tuscaloosa, AL, USA}

\affiliation[n]{Departamento  de  Astronomia,  Instituto  de  Astronomia,  Geofísicae  Ciências  Atmosféricas,  Universidade  de  São  Paulo,  São  Paulo, Brazil}

\affiliation[o]{Núcleo de Astrofísica e Cosmologia \& Departamento de Física, Universidade Federal do Espírito Santo, 29075-910, Vitória, ES, Brazil }
\affiliation[p]{INAF -- Osservatorio Astronomico di Trieste, via Tiepolo 11, 34131 Trieste, Italy}
\affiliation[q]{IFPU -- Institute for Fundamental Physics of the Universe, via Beirut 2, 34151, Trieste, Italy}

\affiliation[r]{Instruments4, 4121 Pembury Place, La Cañada-Flintridge, Ca 91011, USA}

\affiliation[k]{Abante Asesores, Plaza de la Independencia, 6, 28001 Madrid, Spain}

\abstract{We consider a cosmological model where dark matter and dark energy feature a coupling that only affects their momentum transfer in the corresponding Euler equations. We perform a fit to cosmological observables and confirm previous findings within these scenarios that favour the presence of a coupling at more than $3\sigma$. This improvement is mainly driven by cluster counts from Planck  Sunyaev-Zeldovich data that we include as a certain prior. We subsequently perform a forecast for future J-PAS data and find that clustering measurements will permit to clearly discern the presence of an interaction within a few percent level with the uncoupled case at more than $10\sigma$ when the complete survey, covering $8500$ sq. deg., is considered. We found that the inclusion of weak lensing measurements will not help to further constrain the coupling parameter. For completeness, we compare to forecasts for DESI and Euclid, which provide similar discriminating power.}

\begin{document}
\maketitle

\section{Introduction}
\label{sec:intro}
The upcoming generation of cosmological observational campaigns will crucially impact our understanding of the universe and, in particular, the physics of its dark sector. Among the available data in the near future, the on-going and planned galaxy surveys will contribute a very rich source of information able to put stringent constraints on the statistical properties of the large scale structures in the universe. Three different kinds of galaxy surveys are currently operating according to the method used for the determination of redshifts. On one hand, spectroscopic surveys that can perform high-precision redshifts measurements from high-quality spectra of a pre-selected sample of galaxies. Examples include the current BOSS~\cite{2013AJ....145...10D} and the future  spectroscopic Euclid~\cite{EUCLID} and DESI~\cite{DESI}. Second, photometric surveys that obtain photo-spectra using photometry with a reduced number of filters. These surveys can build larger catalogues of objects, but with poorer redshift accuracies than the spectroscopic ones. The best example of a current photometric survey is DES~\cite{Abbott:2005bi} and  among the future ones we have the photometric Euclid survey~\cite{EUCLID} and LSST~\cite{Mandelbaum:2018ouv}. Finally, there is a third class of surveys, the so-called spectro-photometric surveys which produce high-quality pseudospectra by combining photometry in many different  frequencies obtained through the combination of broad, medium and narrow band filters. From those pseudospectra, precise redshifts can be obtained for a high number of sources.  Among the current and future spectro-photometric surveys we have PAU~\cite{Marti:2014nha},  J-PLUS~\cite{Cenarro:2018uoy} and the future J-PAS~\cite{Benitez:2014ibt}. 

A very relevant question that galaxy surveys will be capable of unveiling is the existence and nature of a coupling among the dark components. In this respect, many possibilities have been explored in the literature~\cite{Wetterich:1994bg,Amendola:1999er,Farrar:2003uw,Wang:2016lxa,Pourtsidou:2013nha,Koivisto:2015qua,PhysRevD.64.063501,Benetti:2019lxu,Boehmer:2015kta,Salzano:2021zxk,Boehmer:2015sha}, most of which modify the background evolution of the universe and, consequently, they are already constrained by geometrical tests. In this work, we are interested in another class of couplings with the remarkable property of leaving the background evolution unmodified and only affecting the perturbations sector. In particular, they impact the clustering of dark matter so that the aforementioned galaxy surveys will provide a crucial tool to test these couplings by means of confronting the inferred statistical properties of the dark matter distribution with the collected data. These couplings have been considered in the literature arising from different scenarios, e.g., an elastic Thomson-like scattering between dark matter and dark energy~\cite{Simpson:2010vh,Baldi:2014ica,Baldi:2016zom,Kumar:2017bpv}, from a fluid description of dark matter interacting with dark energy (dubbed pure momentum exchange)~\cite{Skordis:2015yra,Pourtsidou:2016ico,Amendola:2020ldb,Kase:2019veo,Kase:2019mox,Chamings:2019kcl,Jimenez:2020npm} or, from a more phenomenological perspective, as a coupling proportional to the relative velocities of the dark components~\cite{Asghari:2019qld}. In all the cases, the background and the continuity equations of the perturbations are not modified and only the Euler equations of dark matter and dark energy exhibit new terms driven by the coupling. It is remarkable that these couplings provide promising prospects to alleviate the existing tension in the amplitude of the dark matter clustering as measured from low redshift data~\cite{Hildebrandt:2018yau,Asgari:2020wuj,Troxel:2017xyo,Abbott:2021bzy} and the one inferred from CMB observations~\cite{PLANCK2018}. 

In this work, we will focus on the class of models where dark matter and dark energy feature a coupling proportional to their relative velocities that will also serve as proxy for more general scenarios where similar modified equations arise. By the very construction of the coupling and by virtue of the Cosmological Principle, the background evolution is unaffected and only the Euler equations of the perturbations are  modified at first order in cosmological perturbations. For this model we will perform a fit to current cosmological observables to confirm previous findings with updated data, namely that a non-vanishing coupling parameter is favoured by data, with the uncoupled case at more than $3\sigma$. This result is only obtained when including a prior based on cluster counts from Sunyaev-Zeldovich data. We will discuss the consistency of including this data in the form of a prior. In any case, this intriguing result motivates to carry out a forecast with future galaxy surveys to discern to which level of accuracy they will help to discriminate the existence of this coupling in the dark sector. As we will show, they will indeed play a crucial role to probe the existence of a coupling in the dark sector.


\section{Elastic couplings in the dark sector}
\label{sec:themodel}
In this section we will introduce the relevant equations that govern the dynamics of the model under consideration and define its relevant parameters. We will not enter into the details of the model and we refer to~\cite{Asghari:2019qld} for a more comprehensive discussion.

We will assume that the content of the universe can be described by a stress energy tensor of the perfect fluid form
\begin{equation}
T^{\mu\nu}=\sum_{i}\Big[(\rho_i +p_i)u_i^\mu u_i^\nu +p_ig^{\mu\nu}\Big]\;,
\end{equation}
where the sum runs over all the components in the universe, namely baryons, photons and neutrinos together with a dark sector that we assume conformed by cold dark matter and dark energy with a constant equation of  state parameter $w$. While the ordinary matter sector satisfies the usual conservation laws $\nabla_\mu T^{\mu\nu}=0$, the dark sector will be assumed to satisfy the non-conservation equations
\begin{eqnarray}
\nabla_\mu T^{\mu \nu}_{\rm dm}&=&Q^\nu\;,\\
 \nabla_\mu T^{\mu \nu}_{\rm de}&=&-Q^\nu\;,
\end{eqnarray}
where $Q^\nu$ describes the interaction that is proportional to the relative velocity as follows
\begin{equation}
Q^{\nu}=\bar{\alpha}\left(u^\nu_{\rm de}-u^\nu_{\rm dm}\right)\;.
\end{equation}
The parameter $\bar{\alpha}$ measures the strength of the interaction and can depend both on time and space, although, for the sake of simplicity, throughout this work we will mostly assume it constant. From the form of the coupling, it is clear that the background equations are not modified because both dark components have the same background 4-velocities. The perturbed metric in the Newtonian gauge reads
\begin{equation}
\dd s^2=a^2(\tau) \Big[-(1+2 \Psi)\dd \tau^2 + (1-2 \Phi)\dd \vec{x}\,^2\Big] \;,
\end{equation}
and the equations governing the dark sector perturbations, in the absence of anisotropic stress so that $\Phi=\Psi$, are the following:
\begin{eqnarray}
\label{eq:deltaDM}
\deltadm'&=&-\thetadm + 3 \Phi' \;, \\
\label{eq:thetaDM}
\thetadm'&=&-\mathcal{H} \thetadm + k^2 \Phi + \Gamma(\thetade-\thetadm)\;, \\
\deltade'&=&-3 \mathcal{H}\left(\cs^2-w\right)\deltade +3(1+w)\Phi'  -\thetade(1+w)\left(1+9 \mathcal{H}^2\frac{\cs^2-w}{k^2}\right) \;, \\
\label{eq:thetaDE}
\thetade'&=&\left(-1+3\cs^2\right) \mathcal{H}\thetade+k^2\Phi +\frac{k^2 \cs^2}{1+w}\deltade-\Gamma R(\thetade-\thetadm)\;,
\end{eqnarray}
where we have defined the usual density contrast $\delta \equiv \delta\rho/\rho$ and Fourier velocity perturbation $\theta \equiv i \vec k\cdot \vec v$. Throughout this work, the dark energy sound speed will be assumed to be $\cs=1$. Moreover, we have introduced the quantities
\begin{eqnarray}
\Gamma&\equiv& \bar{\alpha}\frac{a}{\rhodm} \;, \\ \label{eq:Scoupling}
R&\equiv&\frac{\rhodm}{(1+w)\rhode} \;. \label{eq:Rcoupling}
\end{eqnarray}
The quantity $\Gamma$ can be interpreted as the interaction rate between dark matter and dark energy. Thus, the relative hierarchy between $\Gamma$ and the Hubble expansion rate $\mathcal{H}$ will determine the relevance of the coupling. For a constant $\bar{\alpha}$, we can see that $\Gamma\propto a^{4}$ so that it grows as the universe expands. Hence, the coupling is irrelevant at sufficiently high redshift and it only affects the evolution of the perturbations at low redshift when $\Gamma\gtrsim\mathcal{H}$. On the other hand, the dark matter-to-dark energy ratio $R$ measures the relative importance of the coupling for dark matter and dark energy according to their relative abundances. Since this quantity can be written as $R=\frac{\Omega_{\rm DM}}{(1+w)\Omega_{\rm DE}}a^{3w}$, we can see that it is large throughout the universe evolution so that the coupling is larger in the dark energy equations. However, it is the effect on the dark matter perturbations that will impact their evolution (see~\cite{Asghari:2019qld}). Finally, since the coupling parameter $\bar{\alpha}$ has mass dimension 5, it is convenient to normalise it appropriately and work with the dimensionless parameter $\alpha$ defined as
\begin{equation}
\alpha\equiv \bar{\alpha} \frac{8  \pi G}{3 H_0^3} \; .
\end{equation}
As we will see below, the normalisation has been chosen so that its natural value is $\mathcal{O}(1)$. 

After this brief review of the theoretical model, we  proceed to numerically analyse the evolution of the perturbations in the next Section.

\section{Numerical analysis}
\label{sec:numerical}

In order to gain some intuition on the observable effects of the coupling, we have solved the system of equations \eqref{eq:deltaDM}-\eqref{eq:thetaDE} numerically. To do so we have modified the publicly available code  \texttt{CLASS}~\cite{Julien:CLASS} for the computation of the linear perturbations, using the Newtonian gauge. As explained in Section~\ref{sec:themodel}, the background cosmology remains unaffected and, therefore, we only need to modify the code by adding the new interacting term in the Euler equations, for both DE and DM. As a consequence, any deviation from the standard model is only due to the modification in the perturbation sector.

The model has also been explored modifying the publicly available code \texttt{CAMB}~\cite{Lewis:1999bs}, in which the evolution of the cosmological perturbations is computed in the synchronous gauge. The comparison of gauge-invariant quantities then ensures the consistency of both numerical implementations. It is worth mentioning, however, that in the case of \texttt{CAMB} some extra considerations must be taken into account to correctly include the new interaction. Although only modifications in the velocity equations of the dark sector are needed also in the synchronous gauge  (see Appendix~\ref{ap:sync}), using the residual gauge freedom of this gauge to set $\thetadm=0$ at all times is no longer possible, precisely because the interaction sources the evolution equation for $\thetadm$. Therefore, a new evolution variable must be introduced to deal with the DM velocity evolution. A potential problem with this approach is that the gauge remains partially unfixed so we could obtain unphysical gauge solutions. This could be avoided by enlarging the content of the universe with a marginal dust component that can be used to completely fix the gauge (this approach has been used in e.g.~\cite{Lesgourgues:2015wza}). We have corroborated however that the obtained numerical solutions correspond to physical modes and any potential gauge mode remains subdominant. 

We will fix a fiducial model for our analysis in this Sec. with the cosmological parameters to $H_0=67.4$\,km/s/Mpc, $\Omega_{\rm b} h^2=0.0224$, $\Omega_{\rm dm}h^2=0.120$, $w=-0.98$ and $\cs^2=1$. We choose as reference model a $w$CDM rather than \LCDM to clearly isolate the effects of the interaction and to avoid any degeneracies with the DE equation of state resulting in a cosmological constant or not,  as the interaction requires $w\neq-1$ and the cosmological constant does not have perturbations. Any other parameter is set to the default value of each code.

Before proceeding to the discussion of the results obtained from the modified numerical codes, it is convenient to warn that we are only considering the linear evolution and our statements in the following refer to that regime. The gravitational collapse induces non-negligible non-linear effects on small scales and the scale at which non-linearities become important shifts towards larger scales throughout the universe evolution so the linear regime becomes valid up to smaller values of $k$ at lower redshift. This non-linear evolution will modify our findings so our statements only hold up to the scale of non-linearity. Beyond that scale, we should resort to proper $N$-body simulations to draw reliable conclusions. Simulations for a similar model were performed in e.g.~\cite{Baldi:2014ica,Baldi:2016zom}, although we find that the different time-evolution of the coupling constant casts doubts on the possibility of making a direct translation of their results to our case.

\subsection{Constant coupling}
\label{sec:ctacoupling}
Let us start by revisiting the scenario studied in Ref.~\cite{Asghari:2019qld}, where the coupling parameter has neither time nor scale dependence: $\alpha{(\tau,k)}=\alpha$.  In the following, we will perform a somehow more extensive analysis of the perturbations evolution from the numerical solutions obtained by using the aforementioned cosmological parameters and study the dependence on the coupling  $\alpha$.

\subsubsection{Matter and CMB power spectra}
\label{sec:Pk-CMB}
The elastic coupling is driven by the relative velocity between DM and DE (eq.~\eqref{eq:thetaDM} and eq.~\eqref{eq:thetaDE}, respectively), therefore, distinctive behaviours with respect to a non interacting model are expected to appear on small scales ($k\gg \mathcal{H}$), since for large scales both DM and DE share the same rest frame. Also, it is important to notice that the background does not differ from the uncoupled $w$CDM model, hence, all the effects are genuinely due to the modified perturbation sector.

\begin{figure}
	\centering
	\includegraphics[scale=0.16745]{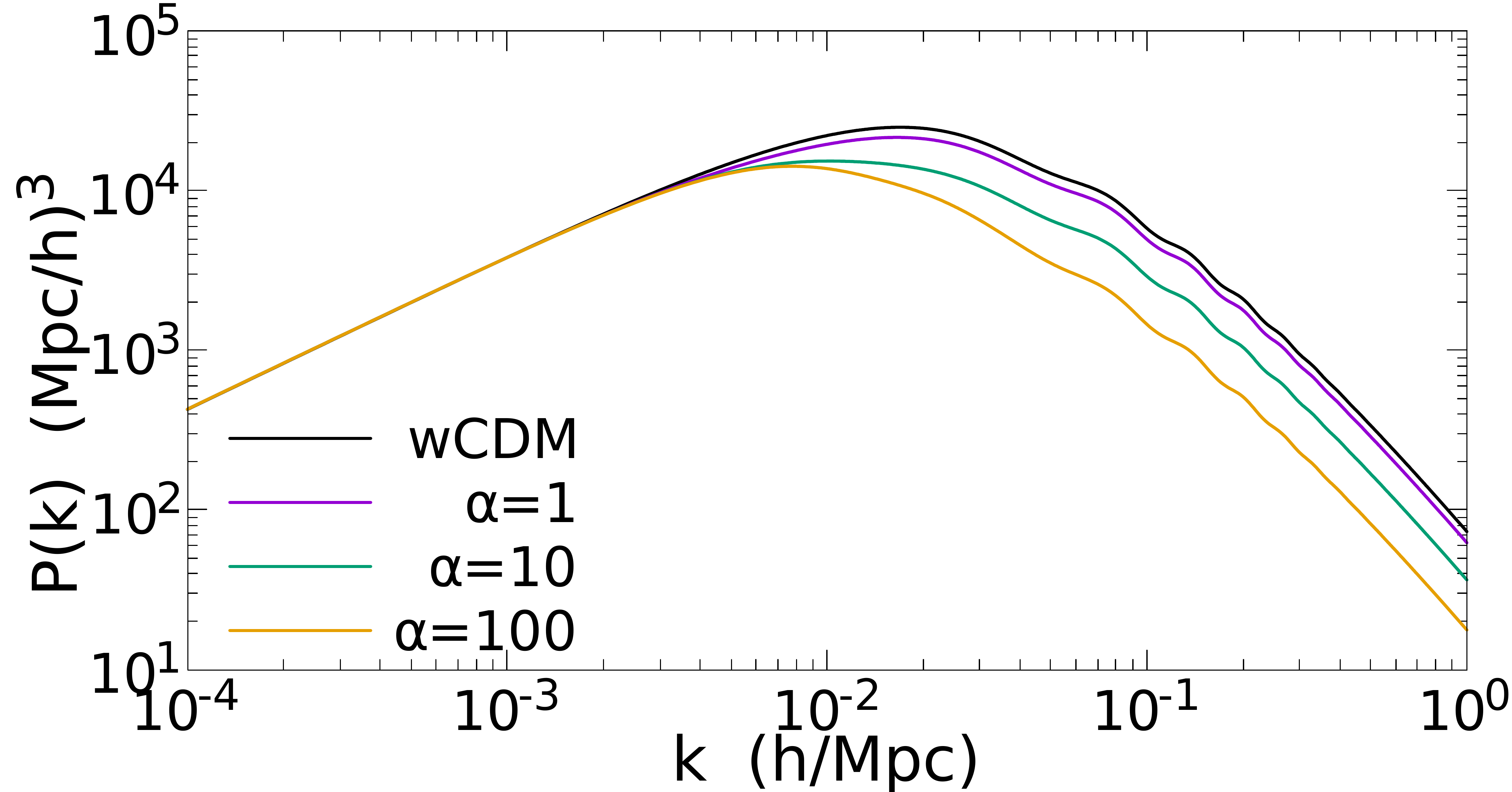}
	\includegraphics[scale=0.16745]{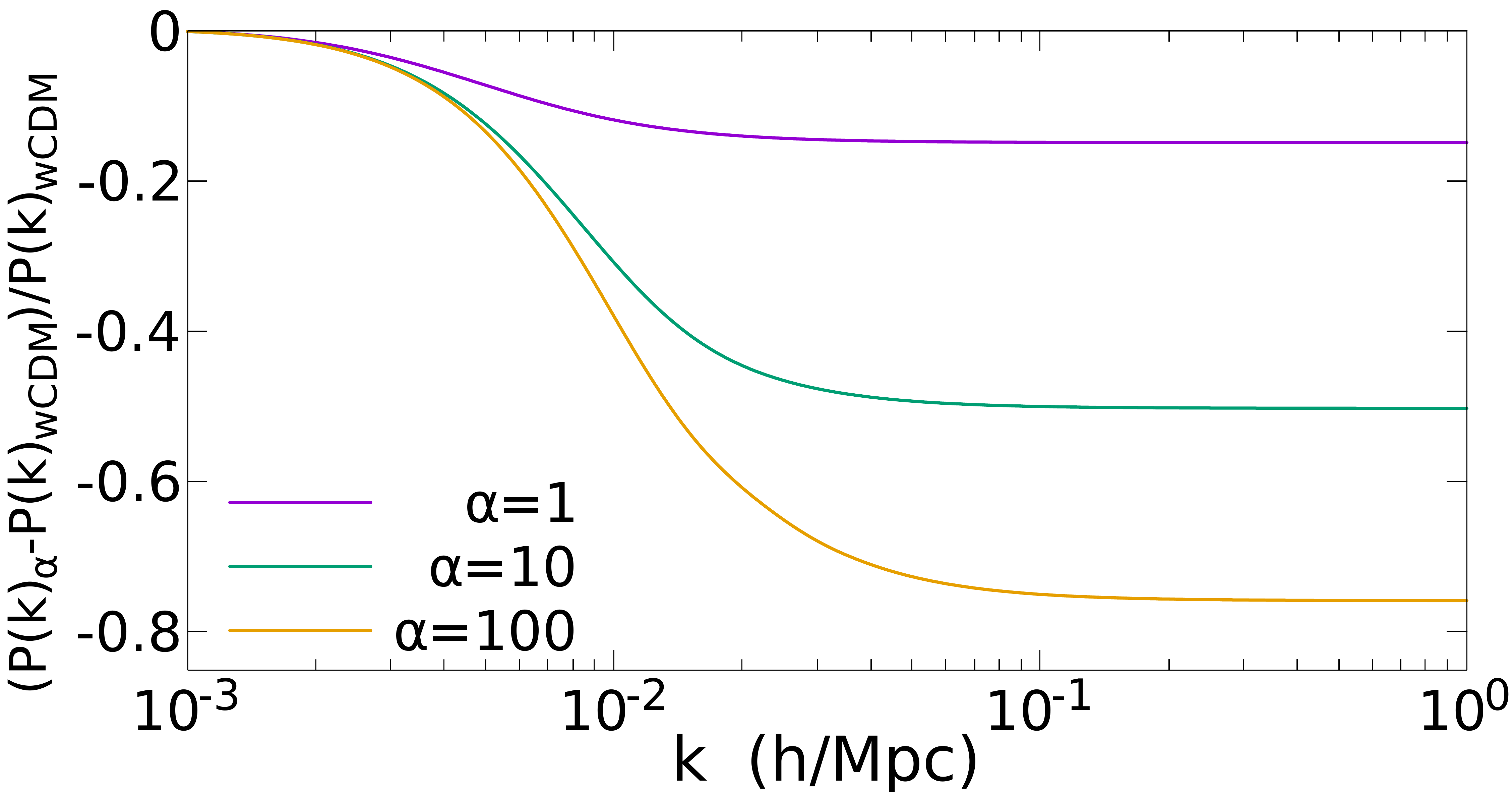}
	\caption{In the left plot, we show the matter power spectrum for several values of the coupling parameter $\alpha$ and for a $w$CDM model with the same cosmological parameters. In the right plot, we show the relative ratio of the matter power spectrum using different values of the coupling parameter $\alpha$.}
	\label{fig:PK}
\end{figure}

In Figure~\ref{fig:PK}, we show the matter power  spectrum for different values of the coupling parameter $\alpha$ as well as its relative ratio with respect to the $w$CDM model for the same cosmological parameters. As commented above, the interaction leaves no imprint on large scales and only the small scales are affected where we see a suppression of power. This suppression exhibits two  different regimes, namely: while for intermediate scales it shows a $k$-dependence, when we go to smaller scales it saturates becoming $k$-independent. The suppression in the matter power spectrum is larger as the value of the coupling parameter $\alpha$ is increased.  A remarkable property of the coupling is its ability to shift the turnover of the matter power spectrum without modifying the matter-radiation equality time, i.e., for fixed matter and radiation density parameters. Notice the resemblance of the suppression to the case of massive neutrinos, which however does not modify the turnover for fixed radiation abundance. In the next Section~\ref{sec:structures}, we will understand these features by looking at the evolution of the perturbations.

\begin{figure}
	\centering
	\includegraphics[scale=0.27]{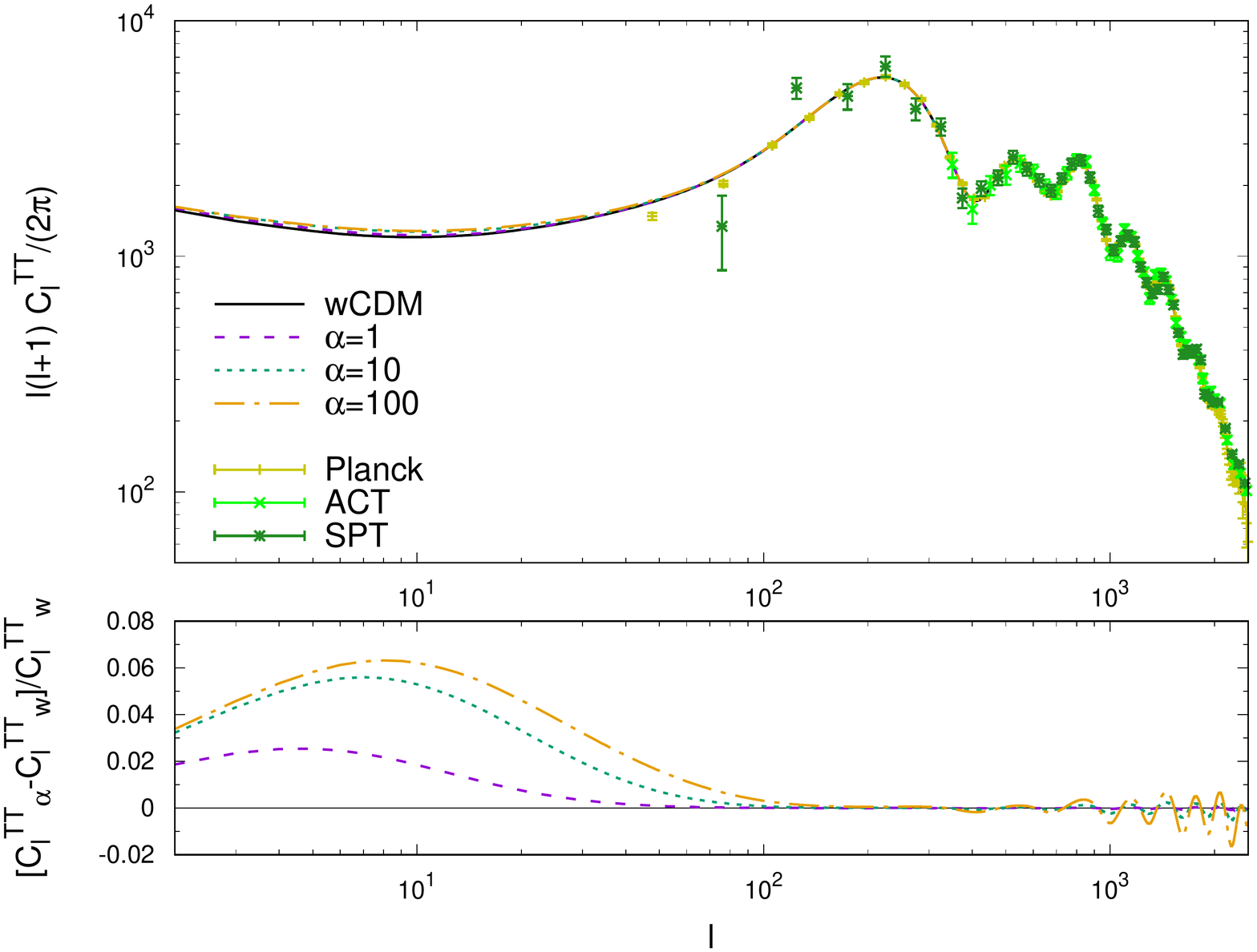} 
	\hspace{0.1cm}
	\includegraphics[scale=0.27]{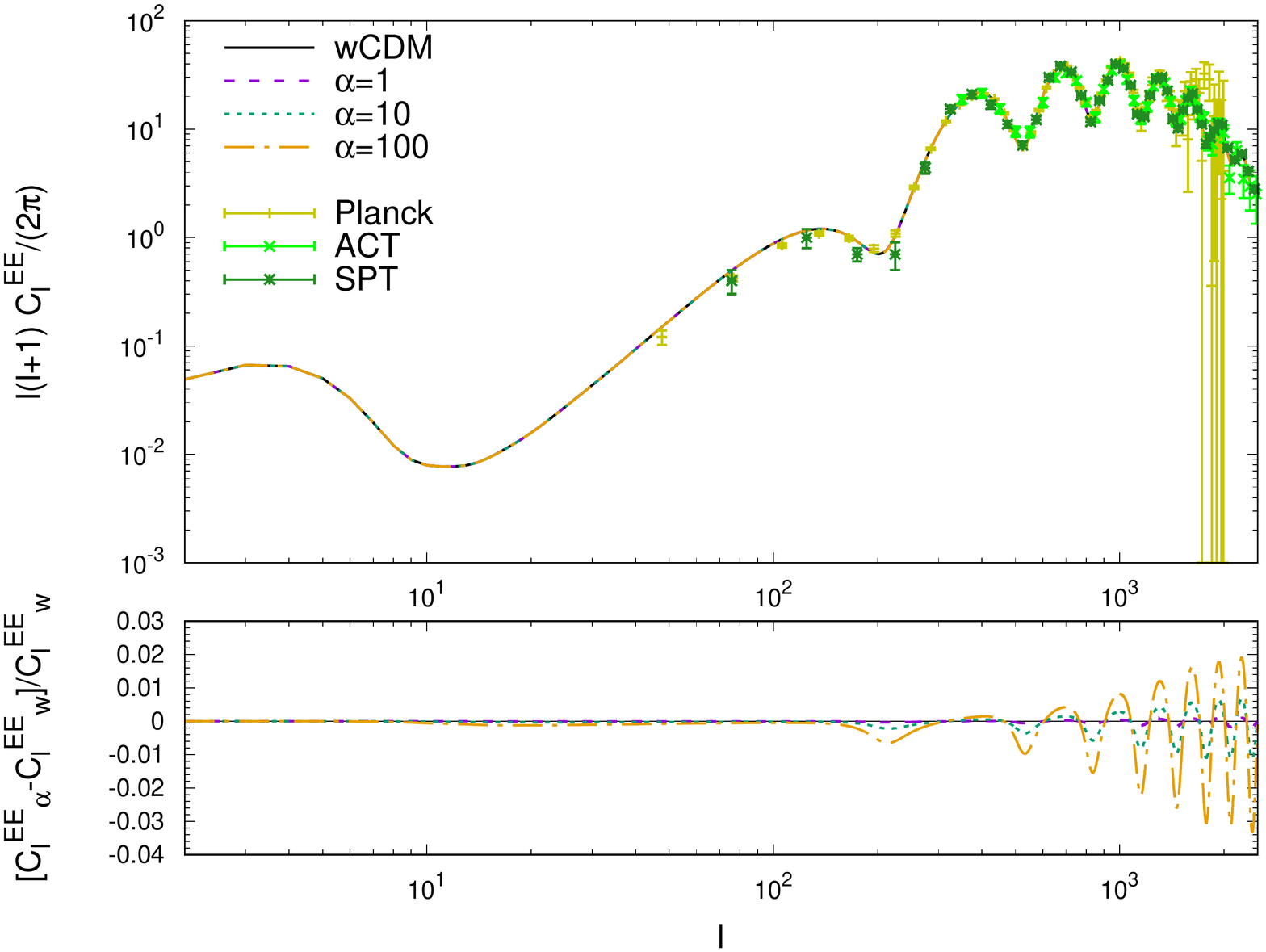}
	\includegraphics[scale=0.27]{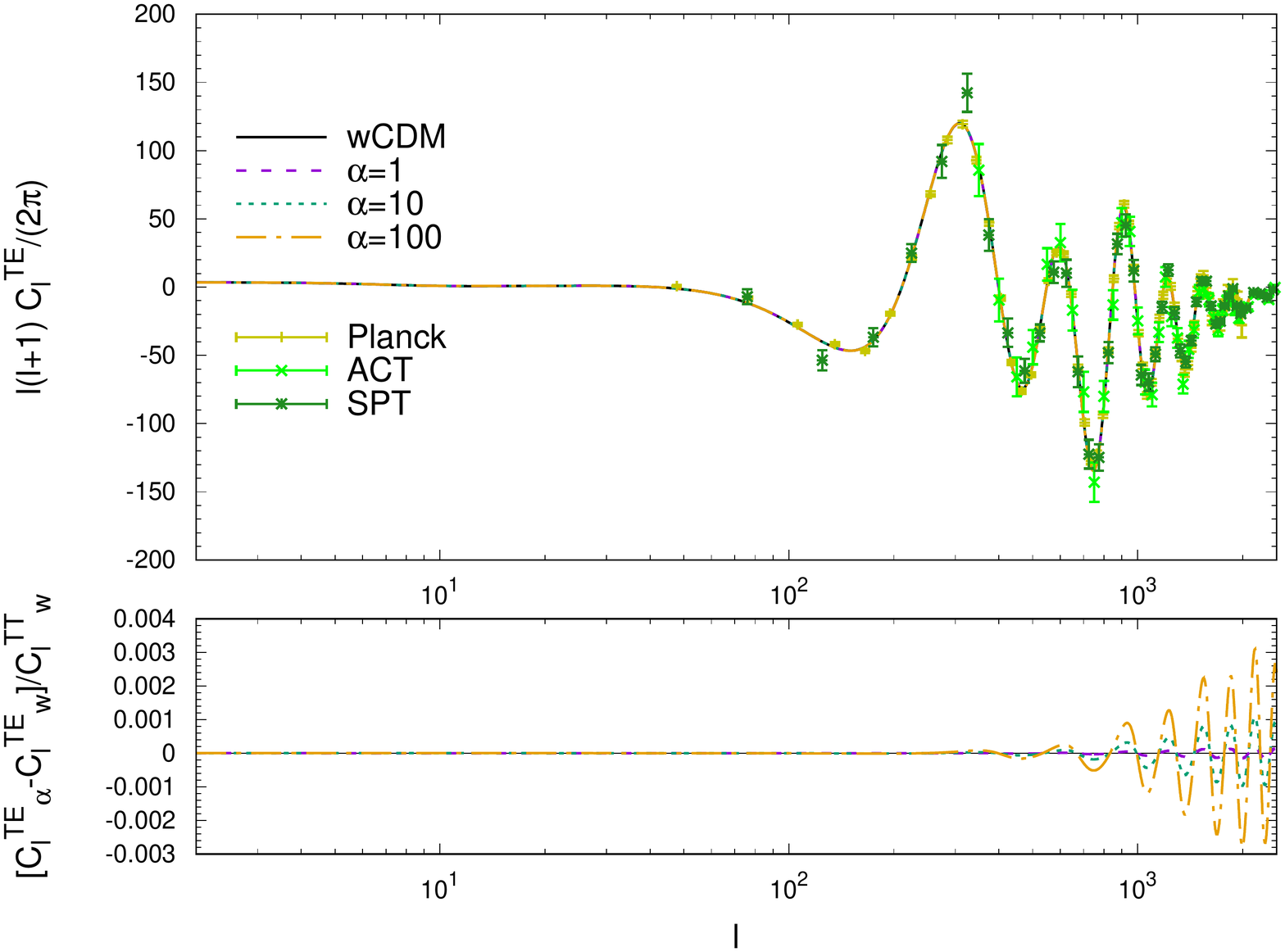}
	\hspace{0.1cm}
	\includegraphics[scale=0.27]{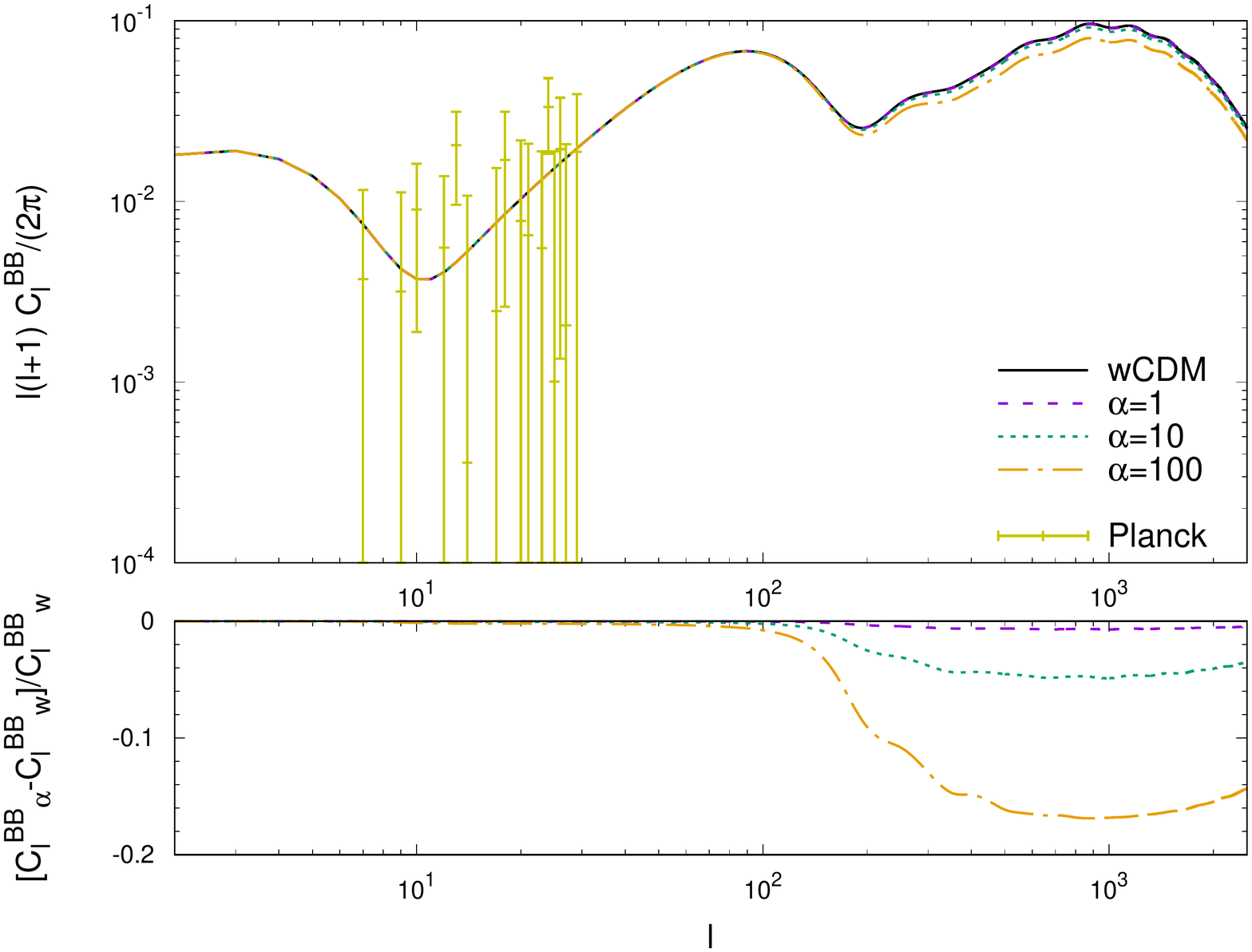}
	\caption{The CMB angular power spectra are presented for different values of the parameter $\alpha$. The normalisation of relative errors is presented in all cases with respect to a $w$CDM model ($\alpha=0$), and temperature and polarisation data from Planck~\cite{Planck2019V},
ACT~\cite{ACT2017} and SPT~\cite{SPTpol2018} are also included.}
	\label{fig:Cl}
\end{figure}

\begin{figure}
	\centering
	\includegraphics[scale=0.27]{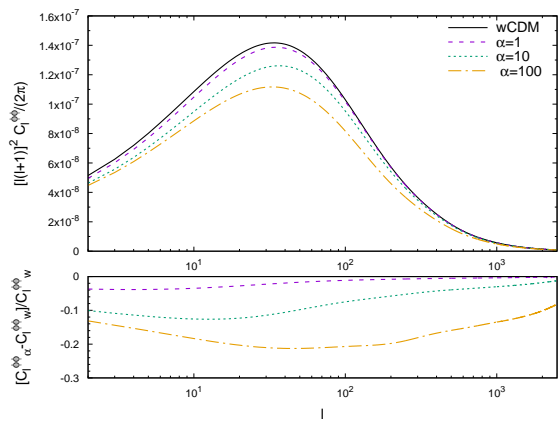}
	\caption{The amplitude of the lensing potential is sensitive to the parameter $\alpha$, leading to modifications on the high-$\ell$ oscillations of both the temperature and polarisation power spectra.}
	\label{fig:lenpot}
\end{figure}

The effects of the elastic coupling on the CMB power spectra for temperature, polarisation and cross-correlations are shown in Figure~\ref{fig:Cl} for different values of the parameter $\alpha$, together with the current available CMB temperature and polarisation data from the Planck Collaboration~\cite{Planck2019V}, the Atacama Cosmology Telescope~\cite{ACT2017} and the South Pole Telescope~\cite{SPTpol2018}. The power spectrum for temperature is most significantly modified at large scales through the Integrated Sachs–Wolfe (ISW) effect, as expected because the interaction is relevant at very late times. The polarisation power spectra are however mainly modified at small scales increasing the amplitude of the high-$\ell$ oscillations due to lensing effects (these oscillations in the high-$\ell$ sector are also visible in the temperature power spectrum) and a non-oscillating correction in the $BB$ power spectrum.  Since the coupling reduces the clustering of matter, the lensing potential decreases as $\alpha$ increases, as shown in Figure~\ref{fig:lenpot}.

\subsubsection{Growth of structures: lowering the value of \texorpdfstring{$ \sigma_8 $}{TEXT} }
\label{sec:structures}
When the interaction is efficient, DM and DE become tightly coupled so they tend to form a locked system, thus preventing the clustering of DM due to the DE pressure that causes the suppression in the matter power spectrum explained above. In order to gain a more thorough understanding of the underlying mechanism, we will analyse how the growth of structures is modified due to the coupling. With this in mind, this section is devoted to the study of the DM density contrast evolution as well as the Newtonian potential $\Phi$. We will then explore the impact on the parameter $\sigma_8$, which will be relevant for the fit to data carried out in Section~\ref{Sec:Fit}.

\begin{figure}
	\centering
	\includegraphics[scale=0.1175]{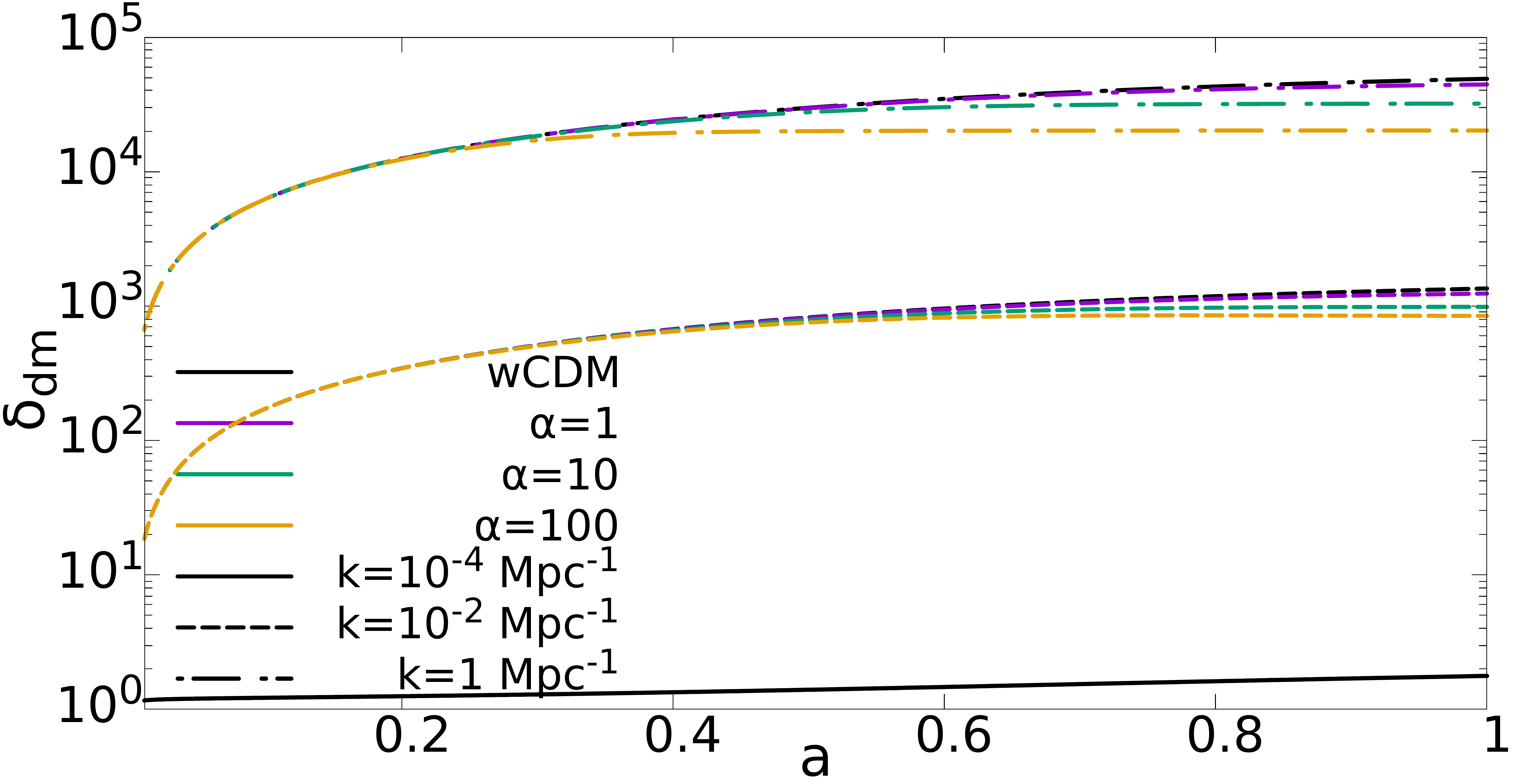}
	\includegraphics[scale=0.1175]{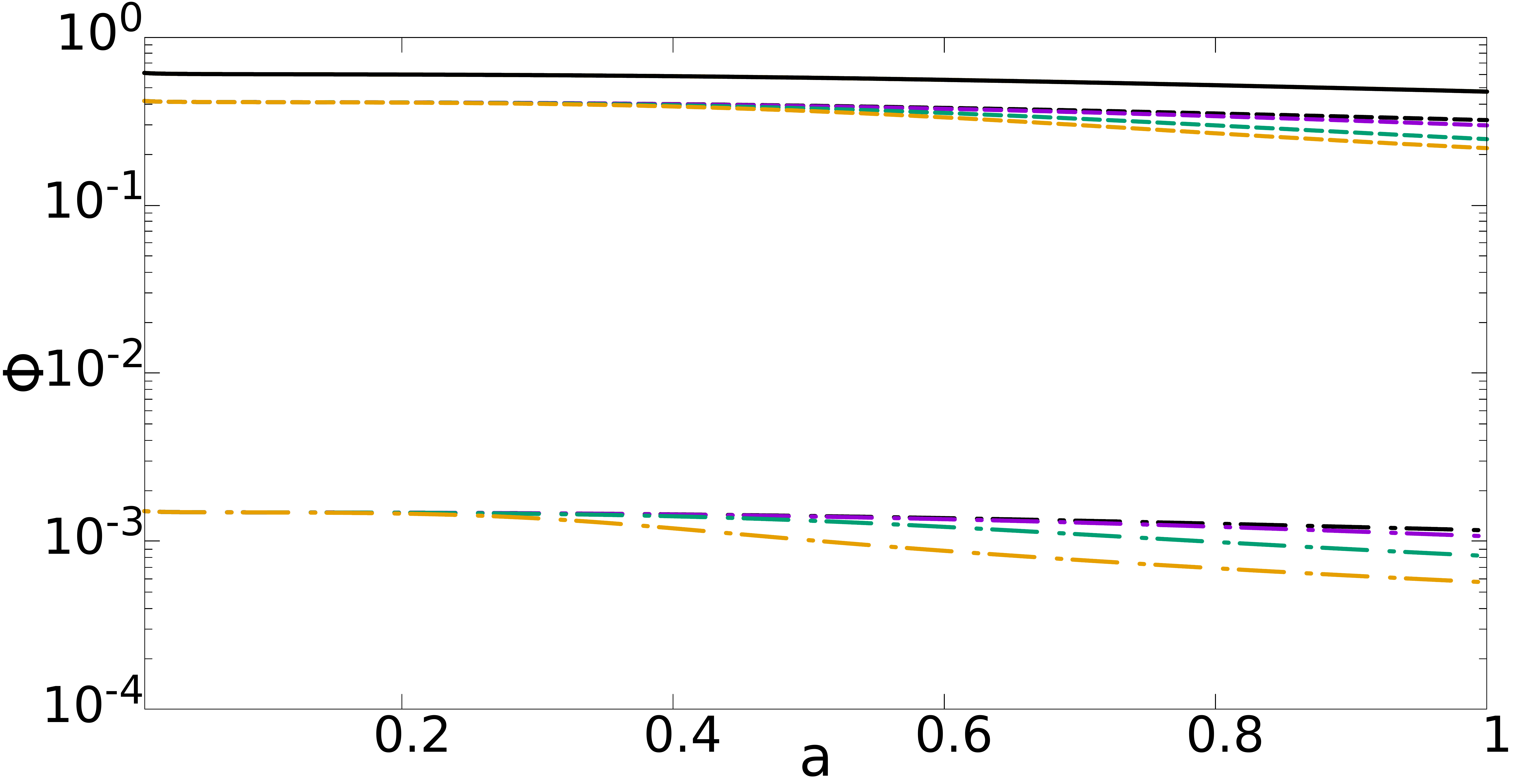}
	\caption{Evolution with the scale factor of the DM density contrast $\deltadm$ (left) and the Newtonian potential $\Phi$ (right) for different values of the coupling constant $\alpha$ and different modes. For these plots, we have used the Newtonian gauge. The black line represents the reference $w$CDM model while the purple, green and yellow lines are the elastic coupling model with $\alpha=1,10$ and $100$ respectively. Solid lines represents the mode $k=10^{-4}$\,Mpc$^{-1}$, dashed lines the mode $k=10^{-2}$\,Mpc$^{-1}$ and dash-dotted lines the mode $k=1$\,Mpc$^{-1}$.}
	\label{fig:delta_dm}
\end{figure}

In Figure~\ref{fig:delta_dm}, we display the evolution of $\deltadm$ and $\Phi$ when the elastic coupling is efficient, that is, when $\Gamma\gg\mathcal{H}$. When the interaction dominates, at low redshift and small scales, the DM density contrast ceases growing and it saturates to a constant value. Concerning  the Newtonian potential, the coupling induces an additional decrease, which will affect the late ISW effect. The freezing of the DM density contrast explains the suppression on the matter power spectrum shown in Figure~\ref{fig:PK} and it also explains why the gravitational potential exhibits a more pronounced decrease at late times. 

\begin{figure}
	\centering
	\includegraphics[scale=0.24]{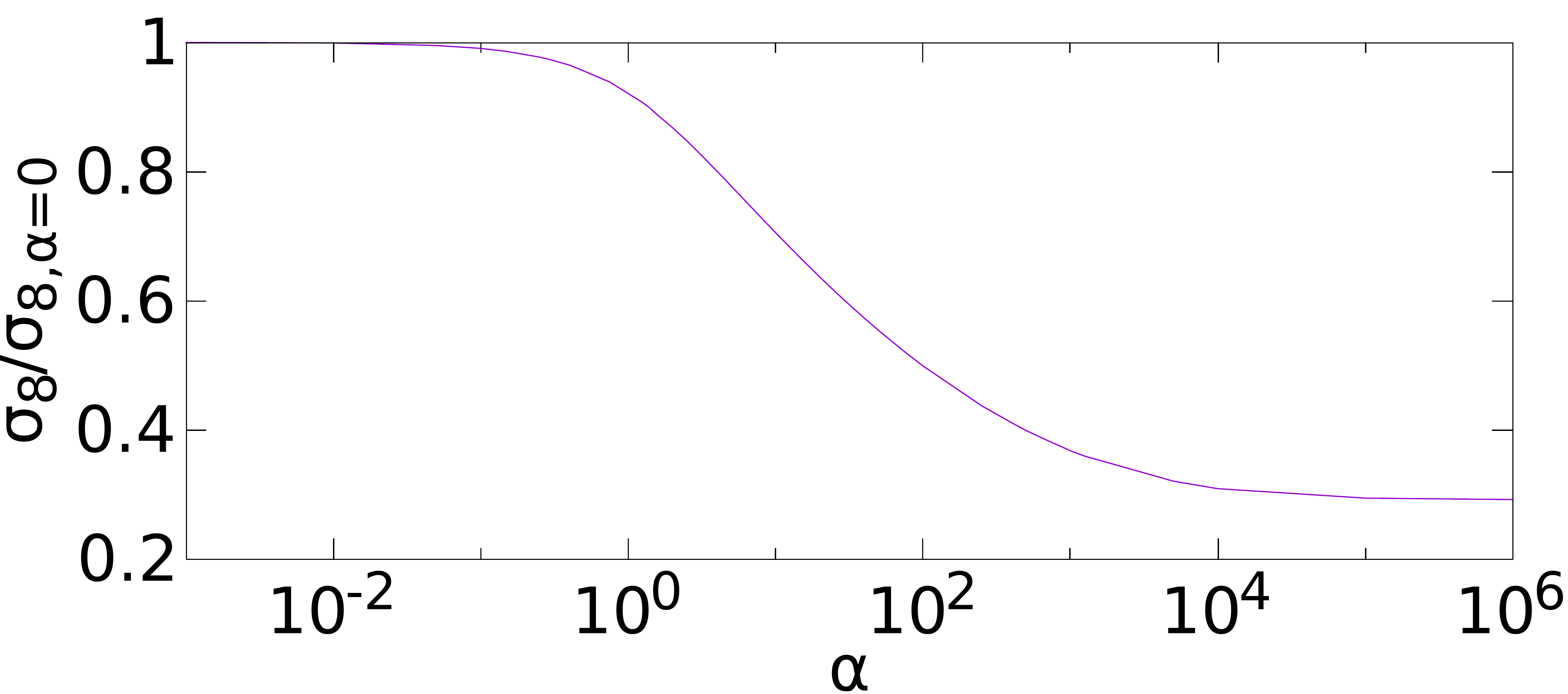}
	\caption{Evolution of the ratio of the value of $\sigma_8$ compared to a non interacting model value $\sigma_{8,\alpha=0}$ using the same cosmological parameters, for different values of the coupling parameter $\alpha$. }
	\label{fig:sigma8}
\end{figure}

A direct consequence of the explained features is that the amplitude of matter fluctuations is reduced. This can be clearly seen in Figure~\ref{fig:sigma8}, where we plot the evolution of the ratio between the values of $\sigma_8$ for the coupled and uncoupled cases, keeping fixed the remaining cosmological parameters. Since the available amount of DM is limited, the suppression in $\sigma_8$ saturates to $\sim30\%$ of the non interacting value as we consider very large values for $\alpha$. The very same reason explains why, in the analogous case of interactions with baryons, this saturation is smaller as shown in~\cite{Jimenez:2020ysu}.

\begin{figure}
	\centering
	\includegraphics[scale=0.235]{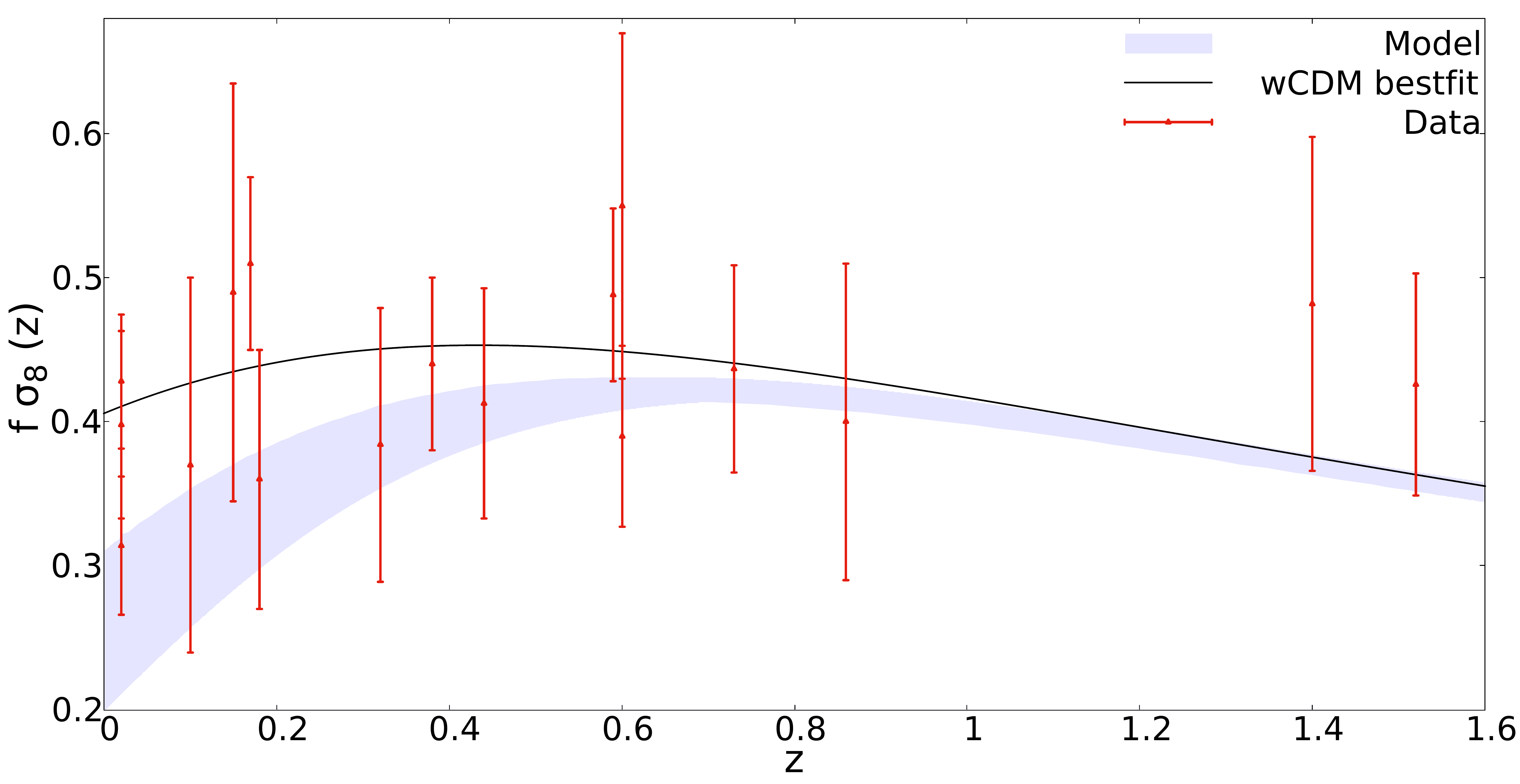}
	\caption{Evolution of $f\sigma_8(z)$ for the reference model $w$CDM (black line), for the interacting model (blue area) and data in Ref.~\cite{Quelle:2019vam} (red). The blue area encloses $f\sigma_8(z)$ evolution for values of the coupling parameter $\alpha$ inside the 1$\sigma$ region of the fit of Table~\ref{tab:alphacta} $\alpha \in [0.675,1.265]$, and for the different modes $k \in [0.007,1]\;h/$Mpc in order to take into account the scale dependence of $f\sigma_8(z)$ for this model. The other cosmological parameters are set to the mean value of the fit of Table~\ref{tab:alphacta}, while the derived parameters are calculated by the code in each case. The data points are only for an illustrative goal, as they are obtained for several scales but assuming a $k$-independent model, while in our case $f\sigma_8(z)$ is $k$-dependent. Thus, before drawing any conclusion from the fit to these data, a more appropriate re-compilation of the points by including the scale-dependence of the model would be necessary.}
	\label{fig:fsigma8}
\end{figure}

\subsubsection{Velocities evolution: new peculiar ones and Dark Acoustic Oscillations}
\label{Sec:velocity}

\begin{figure}
	\centering
	\includegraphics[scale=0.117]{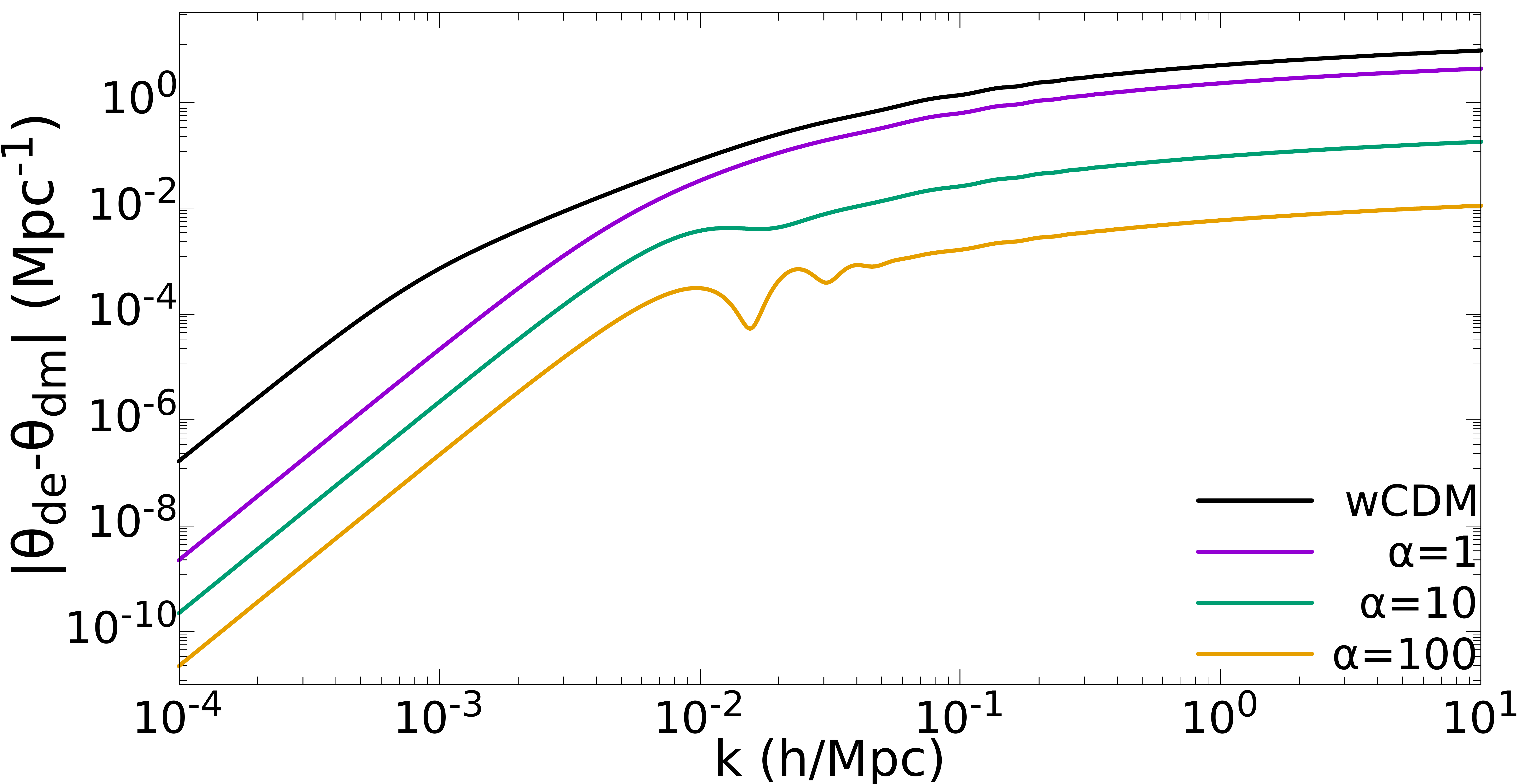}
	\includegraphics[scale=0.117]{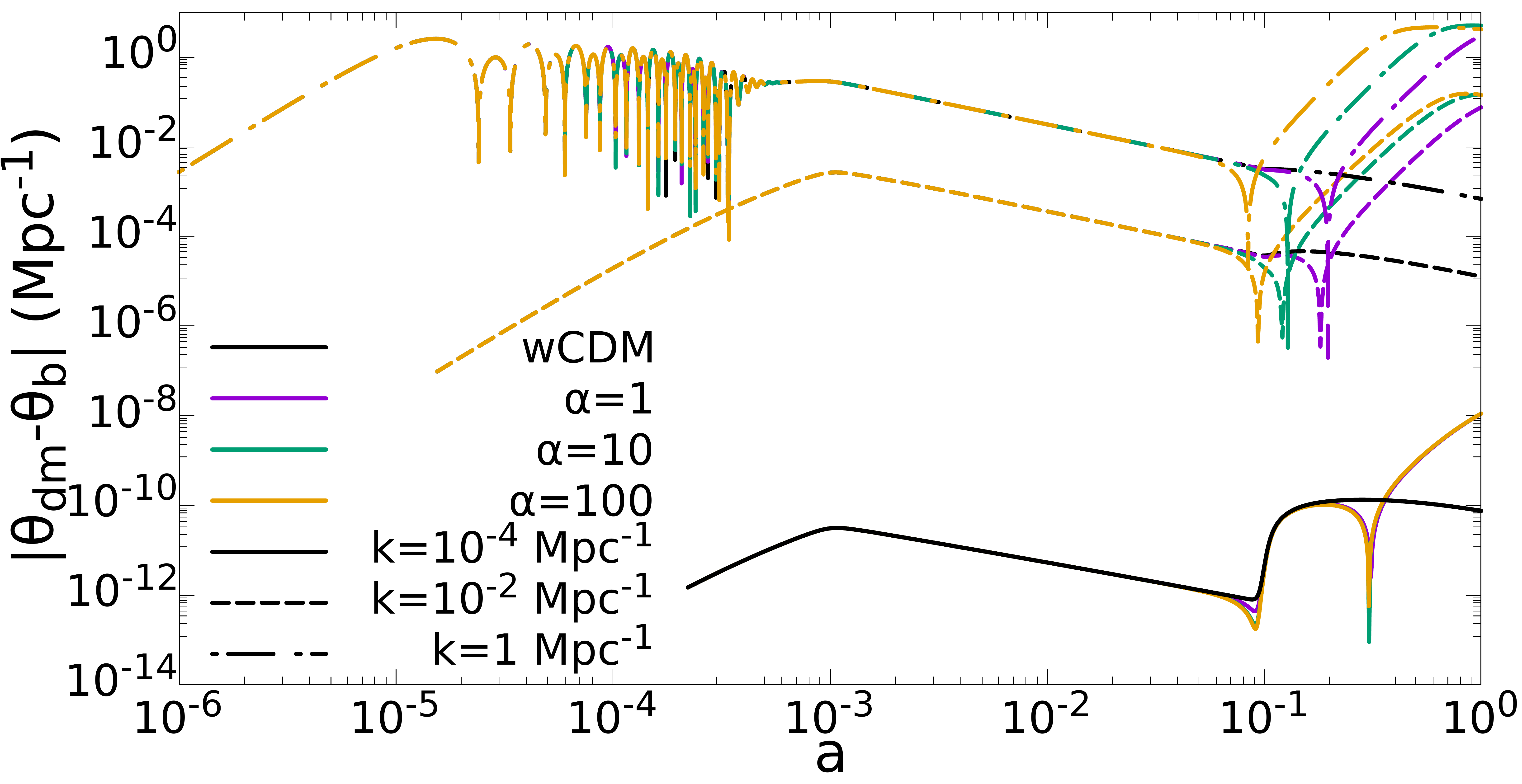}
	\caption{Relative velocity today between DE and DM today for different scales with and without interaction (left) and evolution of the relative velocity between matter components for different values of the coupling constant $\alpha$ and different modes (right).  The black line represents the reference $w$CDM model while the purple, green and yellow lines are the elastic coupling model with $\alpha=1,10$ and $100$ respectively. Solid lines represents the mode $k=10^{-4}$\,Mpc$^{-1}$, dashed lines the mode $k=10^{-2}$\,Mpc$^{-1}$ and dash-dotted lines the mode $k=1$\,Mpc$^{-1}$. }
	\label{fig:theta}
\end{figure}

\begin{figure}
	\centerline{
		\includegraphics[scale=0.165]{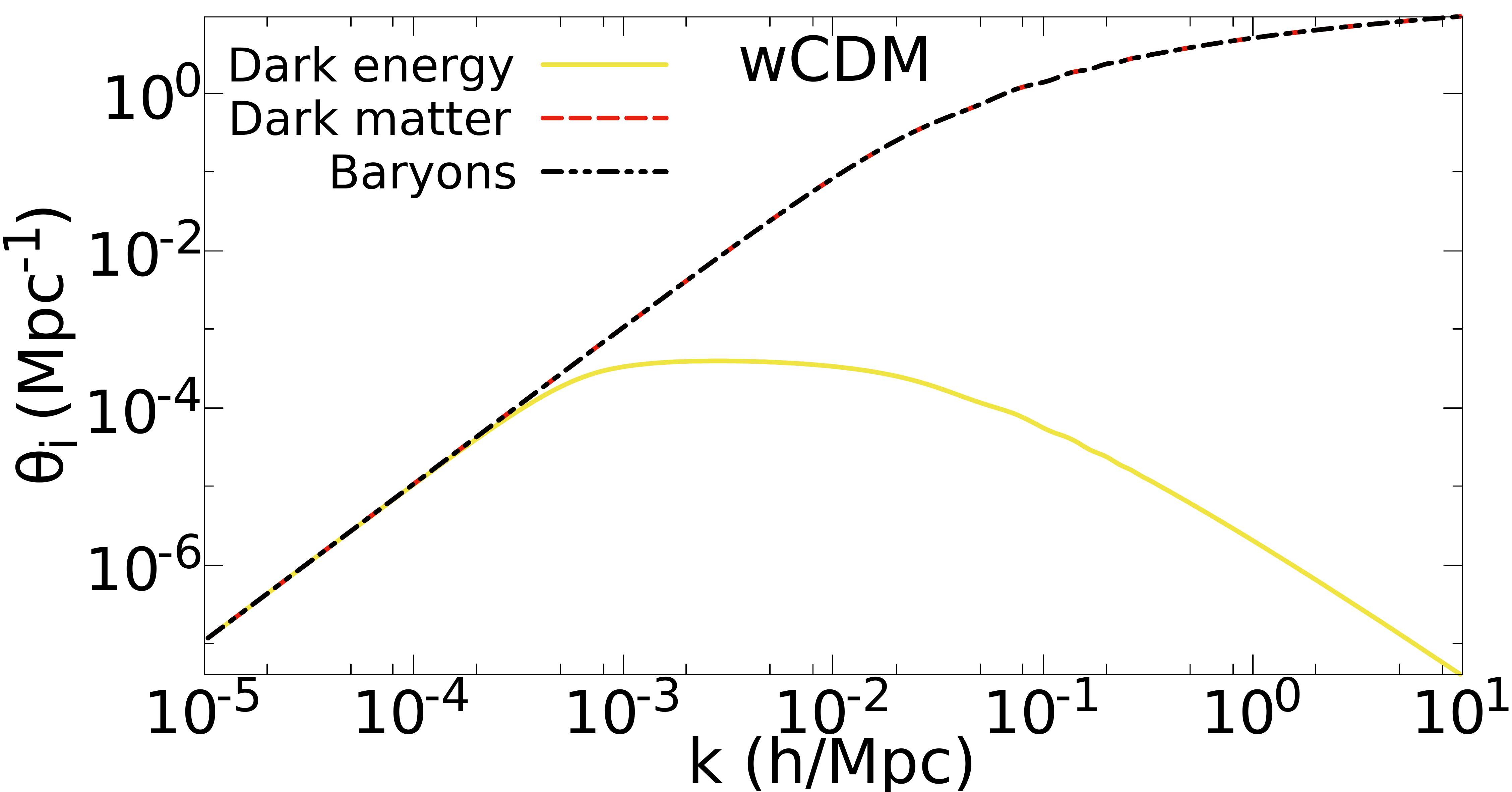}
		\includegraphics[scale=0.165]{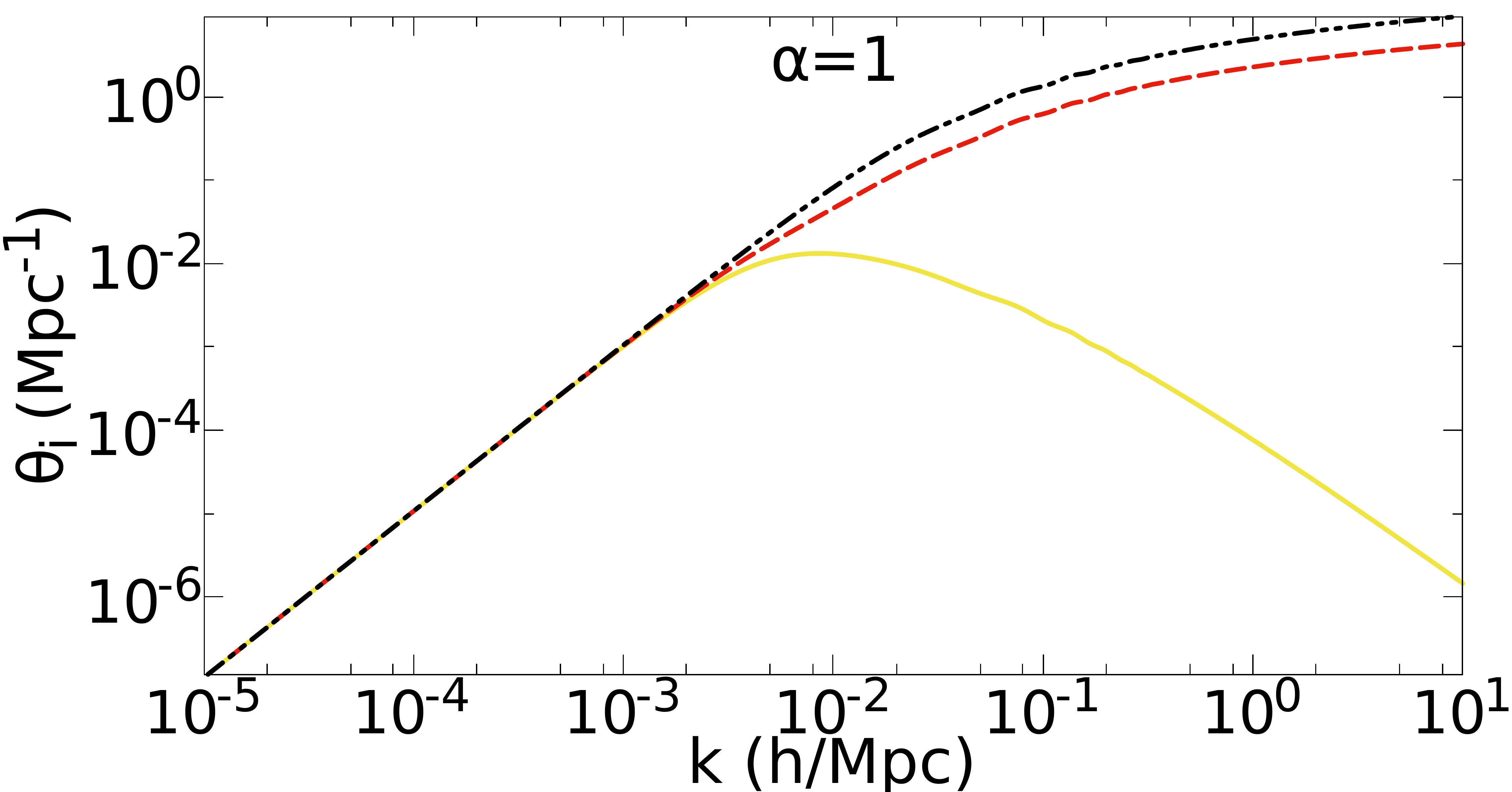}}
		\centerline{
		\includegraphics[scale=0.165]{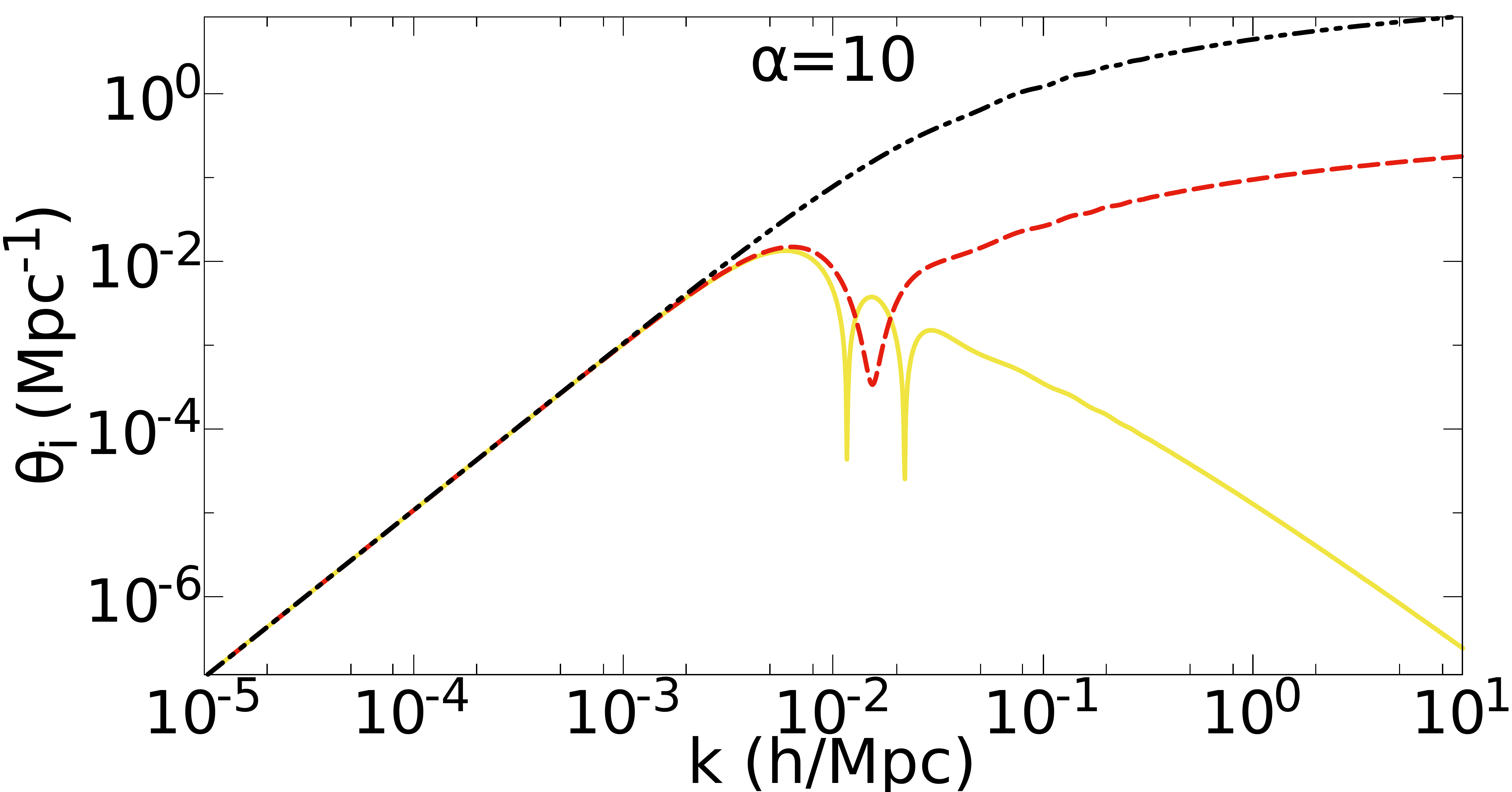}
		\includegraphics[scale=0.165]{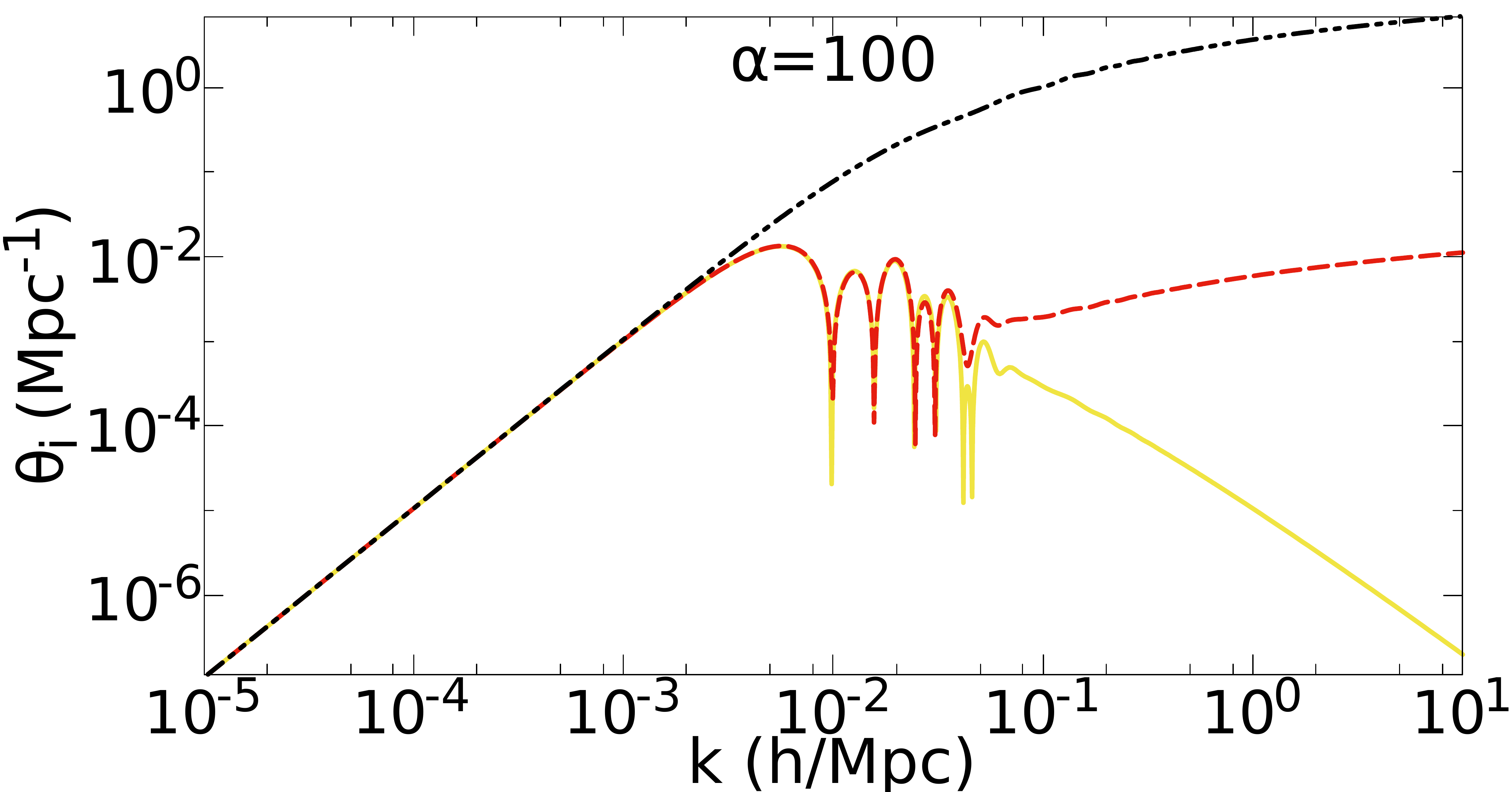}}
	\caption{In these plots, we show the $k$-dependence of today's velocity of DE (yellow line) and DM (red line) for several values of the coupling constant $\alpha$. As a reference for DM, we also plot baryons velocity (black line). We see that large scale behaviour is unaffected, since the interacting term vanishes. On intermediate scales DAO appear, showing the competition between the DE drag and the gravitational pull, resulting in a suppression of DM velocity. Finally, at small scales, DM and DE velocities decouple as consequence of the gravitational pull that dominates over the DE drag. These plots are done using the Newtonian gauge, where matter velocities dominate over DE ones.}
	\label{Fig:velcoupled}	
\end{figure}

The modified Euler equations of DM and DE induce an additional dragging on the DM component sourced by the DE pressure (let's recall that we are assuming $\cs^2=1$) that modifies the evolution of the peculiar velocities. Furthermore, the form of the coupling, similar to the Thomson scattering of baryons and photons before recombination, leads to analogous oscillations in the dark sector that we refer to as Dark Acoustic Oscillations (DAO). These oscillations are more clearly visible in the evolution of the velocities, as shown in Figures~\ref{fig:theta} and~\ref{Fig:velcoupled}.

In the left plot of Figure~\ref{fig:theta} we can see how the relative velocity of DM and DE decreases as the coupling $\alpha$ becomes larger, indicating that the larger $\alpha$ the more efficient is the tight coupling of these dark components. We also see how the oscillations are more noticeable for stronger couplings. 
These oscillations are also manifest in the spectra of the individual velocities shown in  Figure~\ref{Fig:velcoupled}. We can identify the three relevant regimes of the considered scenario. At very large scales, there are no deviations with respect to the $w$CDM model. As we consider smaller scales, the dragging that DE exerts on DM gives rise to the tight coupling that forces both dark components to move coherently and produces the aforementioned oscillations. At even smaller scales, DM tends to fall into the deeper gravitational wells and the DE dragging is not sufficient to prevent it, so both components no longer move coherently. However, the DE pressure still has an effect and delays the falling of DM into the wells, thus causing the appearance of additional peculiar velocities between DM and baryons. As we can see, this regime occurs at smaller scales as we increase the value of $\alpha$.\footnote{It might be convenient to stress again that our analysis is limited to the linear regime. The non-linear evolution of the dark matter clustering in the presence of the friction with dark energy would favour the dissipation of angular momentum that could in turn speed up the clustering.} In this respect, we could think that these velocities would induce an additional velocity bias between the DM and baryons. However, we need to keep in mind that when we use galaxies to trace the underlying DM density field, these correspond to virialised objects within the associated DM haloes. For sufficiently small values of the interaction parameter $\alpha$, we expect the gravitational bound of galaxies to the host DM halo to be sufficiently efficient as to make the galaxies follow the motion of the halo. In other words, galaxies are still good tracers of the DM distributions without additional velocity bias. Let us emphasise however once again that a full account of these effects would require resorting to a proper $N$-body simulation.


\subsection{Screening on small scales}
\label{sec:screencoupling}

The elastic coupling model naturally prevents effects on very large scales as the relative velocities between the universe components are negligible. In this section, we want to explore the possibility of having a screening mechanism suppressing the interaction at some small $k_s$ scale below the horizon. We will model this behaviour in a phenomenological manner by simply introducing a cut-off scale so that the coupling becomes $k$-dependent as follows~\footnote{{Although we have in mind existing screening mechanisms for certain types of theories like e.g. chameleon, symmetron, $K$-mouflage or Vainshtein (see for instance~\cite{Joyce:2014kja}), we do not intend our parameterisation to fully capture or provide a precise characterisation of those screening mechanisms. Rather, we want to explore if cutting-off small scales could improve the compatibility with data.}}
\begin{equation}
\label{eq:alpha_k}
\alpha(k)=\alpha\,e^{-k/k_s}\;.
\end{equation}

\begin{figure}
	\centering
	\includegraphics[scale=0.16745]{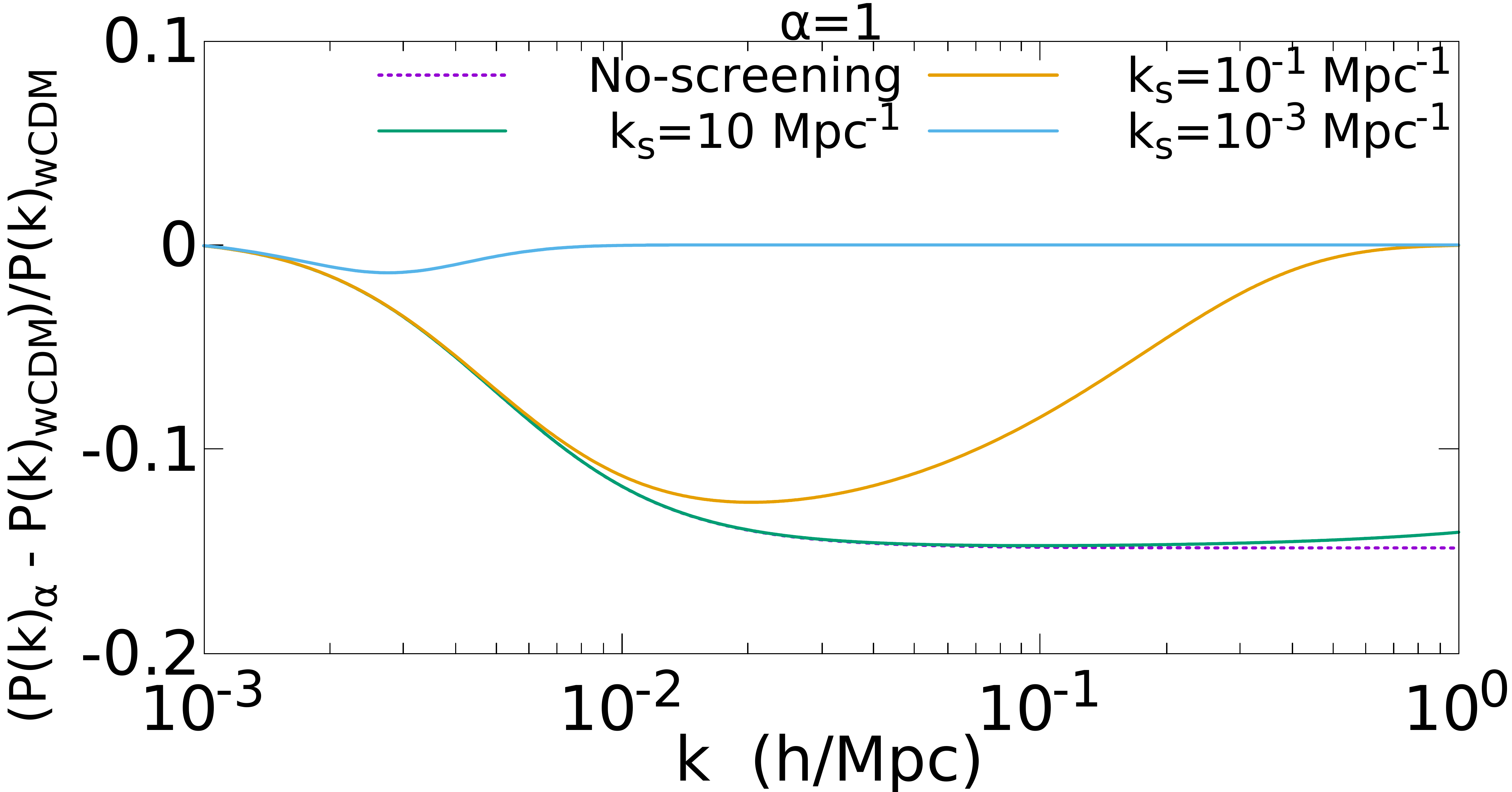}
	\includegraphics[scale=0.16745]{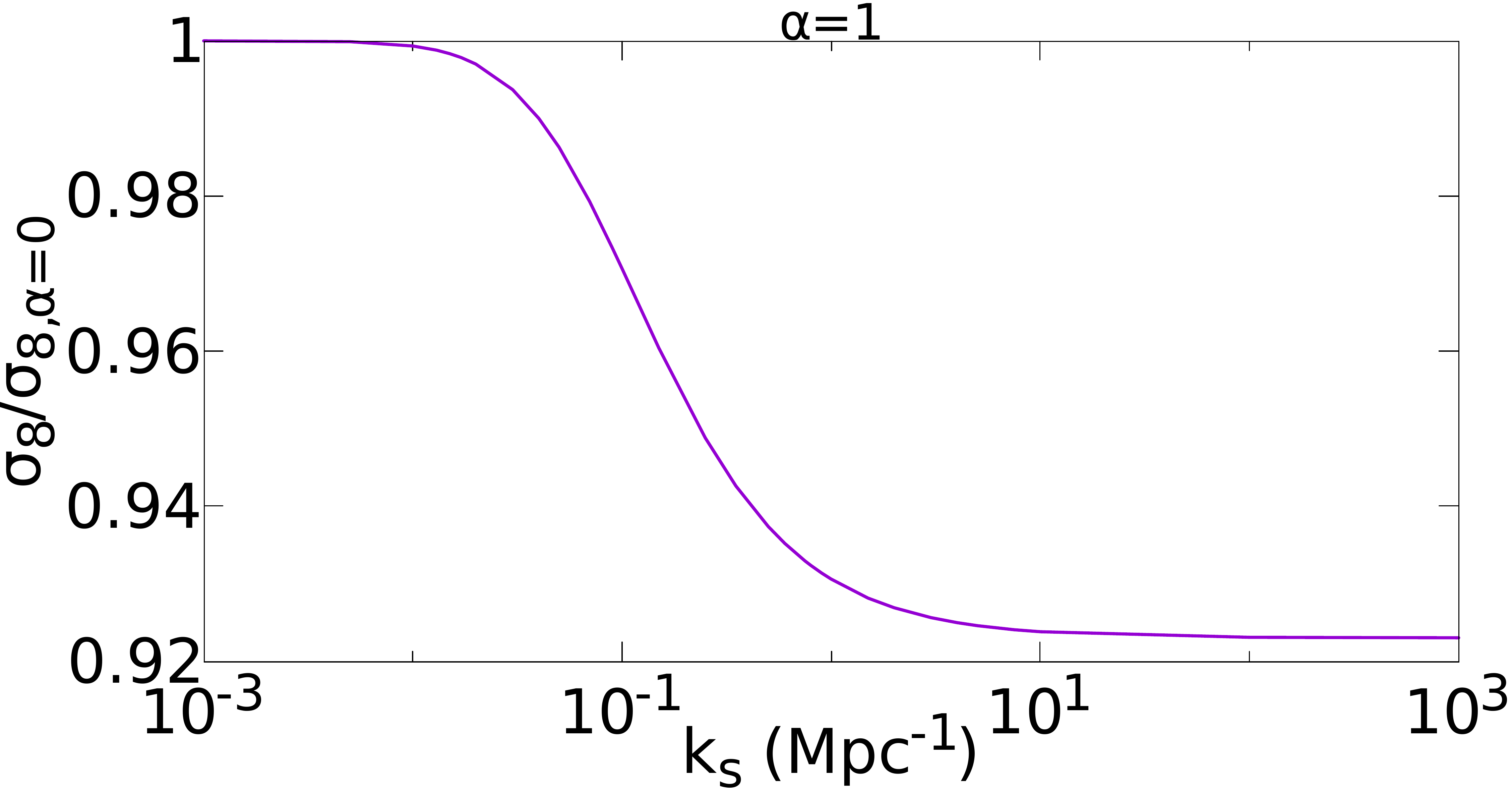}
	\caption{In the left plot, we show the relative ratio of the matter power spectrum using $\alpha=1$ for several values of the screening scale $k_s$, including the previous model with no screening (dotted purple line). In the right plot, we display the ratio of the value of $\sigma_8$ for  $\alpha=1$ and the non interacting value $\sigma_{8,\alpha=0}\;$, as a function of the parameter $k_s$.}
	\label{fig:screening}
\end{figure}

Focusing on the matter power spectrum and the parameter $\sigma_8$ to illustrate the effect of the cut-off we show the results in Figure~\ref{fig:screening}.
In the left panel, we show the relative ratio of the matter power spectrum fixing the parameter $\alpha=1$ for several values of the cut-off scale $k_s$. As expected, we see that those modes smaller than the cut-off scale follow the purple dotted line that represents the previous model of Section~\ref{sec:ctacoupling}, where no screening takes place. For those modes larger than $k_s$ the matter power spectrum tends to behave as a non interacting $w$CDM model, while for scales around the cut-off there is an interpolation between both behaviours.  In the case of the parameter $\sigma_8$, as it is the root mean square of  fluctuation amplitude within spheres of scales $8\,h^{-1}$Mpc, we infer from the right plot of  Figure~\ref{fig:screening} that when the coupling cut-off scales are much smaller than the $\sigma_8$-spheres scales its value is not affected by the cut-off and remains as in Section~\ref{sec:ctacoupling}. When the cut-off scale includes inside it the $\sigma_8$-spheres no lowering on the value of  $\sigma_8$  appears, as expected due to the scale-suppression of the coupling. It is important to remark that very large values of $k_s$ would belong to the non-linear regime where non-linear effects are expected to take on. As before for those cut-off scales represented by $k_s$ close to $8\,h^{-1}$Mpc we have an interpolation between both behaviours.

\section{Fit to the data}
\label{Sec:Fit}
Once we have explored the evolution of the cosmological perturbations and their effects on the CMB and matter power spectra for the considered interacting model, we turn our attention to observations in order to confirm and update the results obtained in~\cite{Asghari:2019qld}. We will also explore the relevance of introducing the aforementioned cut-off to discern if a given screening scale is preferred by data.

We use the public code of Markov chains Monte Carlo called MontePython~\cite{Brinckmann:MP,Audren:MP} applied to our modified \texttt{CLASS} code, to fit several cosmological parameters and our coupling parameters $\alpha$ and $k_s$ with the available data. We use the full Planck 2018 likelihood~\cite{PLANCK2018,Planck2019V} containing data of high-$\ell$ and low-$\ell$ from CMB temperature (TT), polarisation (EE), cross correlation of temperature and polarisation (TE) and the CMB lensing power spectrum, the JLA likelihood with supernovae data~\cite{JLA}, the BAO combined data~\cite{BAO1,BAO2,BAO3}, the PlanckSZ likelihood of Planck Sunyaev-Zeldovich effect~\cite{PlanckSZ} and the weak lensing likelihood CFHTLens~\cite{CFHTL}. The last two datasets are implemented as priors on $S^{\rm SZ}_8=\sigma_8(\Omega_{\rm m}/0.27)^{0.3}$ and $S^{\rm CFHTL}_8=\sigma_8(\Omega_{\rm m}/0.27)^{0.46}$ respectively. The use of PlanckSZ data presents some caveats that we will discuss below.

For our fit, we consider as cosmological parameters the baryon density defined as $100~\Omega_{\rm b} h^2$, the DM density as $\Omega_{\rm dm}h^2$, the scalar spectral index $n_{s}$, the primordial amplitude as $10^{9}A_{s}$, the reionisation optical depth $\tau_{\rm reio}$, the equation of state of DE as $w$, the angular acoustic scale as $100~\theta_{s}$ and the coupling parameters for each model, $\alpha$ for the constant coupling model while $\alpha$ and $k_s$ for the screening model. Also, we consider as derived parameters the redshift of reionisation $z_{\rm reio}$, the Hubble parameter $H_0$, the  matter fluctuation amplitude at $8h^{-1}$Mpc\, as $\sigma_8$ and the total matter density $\Omega_{\rm m}$. The DE equation of state is assumed to be constant and restricted to $w>-1$ in order to avoid instabilities of the interacting model. For the same reason, we assume the coupling parameter $\alpha$ to be positive or, at least, with not a large value if negative (see Appendix~\ref{App:Stability}).

Finally, we remind that any significant difference found with the elastic coupling models will originate from the perturbation sector since we do not change the background cosmology.

\subsection{Constant coupling model \texorpdfstring{$\alpha=const$}{TEXT}}

\subsubsection*{Flat prior on $\log_{10}\, \alpha$}

First, we will start our analysis by considering a flat prior on ${\log_{{10}}\,  \alpha }$ over the range ${\log_{{10}}\,\alpha }\in[-8,4]$. The purpose of this initial fit is to obtain a sense of the order of magnitude of $\alpha$ so we can explore later in more detail around the order of magnitude obtained here. One can argue we are excluding the non-interacting case $\alpha=0$ but further analysis will support our choice. Furthermore, we know that values of $\alpha$ smaller than $10^{-2}$ do not give any appreciable deviation with respect to the non-interacting case, so this first analysis should allow us to infer, as a first approach, if the effects of the  coupling are preferred by data or not.

\begin{figure}
	\centering
	\includegraphics[scale=0.47]{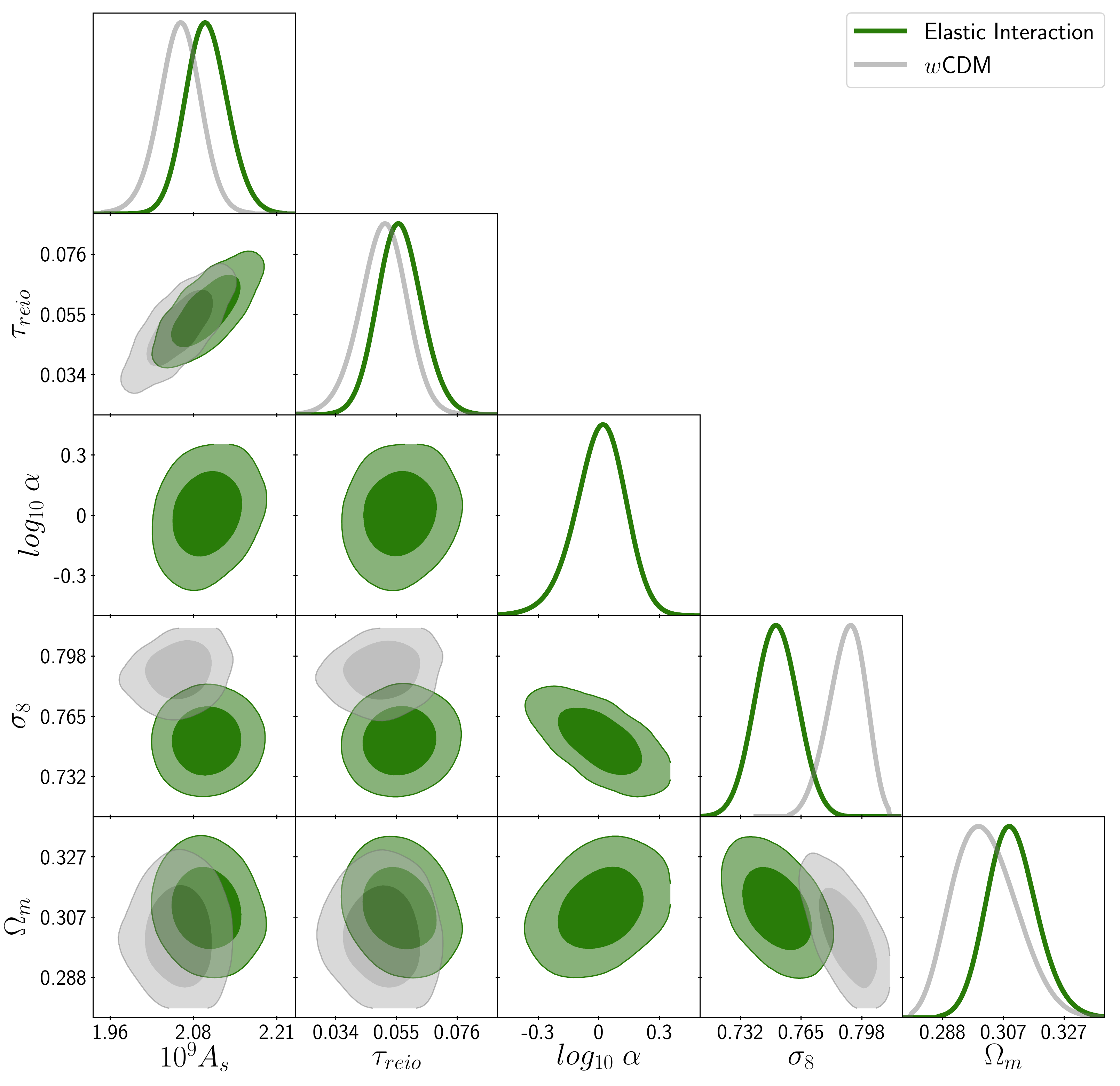}
	\caption{The one-dimensional 1$\sigma$ posteriors distributions and the 2$\sigma$ contours of several parameters for the constant coupling model using a logarithmic analysis on the model parameter (green) and the reference $w$CDM model (grey). }
	\label{fig:contourctalog}
\end{figure}

In Table~\ref{tab:alphactalog}, we show the mean values with 1$\sigma$ and 2$\sigma$ confidence limits for all the parameters considered for the reference $w$CDM model and for the elastic coupling model when the logarithmic prior is used on $\alpha$. In Figure~\ref{fig:contourctalog}, we display the one-dimensional posterior distribution and the two-dimensional contours for some parameters. Although there are significant differences in some cosmological and derived parameters, here we only focus on the coupling parameter $\alpha$ as we will study them in detail below when using a linear flat prior on $\alpha$. With the result of the coupling parameter $\log_{10}\, \alpha = 0.003_{-0.13}^{+0.15}$, we confirm the ideas explained at the end of Section~\ref{sec:themodel}, where the normalisation of the coupling parameter was chosen to get a natural value $\alpha\sim\mathcal{O}(1)$. In view of this result we will perform the analysis with a flat prior on $\alpha$ incorporating the gained knowledge that $\alpha\sim\mathcal{O}(1)$ seems to be the preferred value by data. Furthermore, this analysis will permit to fully include the non interacting value of the coupling parameter $\alpha=0$. We will show the consistency of the results obtained with both priors below, so we defer the discussion to the next section.

\begin{table}
\begin{center}
\renewcommand{\arraystretch}{1.8}
\begin{tabular}{ |c||c|c|c||c|c|c| } 
	\hline
	\hline
	\centering
	&\multicolumn{3}{c||}{$w$CDM model} &\multicolumn{3}{c|}{Elastic Interaction $\log_{10}\, \alpha$}\\ \hline
	Param. & mean$\pm\sigma$ & $2\sigma$ lower & $2\sigma$ upper  & mean$\pm\sigma$ & $2\sigma$ lower & $2\sigma$ upper\\
	\hline \hline \hline
$100\Omega_{\rm b } h^2$ & $2.264_{-0.015}^{+0.015}$ & $2.235$ & $2.294$ &  $2.243_{-0.016}^{+0.016}$ & $2.211$ & $2.274$\\\hline
$\Omega_{\rm dm}h^2$  & $0.1163_{-0.001}^{+0.001}$ & $0.1143$ & $0.1183$ &  $0.1193_{-0.0012}^{+0.0013}$ & $0.1169$ & $0.1219$\\ \hline 
$n_{s}$ & $0.9721_{-0.0043}^{+0.0042}$ & $0.9639$ & $0.9807$ & $0.9662_{-0.0045}^{+0.0043}$ & $0.9576$ & $0.9753$\\ \hline  
$10^{9}A_{s}$ & $2.063_{-0.032}^{+0.035}$ & $1.993$ & $2.133$ & $2.105_{-0.035}^{+0.034}$ & $2.038$ & $2.177$\\ \hline  
$\tau_{\rm reio }$  & $0.0502_{-0.0082}^{+0.0092}$ & $0.03266$ & $0.0686$ & $0.05618_{-0.0087}^{+0.0079}$ & $0.03955$ & $0.07311$\\ \hline  
$w$ & $-0.9478_{-0.043}^{+0.022}$ & $-0.9999$ & $-0.8879$ &  $-0.9788_{-0.021}^{+0.006}$ & $-0.9999$ & $-0.9389$ \\ \hline  
$100~\theta_{s }$  & $1.042_{-0.0003}^{+0.0003}$ & $1.042$ & $1.043$ & $1.042_{-0.00032}^{+0.00032}$ & $1.041$ & $1.043$ \\ \hline  \hline
$z_{\rm reio }$ & $7.141_{-0.86}^{+0.89}$ & $5.298$ & $8.93$ &  $7.837_{-0.82}^{+0.82}$ & $6.202$ & $9.491$ \\ \hline  
$H_0$ \mbox{\small  $[\frac{\rm km}{\rm s\,Mpc}]$ } & $67.88_{-0.96}^{+1.2}$ & $65.71$ & $69.97$ &  $67.58_{-0.65}^{+0.85}$ & $66.02$ & $69.15$\\ \hline  
$\sigma_8$  & $0.7898_{-0.009}^{+0.012}$ & $0.7693$ & $0.8094$ & $0.7516_{-0.012}^{+0.012}$ & $0.7272$ & $0.7764$ \\ \hline  
$\Omega_{\rm m }$  & $0.3018_{-0.012}^{+0.009}$ & $0.2817$ & $0.3226$ & $0.3104_{-0.010}^{+0.008}$ & $0.2933$ & $0.3287$ \\ \hline\hline
$\log_{10}\, \alpha$   & $-$& $-$ & $-$   & $0.003_{-0.13}^{+0.15}$ & $-0.283$ & $0.283$\\ \hline
\hline 
\end{tabular} 
\end{center}
\caption{In this table, we show the mean and $1\sigma$ values and the $2\sigma$ limits for the cosmological and derived parameters for a $w$CDM model (left) and for the interacting model using a logarithmic analysis on  the coupling parameter $\alpha$ (right). }
\label{tab:alphactalog}
\end{table}

\subsubsection*{Flat prior on $\alpha$}

Once we have confirmed, using the logarithmic prior for $\alpha$, that data seems to favour the coupling parameter to be $\alpha\sim\mathcal{O}(1)$, we will perform the same analysis but now using a flat prior on $\alpha$ instead of $\log_{10}\,\alpha$ over the range $\alpha\,{\in}\,[-0.1,100]$, so we are now including the non interacting case. The range has been chosen sufficiently wide as to avoid any border effect in the sampling of the likelihood.

\begin{figure}
	\centering
	\includegraphics[scale=0.47]{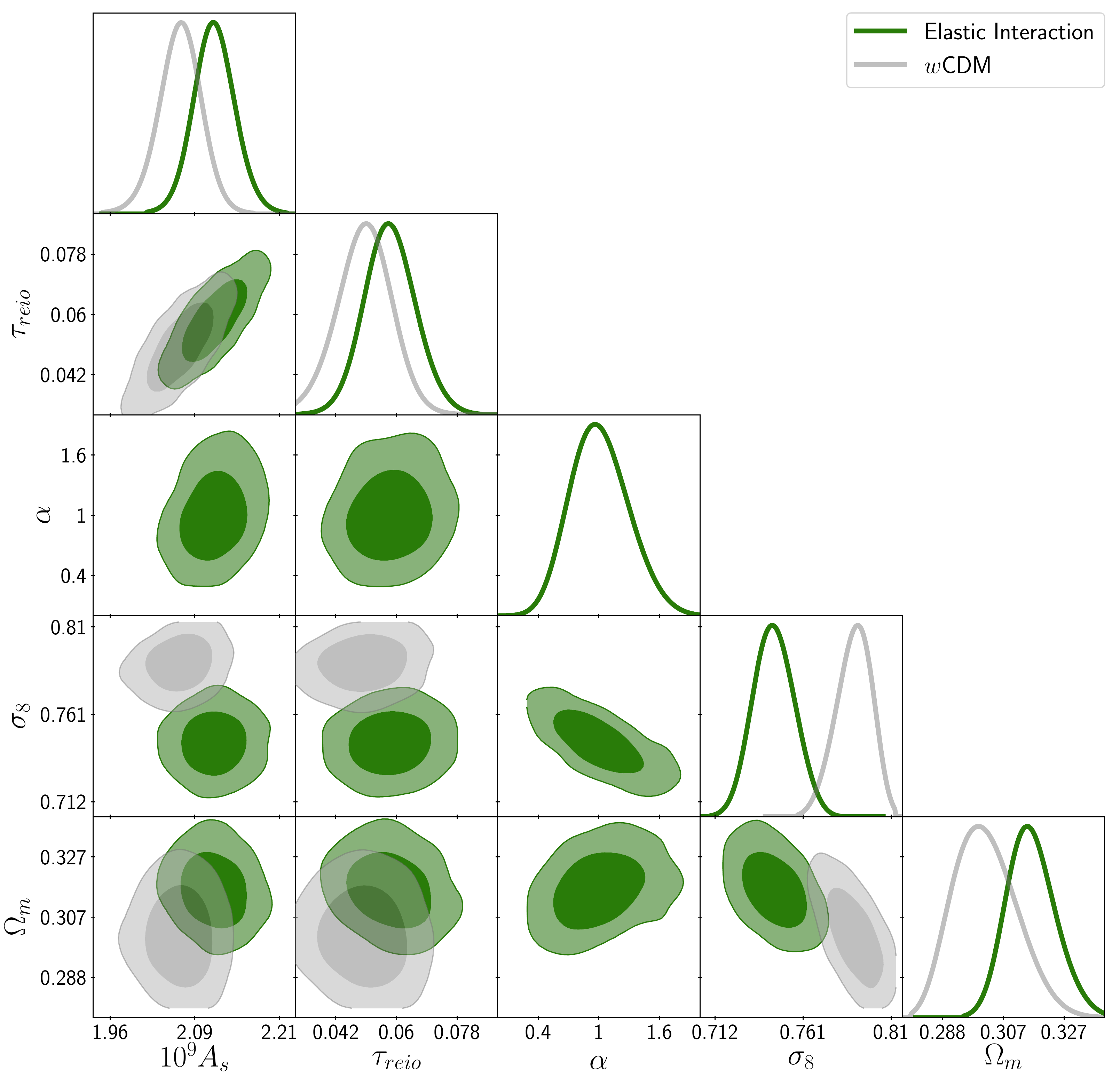}
	\caption{The two-dimensional $1\sigma$ and 2$\sigma$ posterior of several parameters for the constant coupling model. }
	\label{fig:contourcta}
\end{figure}

In Table~\ref{tab:alphacta}, we show the constraints obtained for the reference model $w$CDM and for the interacting model for all cosmological, derived and model parameters. In Figure~\ref{fig:contourcta}, we display for some parameters the one-dimensional posterior distribution and the two dimensional contours obtained with the previously explained datasets. We find a significant increase in the DM density $\Omega_{\rm dm}h^2$ while a slight decrease in the baryons density $100\Omega_{\rm b } h^2$ appears, which combined they are expressed on a small increase of the total matter density $\Omega_{\rm m}$. We also find a small decrease of the value of  the scalar spectral index $n_{s}$ and an increase on the value of the primordial amplitude $10^{9}A_{s}$. 
However, what it is really significant is the lowering in the $\sigma_8$ value if we compare the $w$CDM model and the elastic coupling model. This result can be connected to what we found in the previous sections: the interaction induces a coupling between DE and DM that appears at the late universe. This coupling leads to the freeze of the growth of structures represented by the density contrast of DM in Figure~\ref{fig:delta_dm} and, as a consequence, we have the suppression at small scales, where the coupling is efficient, as shown in Figure~\ref{fig:PK}. The final consequence is a lower value of $\sigma_8$. Even more remarkable than the previous result is the fact that a non-zero value of the coupling strength is obtained at more than $3\sigma$ confidence level as $\alpha \underset{3\sigma}{\in}[0.249,1.899]$. 
These results confirm the ones obtained in Ref.~\cite{Asghari:2019qld} but now with updated Planck 2018 data and including polarisation, and they are also similar to the ones in Ref.~\cite{Jimenez:2020ysu}, where the same coupling is used between DE and baryons. The results obtained here are compatible with those we got using the logarithmic parameter $\log_{10}\, \alpha$ as one would expect.

Finally, we find a significant $\Delta\chi^2=23.4$ improvement with respect to the $w$CDM model. This improvement is obtained by comparing the best fit of the interacting model with the best fit of a non-interacting $w$CDM\footnote{It may be worth noticing that a pure \LCDM will give a $\Delta\chi^2$ at least as big as $w$CDM so our discussion here is also valid for the pure \LCDM case. As a matter of fact, we have explicitly compared to the pure \LCDM and we have found $\Delta\chi^2=24.1$. In this respect, although \LCDM has one fewer parameter, this is not sufficient to compensate for the big $\chi^2$ difference.}. However, this is not enough to claim that the model is preferred by data with a strong statistical significance, since adding more parameters will always lead to an improvement in the $\chi^2$-squared value, therefore we will resort to a more informative criterion. In particular, we will apply the AIC criterion~\cite{Akaike:AIC} based on the value of the quantity defined as
\begin{equation}
    AIC\equiv-2\log \mathcal{L} +2k\;,
\end{equation}
where $\log \mathcal{L}$ is the maximum of the likelihood and $k$ the number of parameters used in the fit. Applying this criterion, we find an improvement of $\Delta AIC=21.4$, which in turn constitutes a strong evidence favouring the interacting model. This confirms that the substantially lower value of the best fit $\chi^2$ for the interacting model does indicate  and the price of adding only one new parameter (two in the case of a pure \LCDM) is not a sufficient penalty.

\begin{table}
\begin{center}
\renewcommand{\arraystretch}{1.8}
\begin{tabular}{ |c||c|c|c||c|c|c| } 
	\hline
	\hline
	\centering
	&\multicolumn{3}{c||}{$w$CDM model} &\multicolumn{3}{c|}{Elastic Interaction $\alpha$}\\ \hline
	Param. & mean$\pm\sigma$ & $2\sigma$ lower & $2\sigma$ upper  & mean$\pm\sigma$ & $2\sigma$ lower & $2\sigma$ upper\\
	\hline \hline \hline
$100\Omega_{\rm b } h^2$ & $2.264_{-0.015}^{+0.015}$ & $2.235$ & $2.294$ & $2.241_{-0.016}^{+0.015}$ & $2.211$ & $2.272$\\\hline
$\Omega_{ \rm dm}h^2$  & $0.1163_{-0.001}^{+0.001}$ & $0.1143$ & $0.1183$ & $0.1194_{-0.0012}^{+0.0012}$ & $0.1171$ & $0.1218$\\ \hline 
$n_{s}$ & $0.9721_{-0.0043}^{+0.0042}$ & $0.9639$ & $0.9807$ & $0.9659_{-0.0045}^{+0.0042}$ & $0.9575$ & $0.9748$\\ \hline  
$10^{9}A_{s}$ & $2.063_{-0.032}^{+0.035}$ & $1.993$ & $2.133$ & $2.115_{-0.034}^{+0.032}$ & $2.048$ & $2.183$\\ \hline  
$\tau_{\rm reio }$  & $0.0502_{-0.0082}^{+0.0092}$ & $0.03266$ & $0.0686$ & $0.05817_{-0.0082}^{+0.0081}$ & $0.04153$ & $0.07468$\\ \hline  
$w$ & $-0.9478_{-0.043}^{+0.022}$ & $-0.9999$ & $-0.8879$ & $-0.9814_{-0.019}^{+0.004}$ & $-0.9999$ & $-0.9426$ \\ \hline  
$100~\theta_{s }$  & $1.042_{-0.0003}^{+0.0003}$ & $1.042$ & $1.043$ & $1.042_{-0.00032}^{+0.00032}$ & $1.041$ & $1.043$ \\ \hline  \hline
$z_{\rm reio }$ & $7.141_{-0.86}^{+0.89}$ & $5.298$ & $8.93$ & $8.051_{-0.78}^{+0.8}$ & $6.459$ & $9.694$ \\ \hline  
$H_0$ \mbox{\small$[\frac{\rm km}{\rm sMpc}]$ } & $67.88_{-0.96}^{+1.2}$ & $65.71$ & $69.97$ & $67.1_{-0.62}^{+0.77}$ & $65.63$ & $68.46$\\ \hline  
$\sigma_8$  & $0.7898_{-0.009}^{+0.012}$ & $0.7693$ & $0.8094$ & $0.7455_{-0.012}^{+0.012}$ & $0.7215$ & $0.7688$ \\ \hline  
$\Omega_{\rm m }$  & $0.3018_{-0.012}^{+0.009}$ & $0.2817$ & $0.3226$ & $0.3166_{-0.0087}^{+0.0075}$ & $0.301$ & $0.3338$ \\ \hline\hline
$\alpha$   & $-$ & $-$ &  $-$  & $1.005_{-0.33}^{+0.26}$ & $0.4374$ & $1.648$\\ \hline
\hline 
\end{tabular} 
\end{center}
\caption{In this table we show the constraints obtained from the data explained in the main text for the $w$CDM model (left) and for the interacting model with the coupling parameter constant (right). }
\label{tab:alphacta}
\end{table}

\begin{figure}
	\centering{
		\includegraphics[scale=0.50]{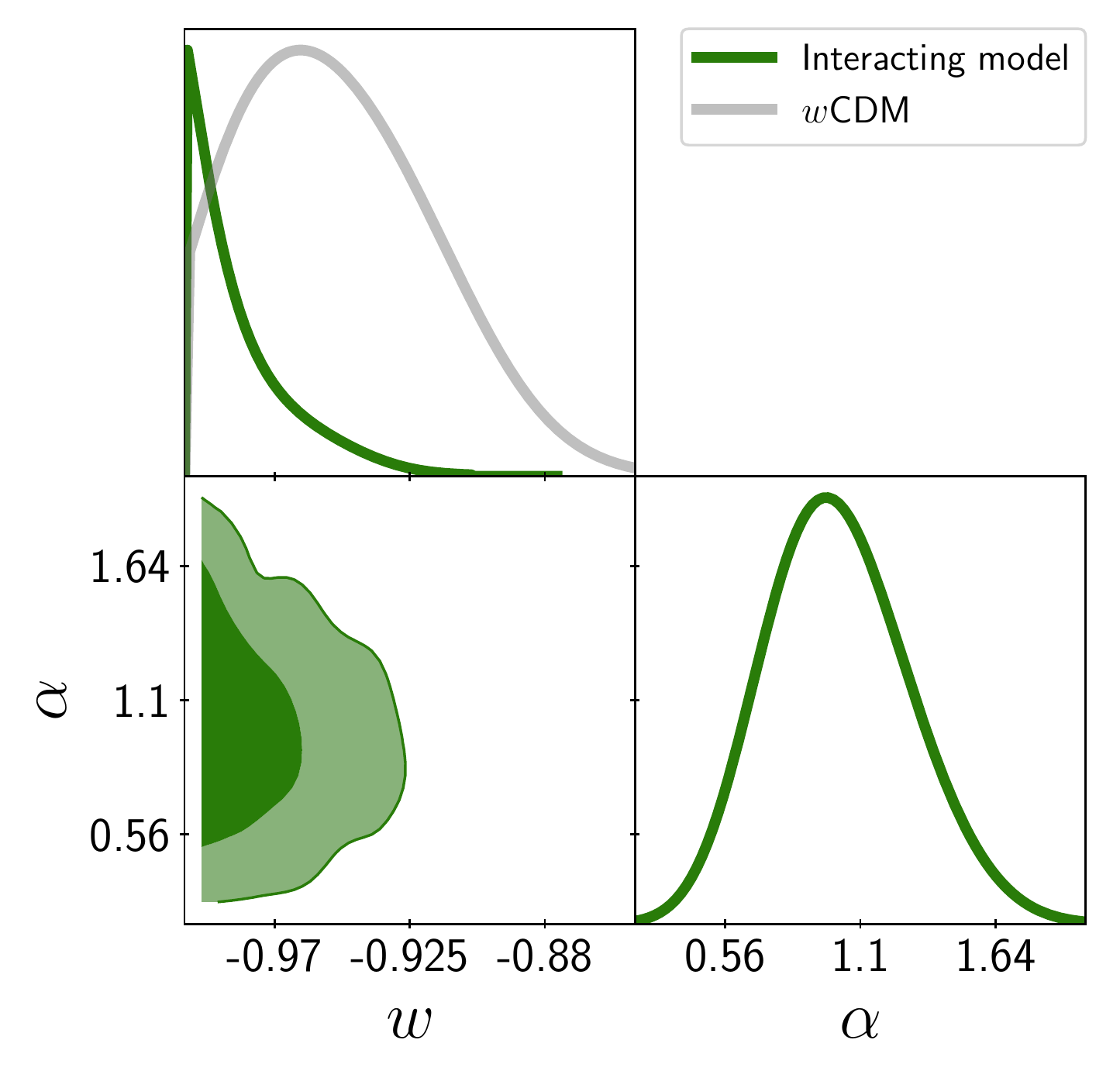}
		\includegraphics[scale=0.35]{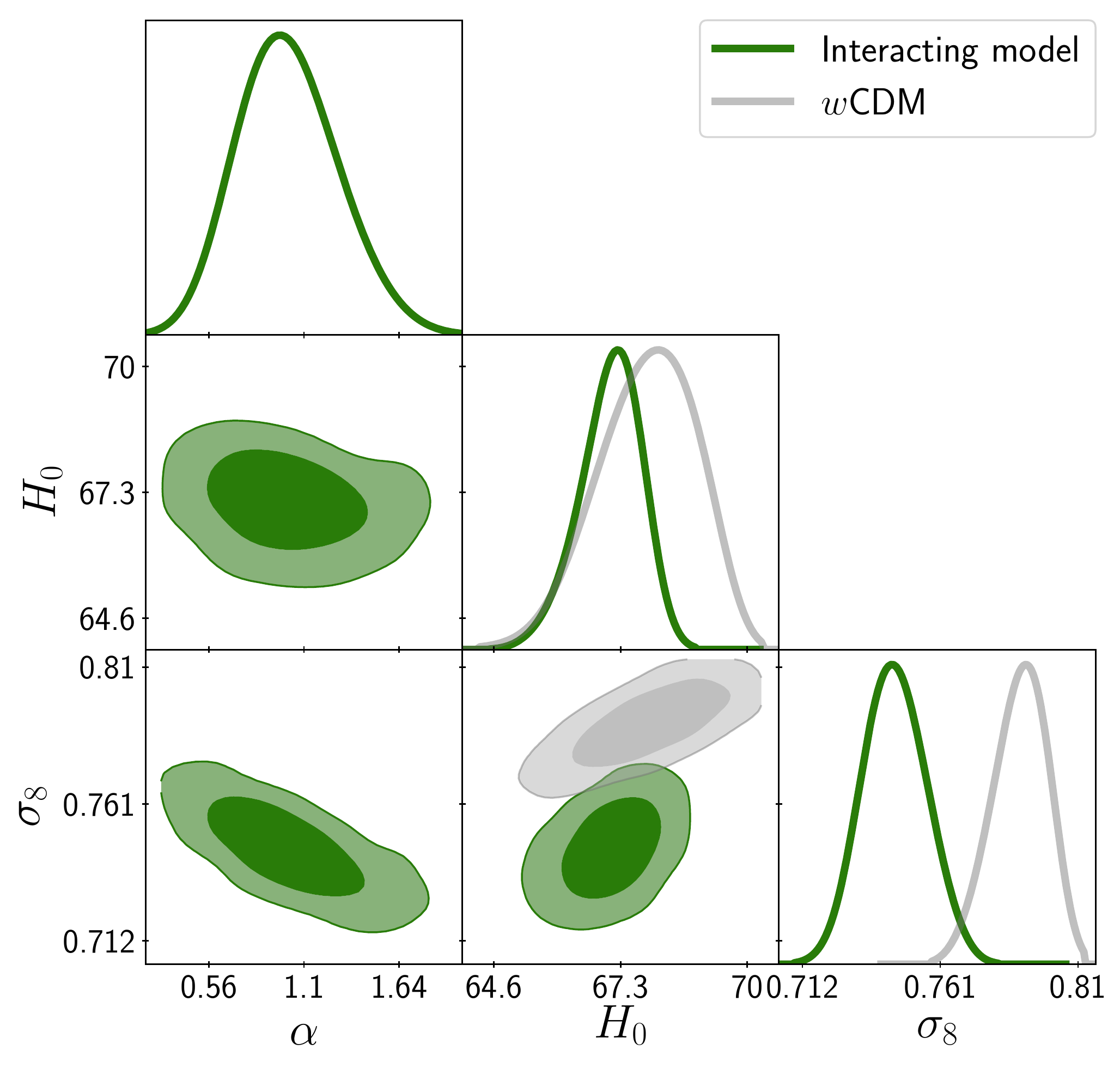}}	
		\caption{The two-dimensional $1\sigma$ and $2\sigma$ posteriors for the $w-\alpha$ plane (left) and for the model parameter $\alpha$, the Hubble constant $H_0$ and $\sigma_8$ (right).}
	\label{Fig:walphaplane}
\end{figure} 

\subsubsection*{Analysis for each data-set}
\begin{figure}
	\centering{
	         \includegraphics[scale=0.5]{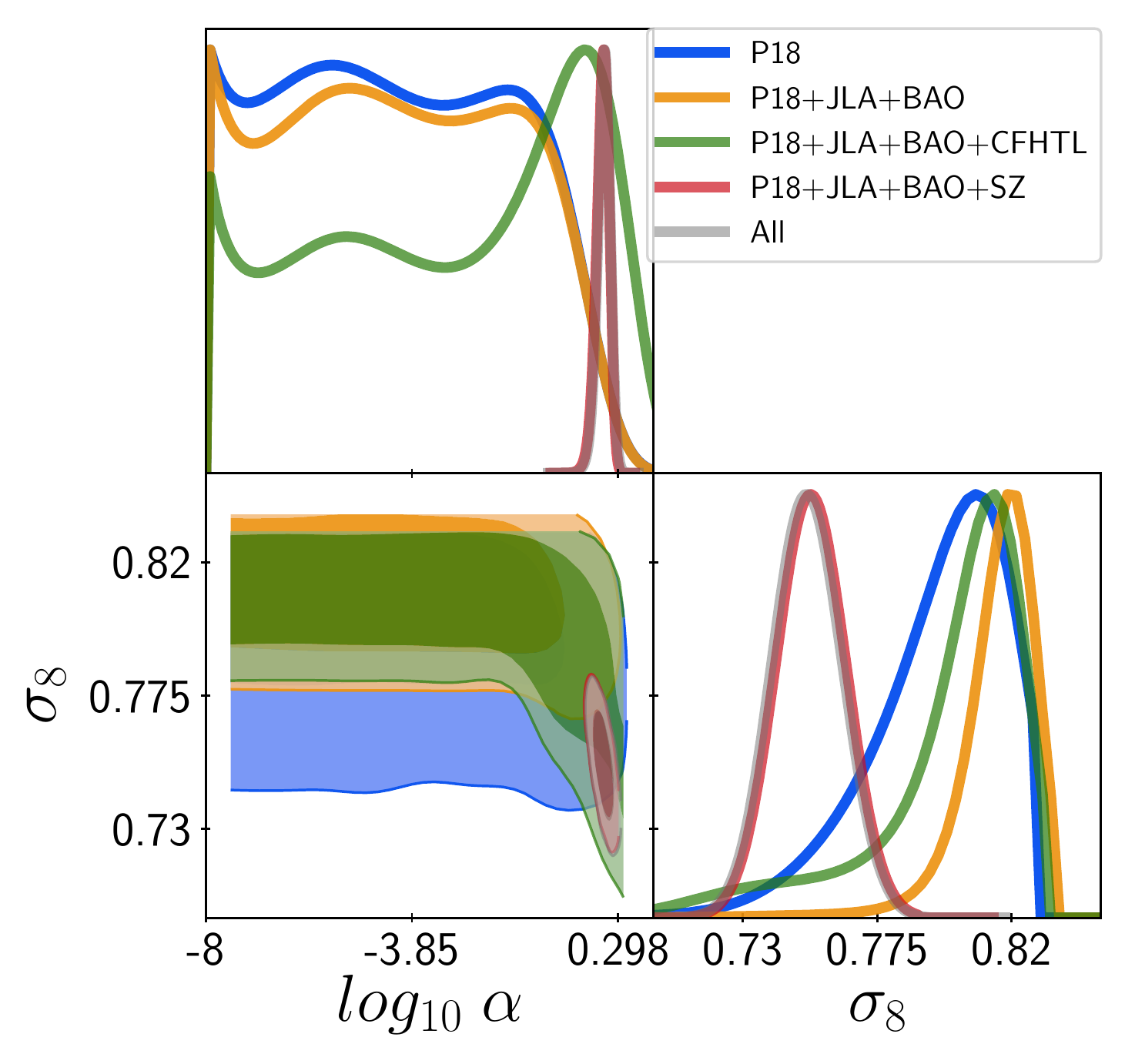}
          	\includegraphics[scale=0.5]{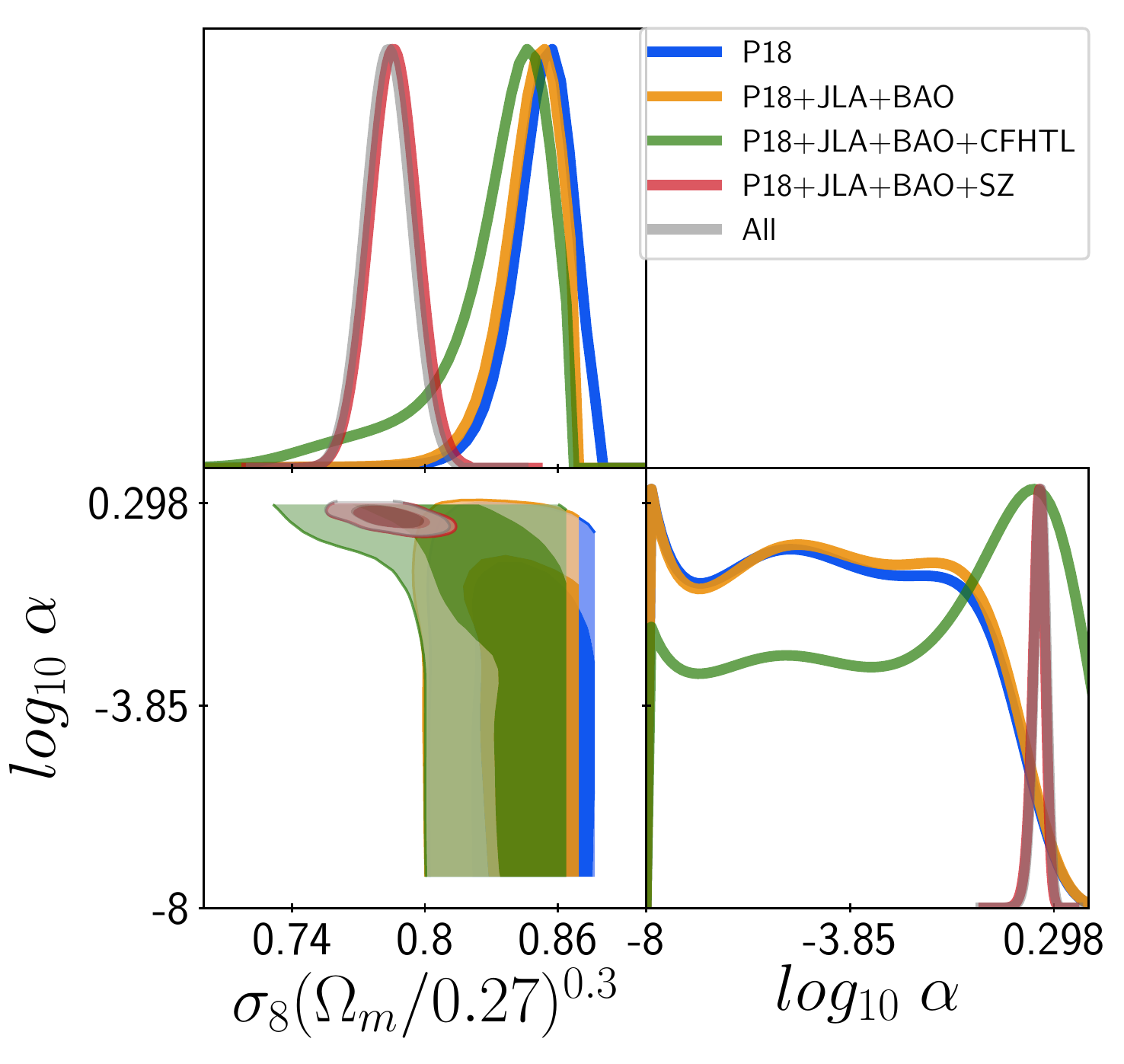}}
	\caption{The two-dimensional $1\sigma$ and 2$\sigma$ posterior of the coupling parameter $\alpha$ and the parameter $\sigma_8$ (left) and for the coupling parameter $\alpha$ and the parameter ${\sigma_8} (\frac{\Omega{_m}}{0.27})^{0.3}$  (right) for the constant coupling model, using different combinations of data-sets: P18 means Planck 2018 likelihood of TT, EE, TE and lensing data (blue), P18+JLA+BAO means Planck 2018 with JLA and BAO likelihood (yellow), P18+JLA+BAO+CFHTL denotes Planck 2018 plus JLA, BAO and CFHTLenS data (green), P18+JLA+BAO+SZ refers to Planck 2018 plus JLA, BAO and Sunyaev-Zeldovich Planck likelihood (red) while All uses all the previous data-sets (grey).}  
	\label{fig:dataset}
\end{figure}

Due to the more than 3$\sigma$ preference and the enormous improvement in the $\chi^2$ for the elastic coupling model, we will analyse the impact of each dataset on the determination of the cosmological parameters and the model parameter $\alpha$ as well as their contribution to lowering the best-fit $\chi^2$. In this analysis we use the combinations of Planck 2018 likelihood of TT, EE, TE and lensing (P18), Planck 2018 with JLA and BAO likelihood (P18+JLA+BAO), Planck 2018 plus JLA and BAO with the CFHTLenS  likelihood (P18+JLA+BAO+CFHTL) and Planck 2018 plus JLA and BAO with the Sunyaev-Zeldovich Planck likelihood  (P18+JLA+ BAO+SZ). We find no substantial change on the previous parameters with the exception of the elastic coupling parameter $\alpha$ and the derived parameter $\sigma_8$, which are shown in Table~\ref{tab:dataset} and Figure~\ref{fig:dataset} . From this analysis, we infer that the Sunyaev-Zeldovich Planck likelihood is the main contributor to the possible detection of the elastic coupling when combined with Planck18, BAO and JLA data, with some additional contribution from the CFHTLenS  likelihood. The Planck 2018 data alone, or with the JLA and BAO data, cannot establish a lower limit to the coupling parameter which would suggest a detection, but they are only able to put an upper limit that is compatible (at least at the 2$\sigma$ level as one can see from Figure~\Ref{fig:dataset}) with the combined data or with P18+JLA+BAO+SZ combined analysis. For completeness, we also show the result for the combination of parameters ${\sigma_8} \left(\frac{\Omega{_m}}{0.27}\right)^{0.3}$. It shows that when the interaction is not preferred by the data set, $\alpha \to 0$, both $\sigma_8$ and ${\sigma_8} \left(\frac{\Omega{_m}}{0.27}\right)^{0.3}$ tend to have the standard model value, while when the value of the coupling parameter $\alpha$ shows the interaction is preferred, $\alpha \to 1$, both $\sigma_8$ and ${\sigma_8} \left(\frac{\Omega{_m}}{0.27}\right)^{0.3}$ lower their values, although they are compatible at least at the 2$\sigma$ level as shown in Figure~\Ref{fig:dataset}. This result is in agreement with the intuitive expectation that the coupling suppresses the growth of structures so the interaction indeed prefers lower values of $\sigma_8$ (for constant $\Omega_{\rm}$) that are in better agreement with low-redshift data.

As we have seen, the detection of the interaction parameter is driven by PlanckSZ data. There are some caveats, however, as to the consistency of using these data as a prior on ${\sigma_8} \left(\frac{\Omega{_m}}{0.27}\right)^{0.3}$. To obtain this value, a translation from temperature profiles to density is necessary as well as a proper identification of the mass bias or mass calibration (see e.g.~\cite{Bolliet:2020kwa} and references therein) that require the use of hydrodynamical simulations properly tackling the non-linear scales and this is done by using $\Lambda$CDM-based simulations. We lack any devoted analysis of the non-linear effects for the model under consideration so it might not be justified to use those data. Our results are thus only valid as long as the potential effects from the interaction are sufficiently small on those scales or, at least, below the precision of the derived values. In any case, it is intriguing that PlanckSZ data as used here not only substantially improves the fit with respect to $w$CDM, but it crucially drives the potential detection of $\alpha$ so our findings are worth being reported here. 

The obtained results motivate to consider two fiducial cosmologies in the forecast analysis that we will perform in Sec.~\ref{Sec:Forecast}, one with a non interacting value of the coupling parameter $\alpha=0$ and with the combined result obtained for it $\alpha=1.005_{-0.33}^{+0.26}$. Using both fiducial cosmologies will allow us to avoid possible bias effects in our forecast since the potential detection of $\alpha\neq0$ is clearly driven by the PlanckSZ data.

\begin{table}
\renewcommand{\arraystretch}{1.8}
 \begin{tabular}{|c||c|c|c|c|c|} 
 \hline\hline\centering
  & P18 & \makecell{ P18+JLA+\\ BAO} & \makecell{ P18+JLA+ \\BAO+CFHTL  }   & \makecell{P18+JLA+ \\ BAO+SZ } & \makecell{All \\data-sets} \\ 
 \hline\hline\hline
 $\log_{10}\, \alpha$ & $-4.2_{-3.5}^{+1.7}$ & $-4.1_{-2.9}^{+2.2}$ & $-3.1_{-1.8}^{+3.8}$ & $-0.01_{-0.12}^{+0.15}$  & $0.003_{-0.13}^{+0.15}$\\ 
 \hline
 $\sigma_8$  & $0.7965_{-0.019}^{+0.028}$ & $0.8123_{-0.02}^{+0.013}$ & $0.798_{-0.014}^{+0.03}$ & $0.7527_{-0.012}^{+0.012}$ & $0.7516_{-0.012}^{+0.012}$\\
 \hline
 ${\sigma_8} (\frac{\Omega{_m}}{0.27})^{0.3}$ &  $0.8497_{-0.029}^{+0.017}$  & $0.8485_{-0.013}^{+0.04}$    &  $0.8313_{-0.021}^{+0.039}$   &  $0.7856_{-0.012}^{+0.012}$     & $0.7836_{-0.011}^{+0.012}$    \\
 \hline
 \hline
\end{tabular}
\caption{In this table, we show constraints on $\alpha$, $\sigma_8$ and the parameter ${\sigma_8} (\frac{\Omega{_m}}{0.27})^{0.3}$ for the elastic model obtained from the different combination of data-sets: Planck 2018 likelihood of TT, EE, TE and lensing data (P18), Planck 2018 with JLA and BAO likelihood (P18+JLA+BAO), Planck 2018 plus JLA and BAO with the CFHTLenS  likelihood (P18+JLA+BAO+CFHTL), Planck 2018 plus JLA and BAO with the Sunyaev-Zeldovich Planck likelihood (P18+JLA+BAO+SZ) and all the data-sets. The other cosmological parameters do not suffer from any substantial change.}
\label{tab:dataset}
\end{table}

\subsection{Screening model \texorpdfstring{$\alpha e^{-k/k_s}$}{TEXT}   }
In this second scenario, we will assume that the coupling strength has the scale dependence explained in Sec.~\ref{sec:screencoupling} which allows to have the UV cut-off scale for the interaction as $\alpha(k)=\alpha e^{-k/k_s}$. We will employ a similar methodology as in the previous Section, i.e., the normalisation for the coupling chosen so that $\alpha\sim\mathcal{O}(1)$ and, consequently, we will set a flat prior over $\alpha\,{\in}\,[-0.1,100]$, while for the scale cut-off parameter we will use a flat prior on $\log_{10}\,k_s{\in}\,[-4,21]$.

In Table~\ref{tab:fitscreening}, we show the best fit value, the mean with $1\sigma$ confidence limits and also $2\sigma$ upper and lower limits for the cosmological and model parameters, both for the reference model $w$CDM and the screening model. The cosmological parameters constraints are essentially indistinguishable from the ones obtained for the constant coupling model. The coupling parameter $\alpha$ presents a slightly higher value but again is more than $3\sigma$ away from a non interacting value. The screening scale $k_s$ is very poorly constrained but clearly prefers very small scales. This points towards no scale suppression preferred by present data. In this respect, we should mention that those small values of $k_s$ are nevertheless beyond the regime of applicability of our analysis. Not only would we expect a different behaviour of the screening on those scales, but non-linear effects will also become important and they should be properly incorporated. Moreover, those very small scales belong to regions not explored by the used data. Finally, we found a significant $\Delta\chi^2=27.1$ improvement with respect to the $w$CDM model. 
As we again find the coupling is strongly preferred with a huge improvement in the value of $\chi^2$ and not a scale cut-off, at least for cosmological scales,  we conclude that the elastic coupling model is preferred by the current data, following the results of~\cite{Asghari:2019qld} with improved data.

\begin{table}
\begin{center}
\renewcommand{\arraystretch}{1.8}
\begin{tabular}{ |c||c|c|c||c|c|c| } 
	\hline
	\hline
	\centering
	&\multicolumn{3}{c||}{$w$CDM model} &\multicolumn{3}{c|}{Elastic Interaction $\alpha e^{-k/k_s}$}\\ \hline
	Param. & mean$\pm\sigma$ & $2\sigma$\,lower & $2\sigma$\,upper  & mean$\pm\sigma$ & $2\sigma$\,lower & $2\sigma$\,upper\\
	\hline \hline \hline
$100\Omega_{b } h^2$ & $2.264_{-0.015}^{+0.015}$ & $2.235$ & $2.294$ & $2.240_{-0.016}^{+0.015}$ & $2.209$ & $2.27$ \\\hline
$\Omega_{\rm dm}h^2$  & $0.1163_{-0.0010}^{+0.0010}$ & $0.1143$ & $0.1183$ & $0.1196_{-0.0011}^{+0.0012}$ & $0.1172$ & $0.1219$\\ \hline 
$n_{s}$ & $0.9721_{-0.0043}^{+0.0042}$ & $0.9639$ & $0.9807$ & $0.9656_{-0.0045}^{+0.0040}$ & $0.9574$ & $0.9747$\\ \hline  
$10^{9}A_{s}$ & $2.063_{-0.032}^{+0.035}$ & $1.993$ & $2.133$ & $2.101_{-0.036}^{+0.031}$ & $2.036$ & $2.169$\\ \hline  
$\tau_{\rm reio }$  & $0.0502_{-0.0082}^{+0.0092}$ & $0.0327$ & $0.0686$ & $0.0552_{-0.0085}^{+0.0076}$ & $0.0392$ & $0.0722$\\ \hline  
$w$ & $-0.948_{-0.043}^{+0.022}$ & $-0.9999$ & $-0.8879$ & $-0.979_{-0.021}^{+0.004}$ & $-0.9999$ & $-0.9427$ \\ \hline  
$100~\theta_{s }$  & $1.042_{-0.00031}^{+0.00030}$ & $1.042$ & $1.043$ & $1.042_{-0.00031}^{+0.00031}$ & $1.041$ & $1.043$ \\ \hline  \hline
$z_{\rm reio }$ & $7.14_{-0.86}^{+0.89}$ & $5.298$ & $8.93$ & $7.75_{-0.84}^{+0.76}$ & $6.131$ & $9.376$ \\ \hline  
$H_0$ \mbox{\small  $[\frac{\rm km}{\rm s\,Mpc}]$ } & $67.88_{-0.96}^{+1.20}$ & $65.71$ & $69.97$ & $67.44_{-0.65}^{+0.73}$ & $65.98$ & $68.92$\\ \hline  
$\sigma_8$  & $0.7898_{-0.0093}^{+0.012}$ & $0.7693$ & $0.8094$ & $0.7483_{-0.012}^{+0.011}$ & $0.7253$ & $0.7723$ \\ \hline  
$\Omega_{\rm m }$  & $0.3018_{-0.012}^{+0.009}$ & $0.2817$ & $0.3226$  & $0.3123_{-0.009}^{+0.008}$ & $0.2953$ & $0.3289$ \\ \hline\hline
$\alpha$  & $-$ & $-$ & $-$ & $1.13_{-0.33}^{+0.26}$ & $0.53$ & $1.78$\\ \hline
$\!\log_{10} k_s \mbox{{\small $[{\rm Mpc}^{-1}]\!$}}$  & $-$ & $-$ & $-$ & $8.5_{-5.0}^{+5.2}$ & $-1.1$ & $16.6$ \\ \hline
\hline 
\end{tabular} 
\end{center}
\caption{In this table we give the obtained constraints for the cosmological and derived parameters for a $w$CDM model (left) and for the interacting model with the coupling parameter constant screened (right). }
\label{tab:fitscreening}
\end{table}

\section{Galaxy surveys forecast}
\label{Sec:Forecast}

The results obtained in the preceding Section show how the scenario with the elastic coupling is statistically preferred by data and potentially rule out the standard model at more than $3\sigma$ for some combinations of cosmological data that include low-redshift measurements of clustering. In the upcoming years, several new galaxy surveys are planned that will further improve on their predecessors, being able to constrain the cosmological parameters with even higher precision. In view of the crucial impact that those measurements might have on the determination of the interaction,  we will next proceed to estimate the discriminating power of these new experiments for the elastic coupling scenarios under consideration in this work. We will focus on the Javalambre Physics of the Accelerating universe Astrophysical Survey (J-PAS)~\cite{Benitez:2014ibt}, but we will also consider DESI~\cite{DESI} and Euclid~\cite{EUCLID} for comparison and to estimate the forecast performance of different planned surveys. This forecast analysis will be done using the publicly available code \texttt{FARO}~\cite{Resco:2020dvn}, which is a galaxy survey forecast code to compute Fisher matrices for galaxy clustering and weak lensing observables. As input for \texttt{FARO} we will need the derivatives of the corresponding observables (to be defined below) with respect to the parameters that we want to forecast ($\alpha$ in our case) evaluated on the fiducial model. These derivatives will be obtained from the numerical outputs of the modified \texttt{CLASS} code that includes the interaction. Before entering into the Fisher analysis employed for our forecast, let us briefly introduce the main survey of interest for us, i.e., J-PAS.

\subsection{The J-PAS survey}
J-PAS is a spectro-photometric survey that will be conducted at the Observatorio Astrof\'isico de Javalambre (OAJ) by the 2.5m diameter Javalambre
Survey Telescope (JST/ T250) equipped with the 5 sq. deg. Field of View  Javalambre Panoramic Camera (JP-Cam). The filter system includes 54 narrow- and 4 broad-band filters covering the
optical range~\cite{2017hsa9.conf.....A} which will allow J-PAS to measure the redshift of several millions of Luminous Red Galaxies (LRG), Emission Line Galaxies (ELG) up to $z=1.3$ and quasars (QSO)
up to $z=3.9$ with a precision of $\delta z=0.003(1+z)$. The survey will cover a maximum area of 8500 sq. deg. in the northern hemisphere. 

The optical and filter systems have already been  tested in the mini-J-PAS survey which covered 1 sq. deg. and whose results have been recently published in~\cite{Bonoli:2020ciz}. These results have confirmed that the nominal photo-z accuracy is achieved and that the expected number density of galaxies allows for excellent cosmological measurements. 
Preliminary forecast analysis~\cite{Costa2019,Salzano:2021zxk,AparicioResco2020} showed that thanks to the large number of ELG and LRG detectable, J-PAS will perform very well in the redshift range $0.3<z<0.6$ and also will cover the high-redshift range thanks to the remarkable number of QSO.

J-PAS will have the unique capability of combining  shear measurements of galaxy weak lensing, with (multitracer) clustering observations from redhsift-space distortions and Alcock-Paczynski effects which will largely improve the precision of cosmological parameters measurements~\cite{Resco:2020dvn}.

\subsection{Fisher formalism for galaxy clustering}
\label{sec:clustering}
After this brief introduction of J-PAS, let us turn to explaining the methodology of our forecast analysis. Following the formalism explained in Ref.~\cite{AparicioResco2020} (where we refer for more details), the galaxy power spectrum for multitracer ($i$ and $j$) analysis can be defined as
\begin{equation}
P_{ij}(k_r,\hat{\mu}_r,z)=\frac{D_{A\;r}^2 E}{D_A^2 E_r}\left(A_i +R \hat{\mu}^2\right)\left(A_j +R \hat{\mu}^2\right)\hat{P}(k) e^{-k_r^2 \hat{\mu}_r^2 \sigma_r^2}\; ,
\end{equation}
where the angular distance $D_A$ and  the Hubble function $E(z)\equiv H(z)/H_0$ build the prefactor due to Alcock-Paczynski effect~\cite{Alcock:ACeffect}, the parameters $A_i$ and $R$ are defined as $A_i=D b_i \sigma_8$ and $R=D f \sigma_8$  where $D$ is the growth factor, $f$ the growth rate, $b_i$ the bias of each tracer and $\sigma_8$ is the amplitude of matter perturbations at $8\,h^{-1}$\,Mpc. Also, $\hat{P}(k)$ is defined as $\hat{P}(k)=P(k)/\sigma_8^2$ and $\sigma_r=\delta z (1+z)/H(z)$, with $\delta z $ the photometric redshift error of each survey. The variables are the redshift $z$, the scale $k$ and the angle of separation with the line of sight $\hat{\mu}$. The subindex $r$ means it is evaluated at the fiducial cosmology. Also because of the Alcock-Paczynski effect they are related as 
\begin{equation}
k=Q k_r\quad \text{and}\quad \hat{\mu}=\frac{E}{E_r Q} \hat{\mu}_r,
\end{equation}
with 
\begin{equation}
Q=\frac{\sqrt{D_A^2 \chi^2 \hat{\mu}_r^2-D_{A \, r}^2 \chi_r^2(\hat{\mu}_r^2-1)}}{D_{A \, r} \chi},    
\end{equation}
being $\chi=\int_0^z\frac{dz}{H(z)}$ the comoving distance.

As shown in Ref.~\cite{AparicioResco2020}, the Fisher matrix for clustering for different tracers $i$ and $j$ centred  at redshift bin $z_a$ is 
\begin{equation}
F^C_{\alpha\beta}(z_a)=\frac{V_a}{8\pi^2}\int_{-1}^{1}{\rm d}\hat{\mu}\int_{k_{\rm min}}^{\infty}{\rm d}k k^2 \dfrac{\partial P_{ij}(k,\hat{\mu},z_a)}{\partial p_\alpha}\bigg{|}_r\:C^{-1}_{j l}\:\dfrac{\partial P_{lm}(k,\hat{\mu},z_a)}{\partial p_\beta}\bigg{|}_r\:C^{-1}_{mi}\: e^{-k^2 \Sigma^2_{\perp}-k^2\hat{\mu}^2(\Sigma_\parallel^2-\Sigma_\perp^2)}\:,
\end{equation}
where $V_a$ is the total volume of the $a$-th bin, $k_{\rm min}$ is fixed to $0.007 h$\,Mpc$^{-1}$, $\{p_\alpha\}$ are the set of cosmological parameters we are doing the forecast for, $C_{ij}$ is the data covariance matrix defined as $C_{ij}=P_{ij}+\frac{\delta_{ij}}{\bar{n_i}}$, with $\bar{n}_i$ the mean galaxy density of tracer $i$ at $z_a$. Finally, the exponential suppression factor is introduced to remove the non-linear scales from the analysis, since properly including those scales necessitates an appropriate characterisation of the non-linear clustering that we lack, and the functions $\Sigma_\parallel$ and $\Sigma_\perp$ are defined as
\begin{eqnarray}
\Sigma_\perp(z) & = & 0.785 D(z)\Sigma_0\;, \\
\Sigma_\parallel(z) & = & 0.785 D(z)\Big[1+f(z)\Big]\Sigma_0\;,
\end{eqnarray}
where $\Sigma_0 = 11 \, \mathrm{Mpc/}h$. We should bear in mind that this parameterisation was obtained for the standard $\Lambda$CDM model so, in principle, it is not directly applicable to our scenario with the elastic interaction because it gives a substantially different linear clustering evolution. Although it is expected that the non-linear clustering will also differ from $\Lambda$CDM, thus casting doubts on the use of the above parameterisation, we do not expect the non-linear scale to differ significantly between both models. Hence, it is justified to use the above parameterisation also for the interacting model. Furthermore, the results of the Fisher analysis for our forecasts are not expected to be very sensitive to these details and it is then sensible to use the standard model values. For the real analysis with data, the difference can be more relevant and care should be taken in order to optimise the amount of information extracted from data.

In our analysis we use a binned distribution in scale $k$ from $0.007 h$\,Mpc$^{-1}$ to $10 h$\,Mpc$^{-1}$ with 20 bins logarithmically equispaced, in the case of the angle of line of sight $\hat{\mu}$ from $-1$ to $1$ with 200 bins equispaced, while the index $i$ and $j$ are the tracers: Luminous Red Galaxies (LRG), Emission Line Galaxies (ELG), Quasars (QSO) and Bright Galaxies (BGS). The redshift distribution depends on the experiment and are explained for the three surveys used for clustering J-PAS, DESI and spectroscopic Euclid in Appendix~\ref{ap:specifications}.

For our fiducial cosmology we consider a flat universe with, $\Omega_{\rm m }=0.310$, $h=0.6774$, $w=-0.98$ and $n_s=0.96$. Due to the more than $3\sigma$ result obtained in the previous sections we consider two fiducial cosmologies, one with $\alpha=0$ and the other with  $\alpha=1$. In addition, as we will consider different tracers for the galaxy clustering power spectra, we need the fiducial values for biases. Since in the early universe dark matter decouples from baryons due to electromagnetic interactions, there is a bias between galaxies and dark matter halos. This bias evolves with the galaxy formation and evolution, and also depends on selection effects of different galaxy surveys. We will consider the following fiducial biases for each tracer~\cite{AparicioResco2020},
\begin{eqnarray}
b_{\rm LRG}&=&\frac{1.7}{D(z)}\;,\\
b_{\rm ELG}&=&\frac{0.84}{D(z)}\;,\\
b_{\rm QSO}&=&0.53+0.289(1+z)^2\;,\\
b_{\rm BSG}&=&\frac{1.34}{D(z)}\;,
\label{eq:biasfix}
\end{eqnarray}
for J-PAS and DESI, while for Euclid $b_{\rm ELG}=\sqrt{1+z}$~\cite{EUCLID}. Since the interacting scenario that we are exploring in this work can potentially modify the bias, we will perform two analysis. Firstly, we will fix the bias to the previous fiducial values. Secondly, we will marginalise over the bias as free parameter in each redshift bin $z_a$, in order to take into account the possible correlations of our desired parameters with the bias. 

The photometric redshift error $\delta z$ for J-PAS is $\delta z=0.003$ for  LRG, ELG and QSO, for DESI is $\delta z=0.0005$ for BGS, LRG and ELG and $\delta z=0.001$ for QSO, while for EUCLID is $\delta z=0.001$ for ELG.

\subsubsection{Clustering results}

\begin{figure}
	\centering
	\includegraphics[scale=0.49]{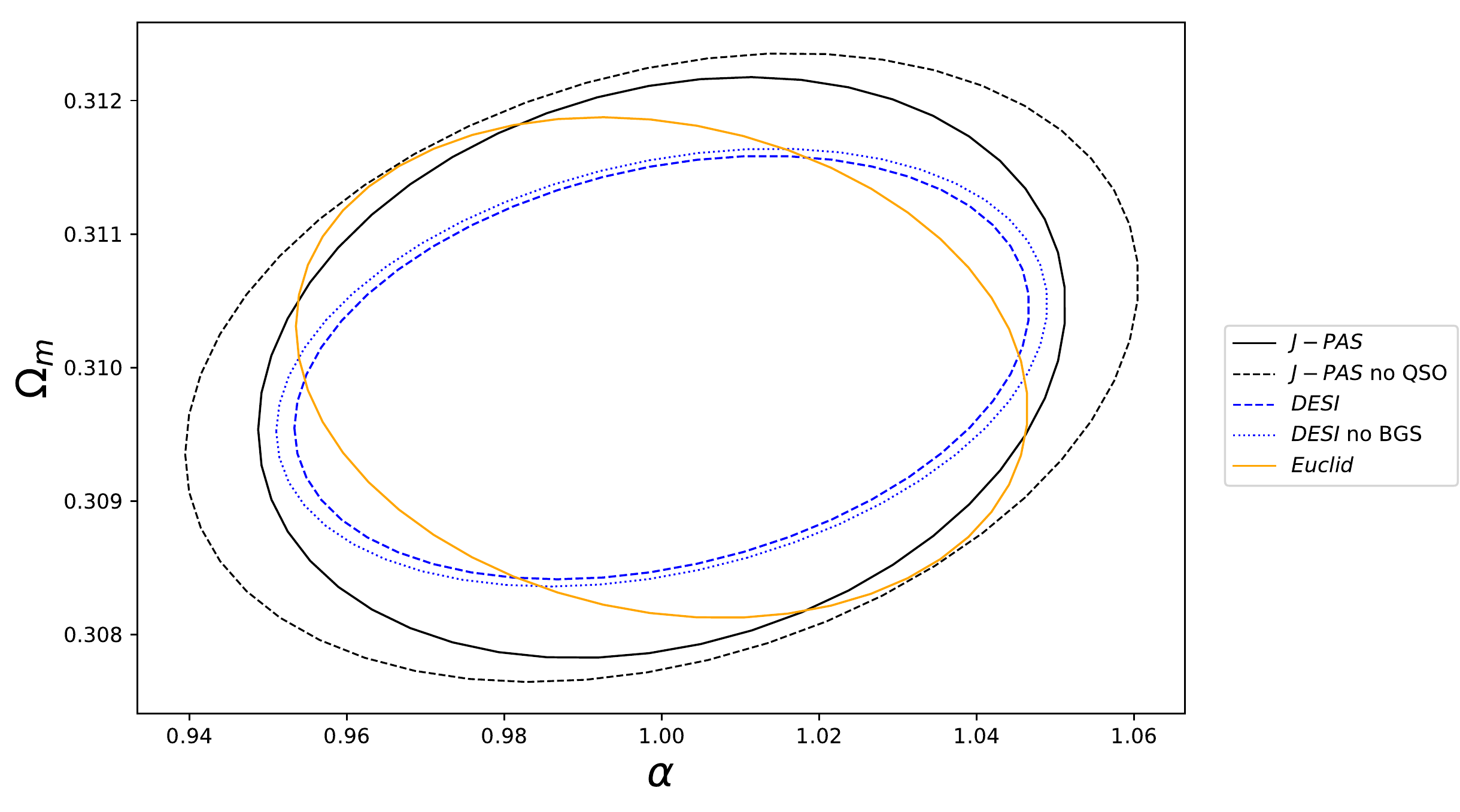}
	\caption{Using as fiducial cosmology $\alpha=1$ and marginalising the bias, the 1$\sigma$ contour errors for the parameters $\Omega_{\rm m }$ and $\alpha$ using clustering for the surveys J-PAS with ELG+LRG+QSO tracers (black solid line) and without QSO (black dashed line), for DESI with LRG+ELG+QSO+BGS (blue dashed line) and without BGS (blue dotted line) and for Euclid using ELG (yellow solid line). }
	\label{fig:clustering}
\end{figure}

In this Section we present the forecast analysis for $\Omega_{\rm m }$  and interaction parameter $\alpha$ for the surveys J-PAS~\cite{Benitez:2014ibt}, DESI~\cite{DESI} and Euclid~\cite{EUCLID} using clustering information. If we consider the results shown in Section~\ref{Sec:Fit} to be reliable, then, our fiducial cosmology is the one with  $\alpha=1$. In Figure~\ref{fig:clustering} , we provide the 1$\sigma$ contour error for the fiducial cosmology $\alpha=1$ that can be measured using the surveys J-PAS using tracers ELG+LRG+QSO or without QSO (black lines), DESI with LRG+ELG+QSO+BGS or without BGS (blue lines) and  Euclid using ELG (yellow line). 

With all the three considered surveys, we have an expected constraint of order $\sim5\%$ on our model parameter $\alpha$ if we consider the fiducial cosmology to have $\alpha=1$. These results would lead to a 10$\sigma$ detection of the interaction and then, if the specifications of each survey are satisfied, such a detection is falsifiable with the data obtained in the next cosmological experiments. In particular, for the J-PAS survey, we can obtain the minimum area necessary to obtain a detection of $\alpha = 1$ at 5$\sigma$. Since the clustering Fisher matrix is proportional to the fraction of sky, we obtain that given a 1$\sigma$ marginalised constraint $\sigma_a$ for a parameter $a$ for a given area $\deg^2_a$, we can calculate the new error for a new area as $\sigma_b = \sqrt{\deg^2_a / \deg^2_b} \, \sigma_a$. Using this relation, we obtain that J-PAS needs only $983$ $\deg^2$ to be able to provide a detection of $\alpha = 1$ at 5$\sigma$.

These results have been obtained marginalising the bias in each redshift bin to prevent effects from having a different clustering evolution in the presence of the interaction. Furthermore, we have fixed the fiducial cosmology to a non-standard model. Thus, in order to test the robustness of our analysis against these choices,  we will now proceed to analyse the change in the expected errors against a change of the fiducial model and the bias marginalisation.

\subsubsection{Forecast marginalising bias}

\begin{figure}
	\centering{
	\includegraphics[scale=0.49]{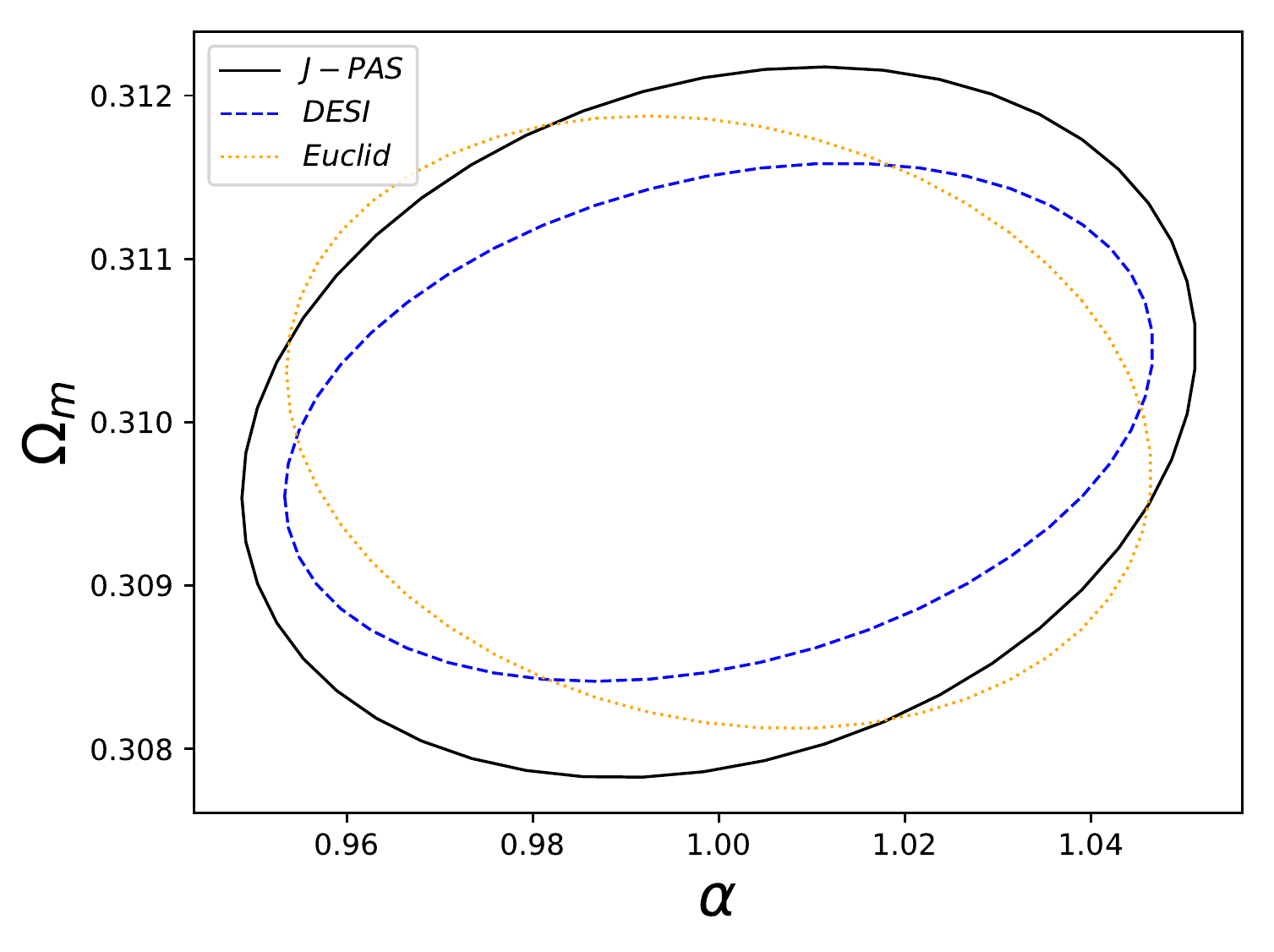}
	\includegraphics[scale=0.49]{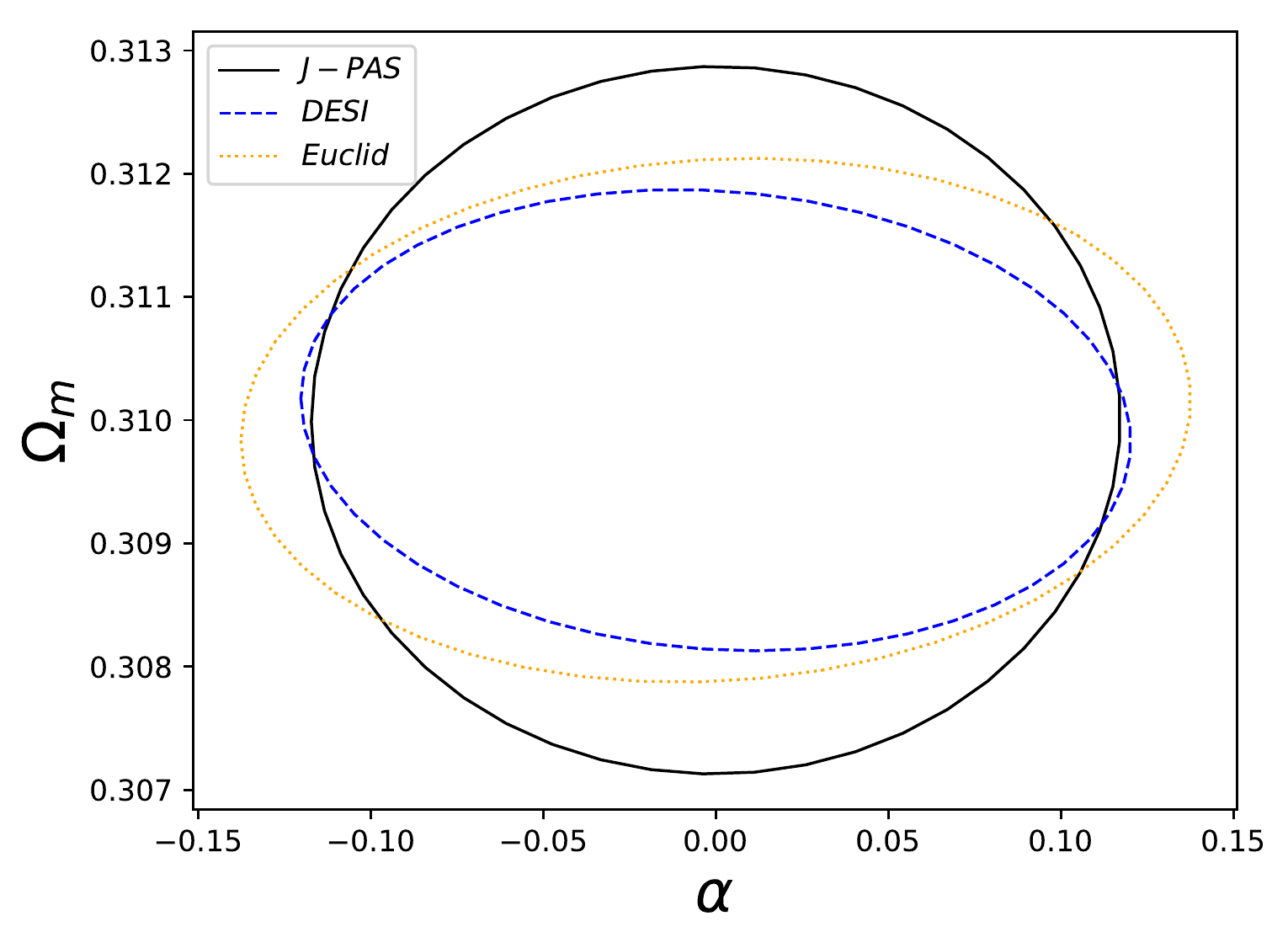}}
	\caption{Using the fiducial cosmology with $\alpha=1$ (left) and with $\alpha=0$ (right) and marginalising the bias, the 1$\sigma$ contour errors for the parameters $\Omega_{\rm m }$ and $\alpha$ using clustering data for the surveys J-PAS with ELG+LRG+QSO tracers (black solid line), DESI with LRG+ELG+QSO+BGS (blue dashed line) and  Euclid using ELG (yellow dotted line).}
	\label{fig:clusteringmarginalized}
\end{figure} 

 In Figure~\ref{fig:clusteringmarginalized}, we show the expected 1$\sigma$ contours for $\Omega_{\rm m }$ and $\alpha$ with a bias marginalisation in each redshift bin, for the fiducial cosmology both the standard one and the bestfit result obtained in Section~\ref{Sec:Fit}. In Table~\ref{tab:alphactam} we show the marginalised errors for the surveys and fiducial cosmologies used. We find an improvement by a factor $\sim$4 on the constraints for the interaction parameter $\alpha$ with respect to the results of Section~\ref{Sec:Fit}. As we have marginalised the bias in each redshift, the obtained errors are higher than the ones  we will obtain in the next section without marginalising. This is expected since the loss of information on the bias reduces the precision of the measurements. 

\begin{table}
	\caption{ Marginalised errors of the parameters $\Delta \Omega_m$ and $\Delta \alpha$ using clustering data and marginalising the bias for J-PAS, Euclid and DESI surveys.} \label{tab:alphactam} 
	\begin{center}
		\begin{tabular}{|c||c|c||c|c|} 
		    \hline
			&\multicolumn{2}{c||}{Fiducial $w$CDM} &\multicolumn{2}{c|}{Fiducial $\alpha=1$}\\ \hline
			\hline 
			Survey& $\Delta \Omega_m$ & $\Delta \alpha$ &$\Delta \Omega_m$ & $\Delta \alpha$\\ \hline 
			J-PAS & 0.0038& 0.155& 0.0029& 0.068\\ \hline
			Euclid & 0.0028&  0.182& 0.0025 &  0.062\\ \hline
			DESI & 0.0025&  0.159& 0.0021&  0.062 \\ \hline
		\end{tabular} \\ 
	\end{center}
\end{table}

\subsubsection{Forecast with standard bias}

\begin{figure}
	\centering{
		\includegraphics[scale=0.49]{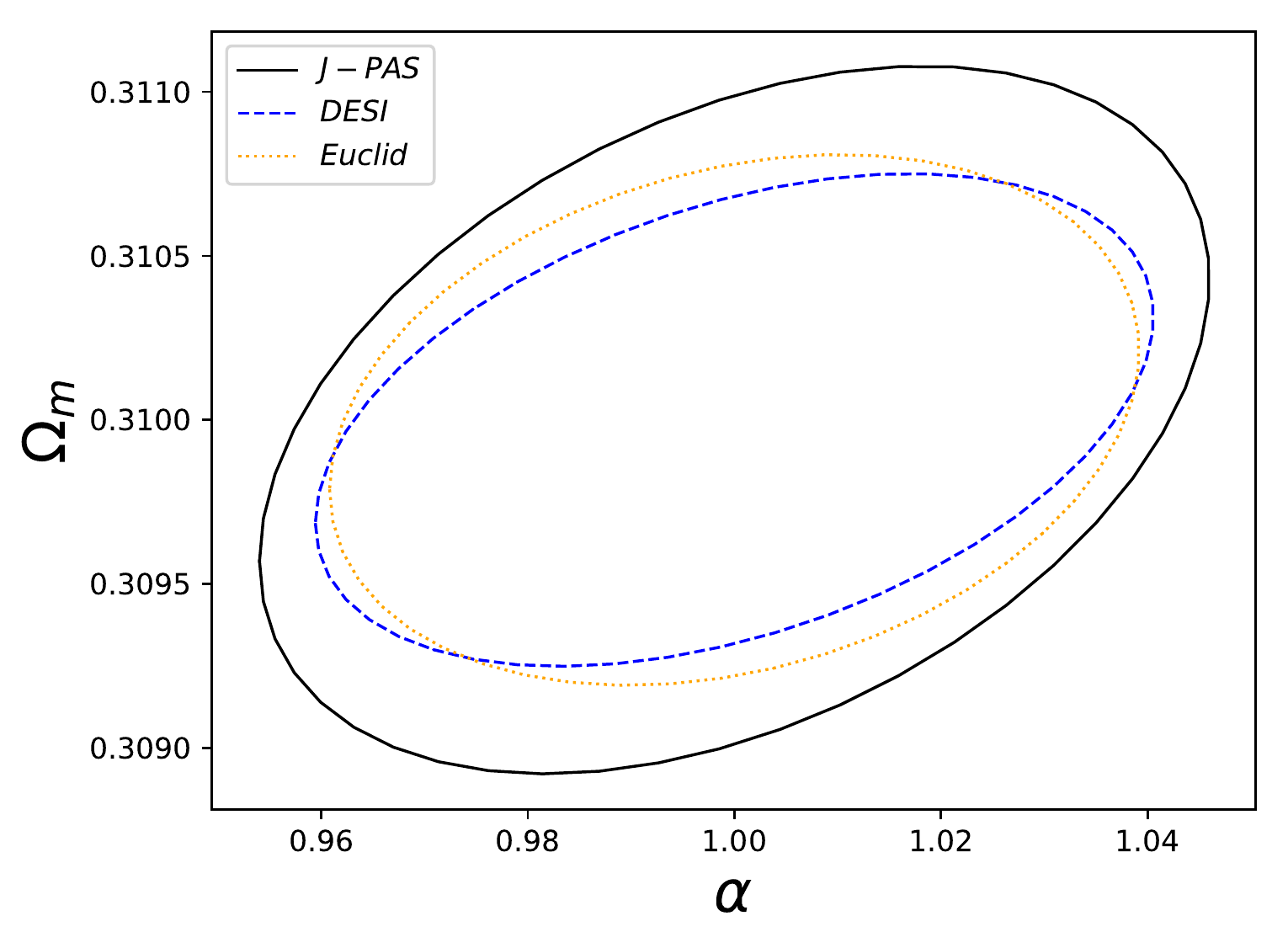}
		\includegraphics[scale=0.49]{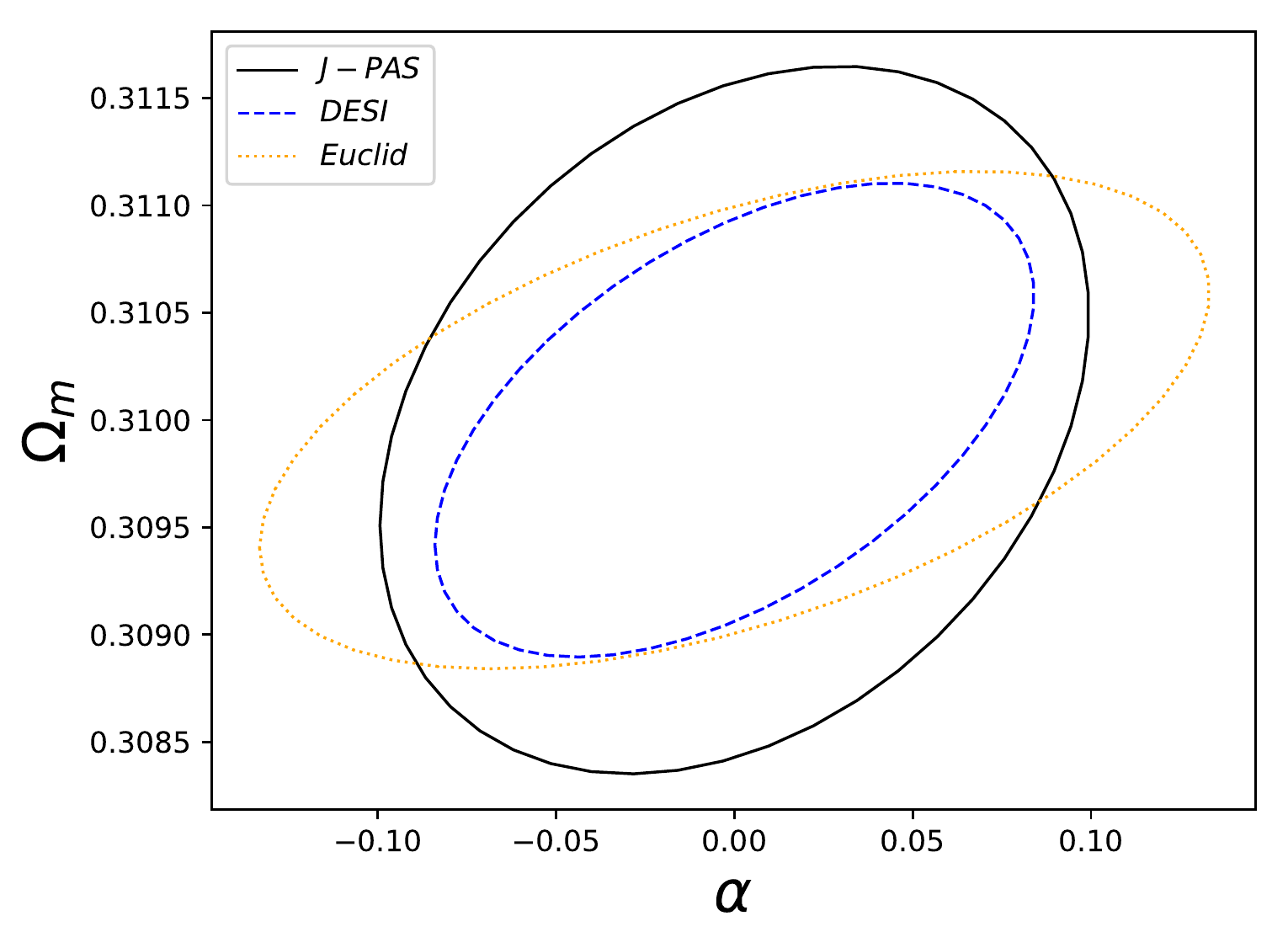}}
	\caption{We show the 1$\sigma$ contour errors for $\Omega_{\rm m }$ and $\alpha$ for the fiducial cosmologies with $\alpha=1$ (left) and with $\alpha=0$ (right) and fixing the bias to the standard model one~\eqref{eq:biasfix}. These results correspond to using clustering data for the surveys J-PAS with ELG+LRG+QSO tracers (black solid line), DESI with LRG+ELG+QSO+BGS (blue dashed line) and  Euclid using ELG (yellow dotted line).}
	\label{fig:clusteringfixed}
\end{figure} 

If we assume that the bias is not substantially modified by the interaction, it is justified to use the bias of the standard model introduced in Section~\ref{sec:clustering}. The benefit of not marginalising the bias is that we do not add error to our parameters $\Omega_{\rm m }$ and $\alpha$ due to the loss of information. As before, in Figure~\ref{fig:clusteringfixed} and in Table~\ref{tab:alphactaLCDM} we present the 1$\sigma$ contour errors and the marginalised errors of the parameters $\Omega_m$ and $\alpha$ using as fiducial cosmology the bestfit one or the standard model for the surveys J-PAS, DESI and Euclid. Again, we find stronger constraints than using the currently available data used in Section~\ref{Sec:Fit}, which can lead to a potential detection of this interaction.

\begin{table}
	\caption{ Marginalised errors of the parameters $\Delta \Omega_m$ and $\Delta \alpha$ using clustering data and bias fixed to the one in $\Lambda$CDM for J-PAS, Euclid and DESI surveys.} \label{tab:alphactaLCDM} 
	\begin{center}
			\begin{tabular}{|c||c|c||c|c|} 
		    \hline
			&\multicolumn{2}{c||}{Fiducial $w$CDM} &\multicolumn{2}{c|}{Fiducial $\alpha=1$}\\ \hline
			\hline 
			Survey& $\Delta \Omega_m$ & $\Delta \alpha$ & $\Delta \Omega_m$ & $\Delta \alpha$ \\ \hline 
			J-PAS & 0.0022& 0.132& 0.0014& 0.061\\ \hline
			Euclid & 0.0015&  0.176& 0.0011& 0.052\\ \hline
			DESI & 0.0015& 0.111& 0.0010& 0.054 \\ \hline
		\end{tabular} \\ 
	\end{center}
\end{table}

\subsection{Fisher formalism for weak lensing}
As in the clustering forecast, we will follow the formalism explained in Ref.~\cite{AparicioResco2020} for the weak lensing analysis. The relevant observable in this case is the convergence power spectrum for a redshift tomography analysis that is given by
\begin{equation}
P_{ij}(\ell)=H_0 \sum_{a}\frac{\Delta z_a}{E_a} K_i(z_a) K_j(z_a)  \Omega_m^2 D(z_a)^2 \sigma_8^2 \hat{P}\left(\frac{\ell}{\chi(z_a)}\right)   \; ,
\end{equation}
where $\hat{P}(k)=P(k)/\sigma_8^2$ and $D(z_a)$ is the growth factor, the index $i$ and $j$ refer to the bin in redshift and $\ell=k\chi(z)$. The functions $K_i$ are defined as
\begin{equation}
K_i(z)=\frac{3H_0}{2}(1+z)\int_{z}^{\infty} \left(1-\frac{\chi(z)}{\chi(z')}\right)n_i(z')dz'\;,
\end{equation}
where $n_i(z)$ is the density function in the $i$th-bin of redshift such that considering photometric redshift error $\sigma_i=\delta z(1+z_i)$ 
\begin{equation}
n_i(z)\propto \int_{\bar{z}_{i-1}}^{\bar{z}_i}z'^2 e^{-(z'/z_p)^{3/2}}e^{-\frac{(z'-z)^2}{2\sigma_i^2}}dz\; ,
\end{equation}
normalised for each $n_i$ as $\int_{0}^{\infty}n_i(z)dz=1$. $z_p$ is defined as $z_p=z_{\rm mean}/\sqrt{2}$, with $z_{\rm mean}$ the mean redshift of each survey. 
Finally, the Fisher matrix for weak lensing is 
\begin{equation}
F^C_{\alpha\beta}=f_{\rm sky}\sum_{\ell}\Delta \ln \ell\frac{(2\ell+1)\ell}{2}\;\mathrm{Tr}\left[\frac{\partial \mathbf{P}}{\partial p_\alpha} \mathbf{C}^{-1}\frac{\partial \mathbf{P}}{\partial p_\beta} \mathbf{C}^{-1}\right]   \:,
\end{equation}
with $C_{ij}=P_{ij}+\gamma_{int}^2\hat{n}_i^{-1}\delta_{ij}$, where the intrinsic ellipticity  is $\gamma_{int}=0.22$ and the number of galaxies per steradian in the $i$th-bin $\hat{n}_i$ is defined as
\begin{equation}
\hat{n}_i=n_\theta\frac{\int_{\bar{z}_{i-1}}^{\bar{z}_i} n_i(z) dz}{\int_{0}^{\infty}n_i(z)dz}\;,
\end{equation}
where $n_\theta$ is the areal galaxy density, which for J-PAS is $n_\theta=12.32$ and for Euclid $n_\theta=30$.

The $i$th-bin distribution depends on the survey used for the lensing analysis (J-PAS and photometric Euclid) and they are explained in Appendix~\ref{ap:specifications}, while for the scale distribution $\ell$ we use $\Delta \ln \ell=0.1$ from $\ell_{\rm min}=5$ and $\ell_{\rm max}=906$. The mean redshift is $z_{\rm mean}=0.5$ for J-PAS and $z_{\rm mean}=0.9$ for Euclid, while the photometric redshift error is $\delta z=0.03$ for J-PAS and $\delta z=0.05$ for Euclid. 
As before, our two fiducial cosmologies are a flat universe with $\Omega_{\rm m }=0.310$, $h=0.6774$, $w=-0.98$, $n_s=0.96$ and both values of the coupling parameter  $\alpha=0$ and  $\alpha=1$ respectively. 

\subsubsection{Results weak lensing }

\begin{figure}
	\centering{
		\includegraphics[scale=0.50]{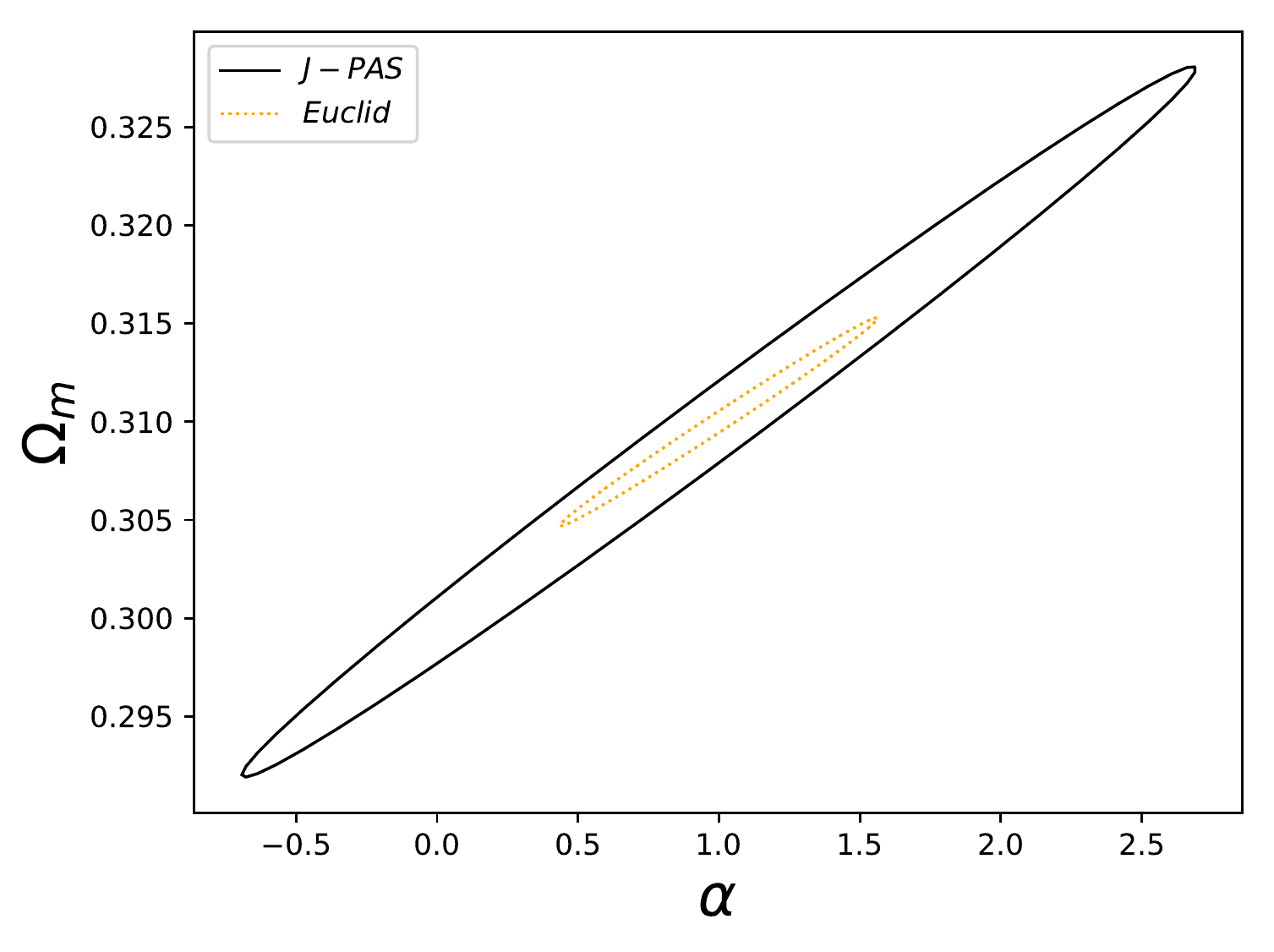}
		\includegraphics[scale=0.50]{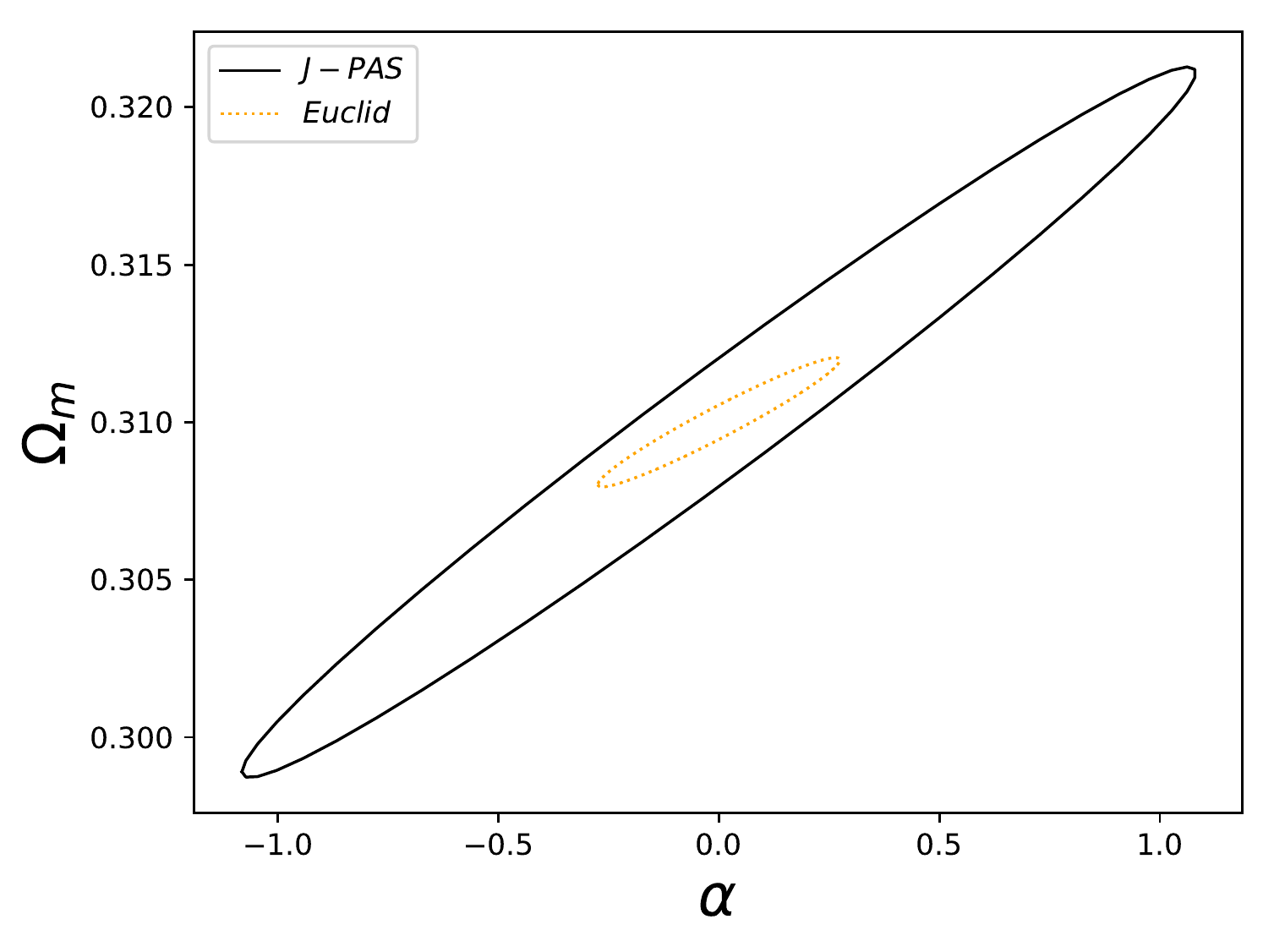}}
	\caption{Using the fiducial cosmology with $\alpha=1$ (left) and with $\alpha=0$ (right), the 1$\sigma$ contour errors for the parameters $\Omega_{\rm m }$ and $\alpha$ using lensing data for the surveys J-PAS with ELG+LRG tracers (black solid line),  and  Euclid using ELG (yellow dotted line).}
	\label{fig:lensing}
\end{figure} 

In this section we present the expected constraints on the cosmological parameter $\Omega_{\rm m }$  and on the model parameter $\alpha$, for the surveys J-PAS~\cite{Benitez:2014ibt} and Euclid~\cite{EUCLID} using lensing information. In Figure~\ref{fig:lensing} we give the expected 1$\sigma$ contour errors for both surveys while in Table~\ref{tab:alphactalens} we show the marginalised errors, using as fiducial cosmology the bestfit result of Section~\ref{Sec:Fit} and the standard model. Although the constraints for $\Omega_{\rm m }$ can be competitive, the constraints on $\alpha$ in the best scenario for Euclid and the standard cosmology are worse than the clustering ones or the results obtained in Section~\ref{Sec:Fit}. J-PAS is even less capable of constraining the model parameter $\alpha$. 

The poor results obtained with lensing can be explained by the effect the model has in the growth function $D(z)$ and the lensing potential $\Psi$. For the values of $\alpha$ considered here, the impact on  $D(z)$ and $\Psi$, in combination with the inability of this model to modify the slip parameter, are not enough to have competitive results for the errors with these surveys, specially for the case of J-PAS.

\begin{table}
\renewcommand{\arraystretch}{1.8}
	\caption{ Marginalised errors of the parameters $\Delta \Omega_m$ and $\Delta \alpha$ using lensing data for J-PAS and Euclid surveys.} \label{tab:alphactalens} 
		\begin{center}
			\begin{tabular}{|c||c|c||c|c|} 
		    \hline
			&\multicolumn{2}{c||}{Fiducial $w$CDM} &\multicolumn{2}{c|}{Fiducial $\alpha=1$}\\ \hline
			\hline 
			Survey& $\Delta \Omega_m$ & $\Delta \alpha$ & $\Delta \Omega_m$ & $\Delta \alpha$\\ \hline 
			J-PAS & 0.0149& 1.43& 0.0239& 2.24\\ \hline
			Euclid & 0.0027&  0.36 & 0.0070 &  0.74\\ \hline
		\end{tabular} \\ 
	\end{center}
\end{table}

\section{Discussion and conclusions}

In this work we have explored a cosmological scenario where dark matter and dark energy are coupled through their Euler equations, which arises from an interaction with no energy transfer. We have confirmed that these scenarios provide promising candidates for alleviating the existing tension in cosmological data when compared to the amplitude of the fluctuations inferred from CMB and those measured by low redshift galaxy surveys. It is remarkable that the latest cosmological data combined strongly favour the existence of an interaction with the non-interacting case excluded at more than $3\sigma$. However, this detection is driven by PlanckSZ data, whose inclusion might not be well justified as we have explained. In any case, the obtained result clearly calls for a more detailed examination of these scenarios. 

We have then proceeded to analyse the ability of future galaxy surveys to unveil the true presence of the interactions. Since one of the main effects of the interacting model is precisely a suppression of the clustering, these surveys might prove to be ideal for constraining these scenarios. Thus, we have used two observables based on clustering and weak lensing and we have compared the forecasts for J-PAS, DESI and Euclid. The analysis from clustering shows that the interaction will be definitely detected by these surveys and, despite their different specifications, all three of them will provide similar precision on the measurement of the interaction parameter, at the level of a few percent, with the non-interacting model clearly excluded. On the other hand, the weak lensing forecast provides a much poorer constraining power than clustering. The effects of this elastic model in the growth of structures that, at the end, modify the weak lensing are not enough to have competitive results with the surveys studied. Hence, clustering data will be the key for constraining the coupling parameter $\alpha$.

\paragraph{Codes:} Modified versions of the codes \texttt{CLASS} and \texttt{CAMB} for the computation of the
evolution of linear perturbations are available on request.

\paragraph{Acknowledgments:} 
We thank Wilmar Cardona for useful discussions on the fit to different datasets. JBJ, DB, DF and FATP acknowledge support from the {\it Atracci\'on del Talento Cient\'ifico en Salamanca} programme, from project PGC2018-096038-B-I00 by Spanish {\it Ministerio de Ciencia, Innovaci\'on y Universidades} and {\it Ayudas del Programa XIII} by USAL. DF acknowledges support from the programme {\it Ayudas para Financiar la Contratación Predoctoral de Personal Investigador (ORDEN EDU/601/2020)} funded by {\it Junta de Castilla y Leon} and {\it European Social Fund}. This work has been supported by the MINECO (Spain) projects FIS2016-78859-P and PID2019-107394GB-I00 (AEI/FEDER, UE). 
LRA  acknowledges  financial support from CNPq (306696/2018-5) and FAPESP (2015/17199-0). 
JA is supported by CNPq (Grants No. 310790/2014-0 and 400471/2014-0) and FAPERJ (Grant No. 233906). 
SB acknowledges PGC2018-097585-B-C22, MINECO/FEDER,  UE  of  the  Spanish  Ministerio  de  Economia,  Industriay  Competitividad.  
CEFCA  researchers  acknowledge  support  from  the  project PGC2018-097585-B-C21.
SC is supported by CNPq (Grants No. 307467/2017-1 and420641/2018-1). 
R.A.D.  acknowledges  support  from  the  Conselho Nacional  de  Desenvolvimento  Científico  e  Tecnológico  -  CNPq  through  BPgrant  308105/2018-4,  and  the  Financiadora  de  Estudos  e  Projetos  -  FINEP grants  REF.  1217/13  -  01.13.0279.00  and  REF  0859/10  -  01.10.0663.00 and  also  FAPERJ  PRONEX  grant  E-26/110.566/2010  for  hardware  funding support  for  the  J-PAS  project  through  the  National  Observatory  of  Braziland  Centro  Brasileiro  de  Pesquisas  Físicas.
VM thanks CNPq (Brazil) and FAPES (Brazil) for partial financial support. This project has received funding from the European Union’s Horizon 2020 research and innovation programme under the Marie Skłodowska-Curie grant agreement No 888258.
CMdO   acknowledges   support   from   Brazilian agencies CNPq (grant 312333/2014-5) and FAPESP (grant 2009/54202-8.
IAA  researchers acknowledge  financial  support  from  the  State  Agency  for  Research  of  the Spanish  MCIU  through  the  “Center  of  Excellence  Severo  Ochoa”  award  to the  Instituto  de  Astrofísica  de  Andalucía  (SEV-2017-0709).
LSJ acknowledges  support  from  Brazilian  agencies  CNPq  (grant  304819/2017-4) and  FAPESP  (grant  2012/00800-4).\\

This paper has  gone through internal  review  by the J-PAS collaboration. Based on observations made with the JST/T250 telescope and JPCam at the Observatorio Astrofísico de Javalambre (OAJ), in Teruel, owned, managed, and operated by the Centro de Estudios de Física del Cosmos de Aragón (CEFCA). We acknowledge the OAJ Data Processing and Archiving Unit (UPAD) for reducing and calibrating the OAJ data used in this work.
Funding for the J-PAS Project has been provided by the Governments of Spain and Aragón through the Fondo de Inversión de Teruel, European FEDER funding and the Spanish Ministry of Science, Innovation and Universities, and by the Brazilian agencies FINEP, FAPESP, FAPERJ and by the National Observatory of Brazil. Additional funding was also provided by the Tartu Observatory and by the J-PAS Chinese Astronomical Consortium.
\\
\appendix
\section{Equations in synchronous gauge}
\label{ap:sync}
The synchronous gauge is defined by the perturbed line element 
\begin{equation}
\dd s^2=a^2(\tau) \Big[-\dd \tau^2 + (\delta_{ij}+h_{ij}) \dd x^i \dd x^j \Big] \;,
\end{equation}
where the perturbed spatial metric is written in terms of the scalar perturbations $h$ and $\eta$ as $h_{ij} = {\rm diag}(-2\eta, -2\eta, h+\eta)$. Following our previous notation, the conservation equations governing the dark sector perturbations read
\begin{eqnarray}
\deltadm'&=&-\left(\thetadm + \frac{1}{2}h'\right) \;, \\
\thetadm'&=&-\mathcal{H} \thetadm + \Gamma(\thetade-\thetadm)\;, \\
\deltade'&=&-3 \mathcal{H}\left(\cs^2-w\right)\deltade  -(1+w)\left(\thetade + \frac{1}{2}h'\right) + 9 (1+w)\mathcal{H}^2\frac{\cs^2-w}{k^2}\thetade \;, \\
\thetade'&=&\left(-1+3\cs^2\right) \mathcal{H}\thetade +\frac{k^2 \cs}{1+w}\deltade-\Gamma R(\thetade-\thetadm)\;,
\end{eqnarray}
where a constant DE equation of state parameter is assumed and $\Gamma$ and $R \Gamma$ conserve the same previous definitions.

\section{Brief discussion on stability}
\label{App:Stability}
In this appendix we will give a brief description of the potential appearance of instabilities  in the evolution of the perturbations due to the interaction. To that end, we first notice that the equations for the evolution of the perturbations can be combined to give a system of two second order equations involving only the density contrasts. If we consider sub-Hubble modes $k\gg\mathcal{H} $ during the matter dominated epoch, the resulting equations can be written as
\begin{eqnarray}
\left[\begin{array}{c}
      \deltadm''\\
      \deltade''
\end{array}  \right]
&+&
\left[\begin{array}{cc}
  \mathcal{H}+\Gamma   & -\frac{\Gamma}{1+w}\\
    -(1+w)\Gamma R  & \;\;(1-3w)\mathcal{H}+R\Gamma
\end{array}  \right]
\left[\begin{array}{c}
      \deltadm'\\
      \deltade'
\end{array}  \right]\nonumber\\
&+&
\left[\begin{array}{cc}
 -\frac32 \mathcal{H}^2   & 3\frac{w-\cs^2}{1+w}\mathcal{H}\Gamma\\
   -\frac32(1+w)\mathcal{H}^2  & \;\;\cs^2k^2+3(\cs^2-w)\mathcal{H}R\Gamma
\end{array}  \right]
\left[\begin{array}{c}
      \deltadm\\
      \deltade
\end{array}  \right]=0
\label{eq:deltasAppB}
\end{eqnarray}
where we have used that $R\gg1$, $k^2\Phi\simeq -\frac32 \mathcal{H}^2\deltadm$ and $\mathcal{H}'\simeq-\frac12\mathcal{H}^2$ that hold well inside the matter dominated epoch. The propagation speeds for the high frequency modes with $k^2\gg {\max }\{\mathcal{H}^2,R\Gamma\mathcal{H}\}$ are not modified by the interaction so we still have the vanishing sound speed of the dust component and $\cs$ for the dark energy fluid in that regime. Thus, no risk of UV Laplacian instabilities arise from the interaction, i.e., the characteristic frequencies of the modes will be real and no exponentially growing modes arise.

On the other hand, other instabilities such as tachyonic modes could be triggered by the interaction due to its effects on the friction and the effective masses. Since the interaction is completely negligible on super-Hubble scales because of the common large scale frame of all the fluids, infrarred instabilities do not arise. This is easy to understand because the standard non-interacting evolution is recovered at those scales. The interaction however leads to the appearance of a second horizon for the dark energy fluid given that is given by the relation
\begin{equation}
k_{\star}^2=\left\vert3\left(1-\frac{w}{\cs^2}\right) R\Gamma \mathcal{H}\right\vert.
\end{equation}
This scale is obtained by comparing the two competing terms $\cs^2k^2$ and $3(\cs^2-w)\mathcal{H}R\Gamma$ in the second line of \eqref{eq:deltasAppB}. For sub-Hubble scales that are above this horizon $\mathcal{H}\ll k\ll k_{\star}$, the $k$-dependence of the equations disappear because the only 
$k-$dependent term $\cs^2 k^2$ can be safely neglected and, if we consider the strongly coupled regime\footnote{Notice that there is a wide band of modes in this regime since $k_{\star}\gg\mathcal{H}$.} $\Gamma\gg\mathcal{H}$, the determinant of the mass matrix $\hat{M}$ in the second line of \eqref{eq:deltasAppB} reduces to
\begin{equation}
\det\hat{M}\simeq-\frac92(\cs^2-w)R\Gamma\mathcal{H}^3
\end{equation}
From this mass term we expect to  have a negative eigenvalue corresponding to the unstable mode leading to the Jeans collapse of the DM component so in order to avoid an additional tachyonic mode we need to impose $\det\hat{M}<0$ that is achieved for $\Gamma>0$, i.e., $\alpha>0$. One could argue why would we want to remove the growing modes, especially since we already have the usual Jeans instability. The reason is that the growing modes associated to the interaction will develop on a time scale parameterically determined by $R\Gamma$ that will be much shorter than the cosmological Hubble time scale since we are in the regime $\Gamma\gg\mathcal{H}$. We do not intend to perform a detailed analytical analysis of the evolution of the perturbations here, but simply motivate why we mostly restrict to $\alpha>0$ in our analysis since this guarantees $\Gamma>0$ and, therefore, the avoidance of the discussed instabilities. Furthermore, we have checked also numerically that the region $\alpha<0$ typically leads to instabilities that spoil the possibility of sensible cosmologies unless $\vert\alpha\vert$ is sufficiently small.

\section{Survey specifications for bins and galaxy densities}
\label{ap:specifications}
In the following tables we show the redshift bins distributions and the densities of  Luminous Red Galaxies~(LRG), Emission Line Galaxies~(ELG), Quasars~(QSO) and Bright Galaxies~(BGS) for the differents surveys used, both for clustering (J-PAS, DESI and spectroscopic Euclid) and lensing analysis (J-PAS and photometric  Euclid). Galaxy densities are in units of $10^{-5} h^3\,$Mpc$^{-3}$.

\begin{center}
\begin{tabular}{|c||c|c|c|}
 \hline
 \multicolumn{4}{|c|}{J-PAS} \\ 
 \hline 
 $z$&LRG&ELG&QSO\\\hline
\hline 0.3 & 226.6 & 2958.6 & 0.45 \\
\hline 0.5 & 156.3 & 1181.1 & 1.14 \\
\hline 0.7 & 68.8 & 502.1 & 1.61 \\
\hline 0.9 & 12.0 & 138.0 & 2.27 \\
\hline 1.1 & 0.9 & 41.2 & 2.86 \\
\hline 1.3 & 0 & 6.7 & 3.60 \\
\hline 1.5 & 0 & 0 & 3.60 \\
\hline 1.7 & 0 & 0 & 3.21 \\
\hline 1.9 & 0 & 0 & 2.86 \\
\hline 2.1 & 0 & 0 & 2.55 \\
\hline 2.3 & 0 & 0 & 2.27 \\
\hline 2.5 & 0 & 0 & 2.03 \\
\hline 2.7 & 0 & 0 & 1.81 \\
\hline 2.9 & 0 & 0 & 1.61 \\
\hline 3.1 & 0 & 0 & 1.43 \\
\hline 3.3 & 0 & 0 & 1.28 \\
\hline 3.5 & 0 & 0 & 1.14 \\
\hline 3.7 & 0 & 0 & 0.91 \\
\hline 3.9 & 0 & 0 & 0.72 \\
\hline
\end{tabular}
\quad
\begin{tabular}{|c||c|c|c|c|}
 \hline
 \multicolumn{5}{|c|}{DESI} \\ \hline
 $z$&BGS&LRG&ELG&QSO\\
 \hline 
\hline 0.1 & 2240 & 0 & 0 & 0 \\
\hline 0.3 & 240 & 0 & 0 & 0 \\
\hline 0.5 & 6.3 & 0 & 0 & 0 \\
\hline 0.7 & 0 & 48.7 & 69.1 & 2.75 \\
\hline 0.9 & 0 & 19.1 & 81.9 & 2.60 \\
\hline 1.1 & 0 & 1.18 & 47.7 & 2.55 \\
\hline 1.3 & 0 & 0 & 28.2 & 2.50 \\
\hline 1.5 & 0 & 0 & 11.2 & 2.40 \\
\hline 1.7 & 0 & 0 & 1.68 & 2.30 \\
\hline
\end{tabular}
\label{ap:dataJPASDESI}
\end{center}
\begin{center}
\begin{tabular}{|c||c|}
\hline
 \multicolumn{2}{|c|}{Euclid$_{\mathrm{ph}}$}\\
\hline $z$ & ELG \\
\hline 0.3 & 7440 \\
\hline 0.5 & 6440 \\
\hline 0.7 & 5150 \\
\hline 0.9 & 3830 \\
\hline 1.1 & 2670 \\
\hline 1.3 & 1740 \\
\hline 1.5 & 1070 \\
\hline 1.7 & 620 \\
\hline 1.9 & 341 \\
\hline 2.1 & 178 \\
\hline 2.3 & 88.3 \\
\hline 2.5 & 41.8 \\
\hline
\end{tabular}
\quad
\quad
\begin{tabular}{|c||c|}
 \hline
 \multicolumn{2}{|c|}{Euclid$_{\mathrm{sp}}$}\\

 \hline  $z$ & ELG  \\
\hline \hline 1.0 & 68.6 \\
\hline 1.2 & 55.8 \\
\hline 1.4 & 42.1 \\
\hline 1.6 & 26.1 \\
\hline
\end{tabular}
\label{fig:dataEUCLID}
\end{center}
\bibliographystyle{utcaps}

\bibliography{biblio.bib}


\end{document}